\documentstyle[aps,epsfig,amsfonts,epsf]{revtex}

\def\comment#1{}

\def\cm#1{}

\newcommand{\p}{\partial}
\newcommand{\f}[2]{\frac{#1}{#2}}
\newcommand{\be}{\begin{eqnarray}}
\newcommand{\ee}{\end{eqnarray}}
\newcommand{\beqn}{\begin{eqnarray}}
\newcommand{\eeqn}{\end{eqnarray}}

\newcommand{\Zm}{Z_{m^2}}
\newcommand{\Zg}{Z_{u}}
\newcommand{\Zphi}{Z_{\phi}}
\newcommand{\ubar}{\bar{u}}
\newcommand{\uhat}{\hat{u}}
\newcommand{\Litwo}{\,\mbox{Li$_{2}$}}

\comment{\title{Minimal renormalization without $\epsilon$
expansion and critical
  behavior
  in three dimensions: analytical resummation of various two- and three-loop
  amplitudes and amplitude
  ratios via variational perturbation theory}}

\title{Three-loop critical exponents, amplitude functions and amplitude ratios
 from variational perturbation theory}

\author{H. Kleinert \and B. Van den Bossche\thanks{Alexander von Humboldt
 Fellow}\thanks{Collaborateur scientifique
 du FNRS, On leave from absence of
 Physique Nucl\'eaire Th\'eorique, B5, Universit\'e de Li\`ege Sart-Tilman,
 4000 Li\`ege, Belgium}}

\address{Institut f\"ur Theoretische Physik, Arnimallee 14 D-14195 Berlin,
 Germany}

\begin{document}

\maketitle

\begin{abstract}
We use variational perturbation theory
to calculate various universal amplitude ratios
above and below $T_c$ in
 minimally subtracted $\phi^4$-theory with $N$ components in
three dimensions. In order to
best exhibit the method
as a powerful alternative to Borel resummation techniques, we
consider only to two- and three-loops
expressions where our results
are analytic expressions.
For the critical exponents, we also
extend existing analytic expressions for two loops
to three loops.
\end{abstract}

\section{Introduction}

Recently, quantum mechanical variational perturbation theory
\cite{kleinertpi} has been successfully extended to quantum field
theory, where it has proven to be a powerful tool for determining
 critical exponents  in
 three  \cite{kleinert257,kleinert279,kleinert287}
as well as in $4-\epsilon$ dimensions
 \cite{kleinert263,kleinert295}. The
purpose of this paper it to apply
 this theory to amplitude ratios which can be measured
 experimentally. Their perturbation expansions
  suffer from the same asymptotic nature as those of the critical
  exponents, thus requiring delicate resummation procedures.
These have been the subject of  numerous studies,
of which we can only mention a few, by various groups. There are
two main approaches followed by various authors which we shall
divide according to their method into a Paris school and a Parisi
school.

 The Paris school follows Wilson´s ideas
\cite{wilson71,wilsonfisher} by considering epsilon  expansions
in $D=4-\epsilon$ dimensions, making use of the fact that in the
upper  critical dimension $D_{\rm up}=4$ the theory is scale
invariant. The results are at first power series in the
renormalized coupling constant $g$. For small $\epsilon$, the
coupling constant goes, in the critical limit of vanishing mass,
to a stable infrared fixed point $g\rightarrow g^*$, where
scaling laws are found \cite{zj76}. The position of the fixed
point is found as a power series in $\epsilon$ which makes   critical
exponents and amplitude functions likewise power series in $\epsilon$.
These series diverge. The large-order behavior
\cite{lipatov,zj77} suggests that these series are  Borel
summable \cite{zjbook,ks}. The exact $\epsilon$-expansions of the
critical exponents are known up to five loops
\cite{larin83,kleinert91}. They have been resummed with the help
of Borel transformations and analytic mapping methods in
Refs.~\cite{shirkov80,zj80,zj98}.

\comment{Several other methods of resummation are however
available. We have already mentioned the variational perturbation
theory \cite{kleinert263}. We should also add the Janke-Kleinert
\cite{jankekleinert} resummation and its extension due to Jasch
and Kleinert \cite{jaschkleinert}. All the techniques are
reviewed and their output compared in the book \cite{ks}, which
is the state of the art concerning the critical behavior of the
$\phi^4$ theory, as far as the critical exponents are concerned.
(This book is not only dedicated to the O$(N)$ symmetry but also
investigates the critical properties of a mixed O$(N)$-cubic
symmetry interaction.) }

The Parisi school follows Ref. \cite{parisi80} in studying
 perturbation expansions
directly in $D=3$ dimensions
\cite{nickel77,zj97,muenster96,muenster96b,ber85,ber87}. In the
original works, renormalization conditions are used according to
which renormalized correlation functions should behave for small
momenta like $G(p)\approx(p^2+m^2)^{-1}$. Recently, these
normalization conditions have been replaced
 by dimensional
regularization near $D=3$ to remove divergences (see for instance
\cite{muenster96b} which uses a regularization in
$D=3-\varepsilon$ dimensions). Contrary to the
$\epsilon$-expansions around $D_{\rm up}=4$, the system is no
treated near the dimension of  naive scale-invariance, and the
scaling properties are no longer obvious order by order in $g$.
In addition, singular terms violating Griffith's analyticity are
introduced which show up by amplitudes having unpleasant logarithmic
dependences on the coupling constant.

The universal amplitude ratios were first discussed in
\cite{aharony76} in the context of Wilson's renormalization group
approach, and by Bervillier \cite{ber76} within the  field
theoretic approach developed in \cite{zj76}. The experimentally
most easily accessible amplitude ratios are formed from the
amplitudes of the leading power behaviors of various physical
quantities in $T-T_c$ above and below the critical temperature
$T_c$. A typical example, and one of the best measured amplitude
ratios, is for the specific heat of superfluid helium above and
below $T_c$. It was obtained in a zero-gravity experiment by Lipa
et al.~\cite{lipa96}, who parameterized the specific heat as
follows (we use the second of the references quoted
in~\cite{lipa96}):
\be
C^{\pm}=A^{\pm}|t|^{-\alpha}(1+D|t|^{\Delta}+E|t|^{2\Delta})+B,
\qquad t=T/T_c-1, \label{lipaparam} \ee
with  $\alpha=-0.01056\pm0.0004, \Delta=0.5,
A^+/A^-=1.0442\pm0.001, A^-=525.03, D=-0.00687, E=0.2152$ and
$B=538.55$ (J/mol K). This parametrization is an approximation to
the Wegner expansion form
\beqn
F=F_{\pm}|t|^{\chi}\Big(1&+&a_{0,1}|t|^{\Delta_0}+a_{0,2}|t|^{2\Delta_0}+
a_{0,3}|t|^{3\Delta_0}+\cdots
\nonumber\\
&+&a_{1,1}|t|^{\Delta_1}+a_{1,2}|t|^{2\Delta_1}+a_{1,3}|t|^{3\Delta_1}+
\cdots\Big) \eeqn
\comment{\nonumber\\
&\vdots&\nonumber\\
\Big)&& \eeqn}
with $\chi$ a combination of critical exponents and $F_{\pm}$
denoting the leading amplitude above and below $T_c$,
respectively. Compared to this general Wegner expansion, higher
powers in $\Delta_0\equiv\Delta$ have been neglected
in~(\ref{lipaparam}), as well as daughter powers $\Delta_i,
i\ge1$. This will be also the case in the present work, where we
shall take into account only one exponent $\Delta$, related to
$\omega$ by the relation $\Delta=\omega\nu$.

Further amplitude ratios are formed from the amplitudes $a_{i,j}$
of the nonleading power behaviors in $T-T_c$, the so-called
confluent terms or crossover functions, these also being universal
quantities \cite{wegner72,nicoll85}. They are known up to  three
loops. None of them  will be examined here.

Apart from critical exponents and  amplitude ratios,  experimental
observations show that the equation of state and the free energy
have a simple scaling form of the Widom type, whose
field-theoretic explanation can be found in various textbooks
 \cite{zj76,zjbook,ks}. For example, the free energy of a system
with magnetization $M_B$ may be represented near $T_c$ by
$F(t,M_B)= |t|^{2-\alpha}f(|t|/M_B^{1/\beta})$, with $\alpha$ and
$\beta$ being critical exponents and $t$ is the relative distance
to the critical temperature. The scaling equation of state has
been calculated in $\epsilon$-expansions to order $\epsilon^2$
for general O$(N)$-symmetry \cite{wilson72} and to order
$\epsilon^3$ for the Ising model  ($N=1$) \cite{wallace73}.

\comment{corresponding\footnote{The case $N=0$ is related to
polymer physics, $N=2$ to the Helium superfluid transitions,
$N=3$ to the classical Heisenberg model describing isotropic
magnets and $N=4$ is related to the Higgs phase transition at
finite temperature. The Ising model $N=1$ describes binary
fluids, liquid-vapor transitions and
antiferromagnets.\label{footnotephysicalmodel}}}

\subsection{Perturbative calculation of amplitude ratios}

Amplitude ratios relate the properties of the disordered phase,
which are easy to calculate, to those of the ordered phase,
which are much harder to derive. Several methods have been
proposed to connect the two phases. One of them is due to
 Bagnuls and Bervillier \cite{ber85},
and was applied further in \cite{ber87}. A similar procedure was
followed in \cite{muenster96,muenster96b} for the amplitude ratio
of correlation lengths, which had been omitted by Bagnuls and
Bervillier. Calculations in three dimensions are usually
numerical \cite{nickel77,ber85,ber87}, although low orders can be
treated analytically (see \cite{muenster96,muenster96b} for
analytic three-loop results). Such analytic studies are important
since they offer insight into the nonanalyticity with respect to
the coupling constant. The amplitude ratio found in
\cite{muenster96,muenster96b} is restricted to the Ising case
$N=1$.  The same is true for \cite{rajantie96},  which includes
all  diagrams up to three loops.

All power series are divergent and require resummation.
Numerically, this has been done for the Ising model in
\cite{zj97} to five loops for the critical exponents, various
amplitude ratios, and the equation of state. Reference \cite{zj97} also
contains comparisons between the results of different groups
(both for $D=3$ and $D=4-\epsilon$), with experiments and with
high-temperature series. For the most up-to-date work, see
 \cite{zj98}, which besides the critical exponents and
amplitude ratios for the Ising model gives also the critical
exponents  for general O$(N)$ symmetry.

An other approach has been  followed by Dohm and collaborators in
Aachen \cite{dohm85,dohm89,dohm90a} who proposed to use an
analytic renormalization scheme in the form of minimal
subtraction when  working in  $D=3$ dimensions.
The use of the minimal subtraction
scheme in field theories at fixed dimensions $2<D<4$  has  one
important advantage:   the renormalization constants are the same
 in both the symmetric phase with $T>T_c$  and  the ordered phase
with $T<T_c$. The renormalization constants are
 power series in the renormalized coupling constant with
coefficients which are poles in $\epsilon$ up to the given order
of the perturbative series

\be
Z=1+\sum_{i=1}^{L}a_{i}g^i, \qquad
a_i=\sum_{j=1}^ib_j\epsilon^{-j}.
\ee

\comment{In contrast to $\epsilon$-expansion around four dimensions,
 $\epsilon$ has here  a fixed value $\epsilon=4-D$.}

The most important property of this scheme is that
the mass does not enter explicitly the expansions, which can therefore be
used on both side of
 $T_c$. Since the critical exponents are
related to the renormalization constants, the mass independence
of the $Z_i$ implies a clear decomposition of the correlation
functions into amplitude functions and power parts. Working in
three dimensions,
 there is a prize to pay:
logarithmic singularities in the coupling constant. They can be removed
using suitable  length scales. This may be the
physical length scale $\xi^+$ above $T_c$, and an other length scale
$\xi_-$ related, in the critical regime,
to the longitudinal mass below  $T_c$. Since they are not exactly equals,
 the Aachen group call the length scale $\xi_-$ a pseudolength. A
precise definition of $\xi_-$ has been given in \cite{dohm85}. With different
collaborators, Dohm has applied this scheme to derive various
critical exponents and renormalization-group functions above $T_c$
\cite{dohm89}, to calculate the heat capacity, the order
parameter and the superfluid density (both above and below $T_c$),
as well as some useful universal combination of observable quantities
 \cite{dohm90a}. So far,
these works have been limited to low orders. The
Ising model is the simplest system,  since it contains no
massless Goldstone modes which cause extra infrared singularities
at intermediate stages of perturbative calculations of the
thermodynamical quantities on the coexistence curve where the
 external magnetic field vanishes.
The infrared singularities are the reason why the analytical equation
 of state and amplitude
functions below $T_c$ have been restricted
\cite{ber76,dohm90a,sndref72} to two loops for general $N$. These
extra infrared singularities, which  cancel at the end of the
calculations, are caused by the physical singularities of the
transverse
\comment{and longitudinal}
susceptibility. Being physical, they remain at the end.
Due to these difficulties, numerical
studies up to five loops below $T_c$, with accurate Borel
resummation are available only for the Ising case
\cite{zj97,ber87,dohm92,dohm98}. Only analytic
three-loop calculations for the thermodynamic quantities below
$T_c$ have become recently available for the general  O$(N)$-system
\cite{dohm99}. Based on these, calculations in which some contributions
were evaluated up to five loops
were done for amplitude ratios at $N=2$ and $N=3$ \cite{dohm98},
 proceeding as follow: Amplitude
functions for the heat capacity were calculated using the three-loop
result of \cite{dohm99} and   five-loop results
for the vacuum renormalization constant \cite{dohm98,kastening98}
and the critical exponent $\alpha$. For $\alpha$, use has been
made  of the values given in \cite{zj80} for $N=1$, of the
value given in the first of Ref.~\cite{lipa96} for $N=2$ (this being
the initial result of the space shuttle experiment, which was
subsequently corrected),
and of the value given in Ref.~\cite{zj85} for $N=3$. Since
then, the works
\cite{kleinert257,kleinert279,kleinert287,kleinert263,kleinert295,ks,zj98}
have appeared and seem to be the best
available references concerning resummed data. Although this is
not the main subject of this paper, it is interesting to see in
which way the new values of $\alpha$ affect the amplitude ratios
 of the heat capacity given in \cite{dohm98}.
This will be done in Section \ref{section4}.

In the following, we shall calculate amplitude ratios with the
help of   Kleinert's variational perturbation theory
\cite{kleinertpi,kleinert257,kleinert279,kleinert287,kleinert263,kleinert295,ks}.
To exhibit the method most clearly, we shall base our
study  on analytical
results only. This will restrict us to the level of three loops.
Working at such low orders, the accuracy of our resummed values cannot compete
with some existing  five-loop calculations. For this reason, we shall
not include nor discuss  error bars in the final results.

To illustrate the method of variational perturbation theory, we
shall first show how to obtain analytic expressions for the
critical exponents, thus extending an earlier two-loop analytic
calculation in Ref. \cite{kleinert263}.
 After this, we apply the procedure to
 amplitude ratios of various experimental quantities. The
 critical exponents are computed directly from the renormalization
 constants of the theory. In the minimal subtraction scheme,
the renormalization constants have only pole terms in $\epsilon$.
\comment{: no
$\epsilon$ expansion is involved in their definition (although
such an expansion may be used for the practical evaluation of
these exponents).}
For the amplitude functions, this is no longer
true: in a $D=4-\epsilon$ approach, they have to be expanded in
$\epsilon$. For this reason it is not a priori clear at which
level the variational method has to be applied. For the purpose of
showing the power of the method to resum amplitude ratios, it is
then better to calculate amplitude ratios in three dimensions.
A resummation of amplitude ratios within the $\epsilon$-expansion method
is postpone to a later publication \cite{kleinertvdb00}. We shall
also consider only the expansions of the
Aachen group, especially their analytical two-loop \cite{dohm97}
and three-loop \cite{dohm99} expansions.
\comment{because, as already
mentioned, there is a clear decomposition of correlation
functions between amplitude functions and exponential parts.}
As a
bonus, since the renormalization constants are the same (apart
for trivial coefficients coming from the respective conventions)
in the minimal subtraction scheme in $D=4-\epsilon$ dimensions
and in fixed $D=3$ dimensions, the critical exponents will be  the same
 in variational perturbation theory.
This will be shown explicitly below.

The paper is organized as follow. In Section \ref{section2}, we
define the model and the conventions. In Section \ref{section3},
we briefly review the strong-coupling approach and apply it to  the
evaluation of the critical exponents at the level of two and three
loops, extending the results of Ref.~\cite{kleinert263}. Section
\ref{section4} is the main part of this paper, where we show how
the strong-coupling limit of various amplitudes and amplitude
ratios are determined. In Section~\ref{maxinfo}, we use the latest
available value for the exponents $\alpha$ and $\nu$  \cite{kleinert257,kleinert279,kleinert287,kleinert263,kleinert295,ks,zj98}
\comment{as given by
Guida and Zinn-Justin \cite{zj98},}
to calculate the amplitude
ratio of the heat capacity and  the univeral
combination $R_C$ (constructed from the leading amplitudes of the heat
 capacity, the order parameter and the susceptibility above $T_c$),
 for $N=0,\cdots,4$,
and to calculate the amplitude ratio of the susceptibilities in
the Ising model ($N=1$).
\comment{Since these values of $\alpha$
and $\nu$ were not used in \cite{dohm98}, we also use them to
evaluate these amplitude ratios using the partial five loops
calculation of the later reference.}
Finally, we draw our
conclusion in Section \ref{section5}. For completeness, we
have added an Appendix containing all
 formulas taken from other publications, and calculations related to them.

\section{Model and conventions}
\label{section2}

The critical behavior of many different physical systems can be
described by an O$(N)$-symmetric $\phi^4$-theory. In
particular, the case $N=0$ describes polymers, $N=1$  the Ising
 transition (a universality class which comprises
binary fluids, liquid-vapor transitions and antiferromagnets),
$N=2$  the superfluid Helium transition, $N=3$
isotropic magnets (transition of the Heisenberg type), and $N=4$
 phase transition of Higgs fields at
finite temperature.  In the presence of an external field $h_B$,
the field energy is given by the Ginzburg-Landau functional
\be {\cal H}= \int d^Dx\left[ \f{1}{2}(\nabla
\phi_B)^2+\f{1}{2}r_0\phi_B^2+u_B(\phi_B^2)^2-h_B.\phi_B \right].
\label{glfunctional}
\ee
To facilitate  comparisons with the results of the Aachen group
\cite{dohm99,dohm97},  we use the same normalizations. The fields
$\phi_B$ and the external magnetic field $h_B$ have $N$
components, $u_B$  is the bare coupling constant, and $r_0$ a
bare mass term, to be specified later. \comment{Finite cut-off
effects are supposed to be to be negligible:
$\Lambda\rightarrow\infty$ and} The integrals are evaluated in
dimensional regularization. In dimension $D=3$, $\phi^4$-theory
is superrenormalizable. This means that only a finite number of
counterterms have to be added in order to make observables finite.
More economically, the divergencies can be removed by a shift of
the mass term and reexpanding in $r_0-r_{0c}$, where $r_{0c}$ is
the critical value
 of $r_0$. In $\epsilon$-expansions, $r_{0c}$ vanishes. Near the critical
 temperature, $r_0$ behaves like  $r_{0c}+a_0t$, where $t$ is the
reduced temperature  $(T-T_c)/T_c$. When working
 near $D=3$ dimensions, it is  possible to use a
simplified shift  $\delta r_0$ that only contains the $D=3$ pole
of $r_{0c}$ (and not the poles at $D_l>3$ with $l=3,4,5,\cdots$,
where $D_l\equiv4-2/l$).  For convenience, we write the
differences as a new mass term: $r_0-r_{0c}=m_B^2$ and
$r_0-\delta r_0={m'}_B^2$. In this way, we arrive to a new bare
theory, with a mass term ${m'}_B^2$ which may be considered as the
physical square mass of the theory.  The introduction of the mass
$m'_B$ makes the theory finite. It has however to be
distinguished from the mass, field and coupling constant
renormalization which still has to be performed:  this
latter renormalization, related to the introduction of the
renormalization constants $Z_i$, is nothing else than a change of variables
 reflecting the fundamental scale-invariance hypothesis of the
renormalization group approach. The distinction between the two
steps -- making the theory finite and renormalizing --
is irrelevant in
$D=4-\epsilon$ dimensions because  $r_{0c}=0$ at $\epsilon=0$:
Finiteness of the theory
and the renormalization program are more intimately related than
in $D=3$ dimensions.
For a thorough discussion
of the difference between the renormalization in $D=4-\epsilon$
and fixed $D=3$ dimensions, see \cite{ber85,ber87}, in particular
p. 7215 in \cite{ber85}.

Within the minimal renormalization scheme, the renormalization
constants $Z$, which are introduced to remove the  poles at $D=4$, are given
by
\beqn
m_B^2&=& m^2 \f{\Zm}{\Zphi},\label{eqmB}\\
A_Du_B&=&\mu^{\epsilon}\f{\Zg}{\Zphi^2}u,\label{eqrenconstant}\\
\phi_B&=&\Zphi^{1/2}\phi,\label{eqphi}
 \eeqn
the quantities on the right-hand-side being the renormalized
ones. In Eq.~(\ref{eqrenconstant}), $\mu$ is an arbitrary reference mass scale
and
\be A_D=\Gamma(1+\epsilon/2)\Gamma(1-\epsilon/2)\bar{S}_D, \mbox{ with }
\bar{S}_D=
\f{2\pi^{D/2}}{\Gamma(D/2)(2\pi)^D}
\label{defAD} \ee
is a convenient geometric factor. The number $\bar{S}_D$ is equal
to $S_D/(2\pi)^D$ where $S_D$ is the surface of a sphere in
$D$-dimensions. Since $A_D$ goes to $\bar{S}_D$ when
$D\rightarrow4$, the renormalization constants have the same form
\cite{kang76} in $D=3$ as in $D=4-\epsilon$, and the resummation
for the critical exponents is identical for the two approaches.
This will be made clear below. For the amplitude calculations,
however, things are different: If the expansions are truncated at
some order, they turn out to depend on the difference between
$A_D$ and $\bar{S}_D$. Rather than saying that the normalization
of $A_D$ is
 a matter of convenience to simplify the $D$-dependence
of lower order results \cite{dohm85,dohm89}, we shall see that
the use of the geometric factor~(\ref{defAD}) improves low-order
results: For example, the one-loop expansion of the amplitude
function for the order parameter is identical to the zero-loop
order  \cite{dohm90a}.

With these conventions and notations, the renormalization
 constants in minimal subtraction are given up to
three loops by \cite{ks,larin83,kleinert91}
\beqn \Zm&=&1+ \f{4(N+2)}{\epsilon}u+8(N+2)
\left[\f{2(N+5)}{\epsilon^2}-\f{3}{\epsilon}\right]u^2
\nonumber\\
&&\mbox{}+8(N+2)\left[\f{8(N+5)(N+6)}{\epsilon^3}
-\f{4(11N+50)}{\epsilon^2}
+\f{31N+230}{\epsilon}\right]u^3,\label{EqZm}\\
\Zg&=&1+\f{4(N+8)}{\epsilon}u+16
\left[\f{(N+8)^2}{\epsilon^2}-\f{5N+22}{\epsilon}\right]
u^2
\nonumber\\
&&\mbox{}+\f{8}{3}\left[
\f{24(N+8)^3}{\epsilon^3}-\f{16(N+8)(17N+76)}{\epsilon^2}
+\f{96\zeta(3)(5N+22)+35N^2+942N+2992}{\epsilon}
\right]u^3,\label{EqZg}\\
\Zphi&=&1-\f{4(N+2)}{\epsilon}u^2-
\f{8}{3}(N+2)(N+8)\left(\f{4}{\epsilon^2}-\f{1}{\epsilon}
\right)u^3. \label{EqZphi}\eeqn
 They are related to that in Ref.~\cite{ks} by the replacement
$u\rightarrow g/12$. This factor comes from the different
coefficient of the coupling term $u\rightarrow g/4!$
in~(\ref{glfunctional}) and the fact that a factor $1/(4\pi)^2$
is absorbed in the definition of $g$ in \cite{ks}, whereas a
factor $A_{D=4}=1/(8\pi^2)$ is included here.

These renormalization constants serve to calculate
all critical exponents including the exponent $\omega$ which characterizes the
approach to scaling.
 This is the subject of Section~\ref{section3}
 in which we illustrate the working  of variational
perturbation theory.

\section{Exact critical exponents up to three loops}
\label{section3}

Variational perturbation theory has been developed for the
calculation of critical exponents in  \cite{kleinert257} and
\cite{kleinert263}
 in $D=3$ and  $D=4-\epsilon$ dimensions, respectively. A review can be found
 in the textbook
\cite{ks}. So we need to recall here only the main steps of the
procedure.

Let $f_L(\ubar_B)$ be the partial sum of order $L$  of a power series
\be f\approx f_L(\ubar_B)=\sum_{i=0}^Lf_i\ubar_B^i.
\label{fntoresum}\ee
In the present context,
\be
\ubar_B=u_B\mu^{-\epsilon}A_D
\label{ubardef}
\ee
with $D=3$ and $\epsilon=1$, i.e., $\ubar_B=u_B/(4\pi\mu)$. The
mass scale $\mu$ will be specified later. As seen from
Eq.~(\ref{eqrenconstant}), this scale leads to a dimensionless
coupling constant  $\ubar_B$. We assume that in
Eq.~(\ref{fntoresum}), the ultraviolet divergencies have been
removed. In $D=3$ dimensions, this is achieved by  working with
$m_B^2$ instead of $r_0$. However, $r_{0c}$ is a nonperturbative
quantity in three dimensions, and working with $m_B^2$ or
${m'}_B^2$ generates nonanalyticities due to the presence of
logarithms of the coupling constant. These will be removed by the
introduction of the correlation length $\zeta_+$ above $T_c$ and
of the length $\zeta_-$ below $T_c$, see \cite{dohm90a}. The mass
scale $\mu$ will be identified with the inverse of these
correlation lengths $\zeta_{\pm}^{-1}$ in the two phases.
  Since the correlations lengths go to
infinity like $|t|^{-\nu}$ as the critical point is approached,
the series have to be evaluated in the limit of an infinite
dimensionless bare coupling constant $\ubar_B$. In the
renormalization group approach, this regime is studied by mapping
the expressions into a regime of finite renormalized quantities
using the renormalization constants~(\ref{eqmB})--(\ref{eqphi}).
If we can find directly the strong-coupling limit, this
renormalization is avoidable. To understand this, consider
 the relation between the renormalized and the bare
coupling constant at the one-loop order
$u=u_B\mu^{-\epsilon}-c/\epsilon(u_B\mu^{\epsilon})^2$,
 where $c$ is a constant.
 At the critical point, $\mu\rightarrow0$, or
 $\ubar_B\rightarrow\infty$, and the series expansion breaks down.
 If we sum a ladder of loop diagrams, we obtain  $1/u=1/(u_B\mu^{-\epsilon})+c/\epsilon$.
Now critical theory can easily be reached to give a renormalized
 $u^*=\epsilon/c$. A strong-coupling expansion in the bare coupling
will turn out to give the same result. From our point of view, the
renormalization group approach is simply a specific procedure of
evaluating  power series in the strong-coupling limit.

In $D=4-\epsilon$ dimensions, the situation is slightly more
involved since
 renormalization is also necessary to obtain UV-finite
quantities, the mass shift $r_0-r_{0c}$ not being sufficient for
this goal  as in the
 superrenormalizable case $D=3$, since $r_{0c}=0$ as
 $\epsilon\rightarrow0$. As far as this paper is concerned, we
 shall make use of the fact that
$D=3$ and $D=4-\epsilon$ dimensions series expansions in term of
renormalized quantities are available in the literature. These
will be converted back to bare expansion, using the inverse of
Eqs.~(\ref{eqmB})--(\ref{eqphi}). For $D=3$ dimensions, this
expresses all physical quantities in powers of $u_B/\mu$. The
mass scale $\mu$ is identified with $\zeta_{\pm}^{-1}$ in the
disordered or ordered phase, respectively. In $D=4-\epsilon$
dimensions, the critical theory is obtained by identifying
$\mu\rightarrow m$ with the renormalized mass $m$ in the
disordered phase. In a subsequent publication
\cite{kleinertvdbBARE}, we will show how to perform directly a
calculation in term of UV-finite bare quantities in
$D=4-\epsilon$. In this way,  the renormalization procedure is
superfluous, our sole problem being the evaluation of the
expansions in the limit of infinite coupling constant.

\comment{\footnote{Some important hints
   from renormalization group analysis are however necessary, namely the fact
   that the critical exponents go to a constant in the strong bare coupling
   limit, with the same approach to scaling $\omega$.}.}

\comment{In the following, we shall first be concerned with the calculation of
  critical
exponents which can be obtained from the disordered phase alone.}

Inverting Eq.~(\ref{eqrenconstant}), we have the expansion
\beqn u&=&\ubar_B\Bigg\{1-\f{4(N+8)}{\epsilon}\ubar_B +8\left[
\f{2(N+8)^2}{\epsilon^2}+\f{3(3N+14)}{\epsilon}
\right]\ubar_B^2\nonumber\\
&&\mbox{}-8\left[ \f{8(N+8)^3}{\epsilon^3}
+\f{32(N+8)(3N+14)}{\epsilon^2}
+\f{96\zeta(3)(5N+22)+33N^2+922N+2960}{3\epsilon}
\right]\ubar_B^3\Bigg\}. \label{gseriesgB}
 \eeqn
The expansion~(\ref{gseriesgB})
has the same strong-coupling limit
 in $D=3$ and $D=4-\epsilon$ dimensions, and it does not matter
 that
$\mu=\zeta_{\pm}^{-1}$ for $D=3$ or $\mu=m$ for $D=4-\epsilon$
since both quantities go to zero in the critical limit with the
same power $|t|^{-\nu}$. With relation~(\ref{gseriesgB}) between
$u$ and $\ubar_B$,  we obtain the bare coupling expansion of the
renormalized square mass and fields:
\beqn m^2&\equiv&Z_r^{-1}m_B^2=m_B^2\Bigg\{1-
\f{4(N+2)}{\epsilon}\ubar_B+4(N+2)
\left[\f{4(N+5)}{\epsilon^2}+\f{5}{\epsilon}\right]\ubar_B^2
\nonumber\\
&&\mbox{}-16(N+2)\left[\f{4(N+5)(N+6)}{\epsilon^3}
+\f{53N+274}{3\epsilon^2}+\f{(5N+37)}{\epsilon}\right]
\ubar_B^3\Bigg\},\label{mseriesgB}\\
\phi&\equiv&\Zphi^{-1/2}\phi_B=\phi_B\left[1+\f{2(N+2)}{\epsilon}\ubar_B^2
-\f{4}{3}(N+2)(N+8)\left(\f{8}{\epsilon^2}+\f{1}{\epsilon}
\right)\ubar_B^3\right]\label{phiseriesgB}. \eeqn
These two expressions are sufficient to calculate the critical
exponents $\nu$ and $\gamma$ and, via scaling relations, all
other exponents. Note that the value of the renormalized coupling
constant at the critical point $u^*$ is not needed to obtain
$\nu$ and $\gamma$. The expansion~(\ref{gseriesgB}) is however
useful for  obtaining an accurate exponent $\omega$  of the
approach to scaling. It was pointed out in
\cite{kleinertwithoutbeta} that $\omega$ can also be deduced from
the expansions of $\nu$ and $\gamma$. However, to reach the same
accuracy, this requires always one more loop compared to the loop
order we are interested in. For this reason, we shall take the
advantage of  Eq.~(\ref{gseriesgB}), whose three-loop order
contains all necessary information to get $\omega$ to that given
order.

\comment{The basic reason for the increase of the loop order is
due to a power counting ($u$ starts like $\ubar_B$) and will be
made clearer below.}

\subsection{Method}
\label{method}

Starting  from Eq.~(\ref{fntoresum}), we follow
\cite{kleinert257,kleinert263,ks} to write
 its strong-coupling limit as
\be f_L(\ubar_B\rightarrow\infty)=\mbox{opt}_{\uhat_B}\left[
\sum_{i=0}^Lf_i \uhat_B^i\sum_{j=0}^{L-i}{-i\omega/\epsilon\choose
j}(-1)^j \right].
\label{fresummed} \ee
The symbol $\mbox{opt}_{\uhat_B}$ denotes optimization with
respect of $\uhat_B$.  This expression holds provided it yields a
nonzero constant. This limit will be denoted  by $f^*$:
\be f(\ubar_B\rightarrow \infty)=
f^*+c_0\ubar_B^{-\omega/\epsilon}+O(\ubar_B^{-2\epsilon/\omega}),\label{fstar}
\ee
where $c_0$ is a constant. The optimalization is supposed to make
$f$ depend minimally on $\uhat_B$. In practice, this amounts to
taking the first derivative to zero (odd orders) or, when it
yields complex results, to taking the second derivative to zero
and selecting turning points.

After having determined the optimum at various order $L$, it is
still necessary to extrapolate the result to infinite order
$L\rightarrow\infty$. The general large-$L$ behavior of the
strong-coupling limit has been derived from an analysis in the
complex plane in \cite{kleinert257,ks}:
\be f_L^*\approx f^*+c_1\exp(-c_2L^{1-\omega}),
\label{largeL}\ee
with constants $c_1$ and $c_2>0$. Knowing this behavior, a
graphical  extrapolation procedure may be used to find
$f^*_{\infty}=f^*$.

To apply the above algorithm to critical exponents, we proceed as follows:
Let $W_L$ be a function obtained from  perturbation theory. It has
an expansion
\be W_L(\ubar_B)=\sum_{i=0}^LW_i\ubar_B^i. \label{functionWL}
\ee
Suppose  that we also know  this function has a leading power
behavior $\ubar_B^{p/q}$ for large $\ubar_B$. The power $p/q$ is
given by a logarithmic derivative
\be \f{p}{q}=\f{d\log W_L}{d\log \ubar_B}. \label{poverq}\ee
The right-hand-side is a power series representation of a
function of the type (\ref{fntoresum}), with $p/q$ being $f^*$
and the approach to $f^*$ in the form of powers
$\ubar_B^{-\omega/\epsilon}$. Equation~(\ref{poverq}) will be used later for
 the determination of the critical exponents. If the series~(\ref{functionWL})
goes to a constant in the strong-coupling limit, the exponent $p$ is
 vanishing, and we are left with
\be
\f{d\log W_L}{d\log \ubar_B}=0.
\label{WLprimezero}
\ee
This equation can be solved for $q$, i.e., for $\omega$. Note
that~(\ref{poverq}) strictly holds for $p\ne0$. However, it can
be shown that this equation may be used also for $p=0$, i.e.,
that~(\ref{WLprimezero}) is a consistent equation for functions
which go to a constant in the strong-coupling limit. This is
explained in Appendix~\ref{appendixmethod}. In the following, we
shall directly use~(\ref{poverq}) and~(\ref{WLprimezero}) for
two- and three-loop expansions where everything can be calculated
analytically. We give below the associated formulas resulting
from Eq.~(\ref{fresummed}). Setting $\rho=1+\epsilon/\omega$, we
find for $L=2$:
\be f_{L=2}^*=\mbox{opt}_{\uhat_B}\left(
f_0+f_1\rho\uhat_B+f_2\uhat_B^2\right)
=f_0-\f{\rho^2}{4}\f{f_1^2}{f_2},
\label{2loops}
\ee
while the three-loop results $L=3$ leads to
\be f_{L=3}^*=\mbox{opt}_{\uhat_B}\left(
f_0+\bar{f}_1\uhat_B+\bar{f}_2\uhat_B^2+f_3\uhat_B^3\right)
=f_0-\f{1}{3}\f{\bar{f}_1\bar{f}_2}{f_3}\left(1-\f{2}{3}r
\right)+\f{2}{27}\f{\bar{f}_2^3}{f_3^2}(1-r), \label{interm3loops}
\ee
where
$\bar{f}_1=f_1\rho(\rho+1)/2,\bar{f}_2=f_2(2\rho-1),
r=\sqrt{1-3\bar{f}_1f_3/\bar{f}_2^2}$.
If the square root is imaginary, the optimal value is given by
the unique turning point. Practically, and this is a virtue of
the analytic result, this square root is always imaginary for
$D=3$, at least as for the exponent $\omega$. The
turning point condition leads to
\be f_{L=3}^*=
f_0-\f{1}{3}\f{\bar{f}_1\bar{f}_2}{f_3}+\f{2}{27}\f{\bar{f}_2^3}{f_3^2},
\label{3loops}
\ee
i.e.,  to same expression as Eq.~(\ref{interm3loops}), but with
$r=0$.  In the case $D=4-\epsilon$ with $\epsilon\rightarrow 0$,
$r$ is real. However, for $\omega$ the $\epsilon$-expansion of $r$
produces higher orders  in $\epsilon$ than the three-loop
approximation admits. Then, in both $D=3$ and  the
$\epsilon$-expansion,~(\ref{3loops}) is the relevant equation. A
word of  caution is nevertheless necessary: The positive root $r$
of
\be  \uhat_B^*=\f{f_2}{3f_3}(-1\pm r) \label{signofRoot}\ee
 has to be chosen in order to match the three-loop result with the
two-loop one in the limit $f_3\rightarrow0$. Doing so, it must be
assumed  that $f_2$ and $f_1$ are nonvanishing. When optimizing
with $f_2=0$,  it is immediate to show that if $f_1 f_3>0$, then
the optimum corresponds to $\uhat_B^*(f_2\rightarrow0)=0$ and
$f_{L=3}^*=f_0$. The other possibility, $f_1=0$, is also
interesting since it occurs in the determination of the exponent
$\eta$. It can be verified that $f_1=0$ implies   taking the
negative root $r=-1$, so that $\uhat_B^*(f_1\rightarrow0)=
-2\bar{f}_2/(3f_3)$ and $f_3^*=f_0+4\bar{f}_2^3/(27f_3^2)$. This
possibility has not been discussed in the  previous  works
\cite{kleinert257,kleinert263,ks}.

\subsection{Critical exponents}

After the introduction to the resummation method to be used in this
work, we can now turn to the actual determination of the critical
exponents.  We start from the definitions within the conventional
renormalization formalism of the functions
\beqn
\gamma_m&=&\left.\f{\mu}{m^2}\f{\p m^2}{\mu}\right|_B,\label{gammam}\\
\gamma_{\phi}&=&\left.\mu\f{\p}{\p\mu}\log\Zphi\right|_B,\label{gammaphi}\\
\beta_u&=&\left.\mu\f{u}{\p\mu}\right|_B, \label{betau}
\eeqn
which, in the critical regime $m_B^2\rightarrow 0$, render  the
critical exponents $\eta_m=\gamma_m^*$ and $
\eta=\gamma_{\phi}^*$ if  the first two equations are calculated
at the fixed point $u^*$ determined by the zero of the third
function $\beta_u$. The derivative of $\beta_u$ at $u^*$ is the
critical exponent of the approach to scaling
$\omega=\left.\p\beta_u/\p u\right|_{u^*}$.

Using the relation between   the bare coupling constant $u_B$ and
the reduced one  $\ubar_B$ given in Eq.~(\ref{ubardef}),
Eqs.~(\ref{gammam}) and~(\ref{gammaphi}) become
\beqn \eta_m&=&-\epsilon\f{d}{d\log\ubar_B}\log
\f{m^2}{m_B^2}=-\epsilon\f{d}{d\log\ubar_B}\log Z_r^{-1}, \label{etam}\\
\eta&=&\epsilon\f{d}{d\log\ubar_B}\log
\f{\phi^2}{\phi_B^2}=2\epsilon\f{d}{d\log\ubar_B}\log
\Zphi^{-1/2},\label{eta}\eeqn
where the renormalization constants $Z_r^{-1}$ and $\Zphi^{-1/2}$
have been explicitly given up to three loops in
Eqs.~(\ref{mseriesgB}) and~(\ref{phiseriesgB}), respectively. The
associated power series expansion in $\ubar_B$ of the exponents
$\eta_m$ and $\eta$ will now be treated with the help of the
formalism described in the previous section, up to two and three
loops.

\subsection{Critical exponents from two-loop expansions}

In order to calculate the two-loop expansions in the critical
strong-coupling limit, we need to know $\omega$ to this order.
\comment{As mentioned  before: it is not necessary to evaluate the
$\beta$ function since each series has the same approach to
scaling \cite{kleinertwithoutbeta}. However, if we do not use the
expansion for $\beta$ or  $u(\ubar_B)$, the series from which it
is derived, we would need  one more loop order
\cite{kleinertwithoutbeta}. It is therefore preferable to
use~(\ref{gseriesgB}).} This will be calculated from
Eq.~(\ref{gseriesgB}). Dividing this series by $\ubar_B$, we know
that the leading power behaviour as $\ubar_B\rightarrow\infty$ is
$-1$ since $u$ is supposed to go to the constant value $u^*$:
$\left.u\ubar_B^{-1}\right|_{\ubar\rightarrow\infty}=u^*\ubar_B^{-1}$.
Calculating the logarithmic derivative of~(\ref{gseriesgB}) and
expanding up to second order in $\ubar_B$, we have
\be \f{d}{d\log\ubar_B}\log
\f{u}{\ubar_B}=\f{-4(N+8)}{\epsilon}\ubar_B+ 16\left[
\f{(N+8)^2}{\epsilon^2}+\f{3(3N+14)}{\epsilon}\right]\ubar_B^2.
\label{dlogubarB2loops} \ee
We now apply formula~(\ref{intermpoverq}). Combining
with~(\ref{2loops}), we identify
\be -1=-\f{\rho^2}{4}\f{\left[-4(N+8)/\epsilon\right]^2}{16\left[
(N+8)^2/\epsilon^2+3(3N+14)\epsilon\right]},
 \ee
i.e.,
\be \f{\rho^2}{4}=1+3\epsilon\f{3N+14}{(N+8)^2},
\label{rhosquareover4} \ee
from which we can deduce $\omega$:
\be
\omega=\f{\epsilon}{\rho-1}=
\f{\epsilon}{-1+2\sqrt{1+3\epsilon(3N+14)/(N+8)^2}}.
\label{resummedomega} \ee
It is  identical to the result  obtained in \cite{kleinert263}.
As a check of~(\ref{resummedomega}), we verify that it reproduces
the well-known $\epsilon$-expansion
\be
\omega_{\epsilon}=\epsilon-3\epsilon^2(3N+14)/(N+8)^2.\label{omegaeps2}
\ee
We refer the reader to \cite{kleinert263,ks} for plots of the
function~(\ref{resummedomega}) as $\epsilon$ goes from 0 to 1,
and for a comparison with the unresummed $\epsilon$-expansion. The
strong-coupling limit of $\omega$ may also be calculated
from~(\ref{omegaovereps}) with an analytic expression different
from~(\ref{resummedomega}), although numerically they are
practically the same, and they certainly have the same
$\epsilon$-expansion \cite{kleinertwithoutbeta}.

This determination of $\omega$ illustrates what we said in
Section~\ref{section2}, that in the minimal renormalization
scheme the critical exponents lead to identical results in $D=3$
and $D=4-\epsilon$ dimensions. This will also be true for the
critical exponents to be calculated in the sequel
\cite{conventions}. For this reason, we shall always keep trace
of $\epsilon$ to facilitate the comparison, although our work is
in $D=3$ dimensions. Only for amplitude ratios to be calculated
later will such a comparison be impossible and $\epsilon$ be set
equal to 1 everywhere.

Knowing $\omega$, we can now determine the exponents $\eta$ and
$\eta_m$.
 According to~(\ref{etam}) and~(\ref{eta}), we take the
logarithmic derivative of~(\ref{mseriesgB})
and~(\ref{phiseriesgB}), reexpand the results up to the second
order in $\ubar_B^2$, and obtain
\beqn \eta_m&=&4(N+2)\ubar_B- 8(N+2)\left[
\f{2(N+8)}{\epsilon}+5 \right]\ubar_B^2,\label{etamgB}\\
\eta&=&8(N+2)\ubar_B^2.\label{etagB}
 \eeqn
Evaluating  $\eta_m$ in the strong-coupling limit in the same way
as $\omega$, i.e., following the algorithm~(\ref{fresummed}), we
find
\be \eta_m=\f{\rho^2}{4}\f{\left[4(N+2)\right]^2}{8(N+2)\left[
2(N+8)/\epsilon+5 \right]}
=\f{(N+2)}{(N+8)+5\epsilon/2}\left[\epsilon+\epsilon^2
\f{3(3N+14)}{(N+8)^2}\right]. \label{etam2loopsdirect} \ee
 For $\eta$, the situation is less clear. In \cite{kleinert257,kleinert263},
 it was argued that the two-loop result cannot
be computed from Eq.~(\ref{etagB}) since no linear term in
 $\ubar_B$ is
 present.  A
direct application of the resummation algorithm would give an
optimum $\uhat_B^*=0$, then a value $\eta=0$ at two-loop order.
This does not lead to the  correct  $\epsilon$-expansion,
according to which the  exponent  start with $\epsilon^2$, i.e.,
with a non-vanishing two-loop contribution.  To apply variational
perturbation theory, it is necessary to modify the procedure. In
Ref.~\cite{kleinert263}, this was done by considering a different
 critical exponent
\be \gamma&=&\nu(2-\eta),\label{expoGamma}
\ee
with
\be \nu&=&\f{1}{2-\eta_m}. \label{expoNu}
\ee
 To obtain their strong coupling limit, we
insert for $\eta_m$ and $\eta$ their perturbative expansions
(\ref{etamgB}) and~(\ref{etagB}), respectively, and  reexpand the
resulting ratios in power of $\ubar_B$ up to the second order.
This gives
\be \gamma&=&1+2(N+2)\ubar_B-4(N+2)\left[
\f{2(N+8)}{\epsilon}-(N-4) \right]\ubar_B^2.
\ee
The critical exponent $\nu$ itself has the expansion
\be
\nu&=&\f{1}{2}+(N+2)\ubar_B-2(N+2) \left[\f{2(N+8)}{\epsilon}
-(N-3) \right]\ubar_B^2.
\ee
The strong-coupling limits are, using $\rho^2/4$ from
Eq.~(\ref{rhosquareover4}),
\beqn \gamma&=&
1+\f{(N+2)}{2(N+8)-\epsilon(N-4)}
\left[\epsilon+\epsilon^2\f{3(3N+14)}{(N+8)^2}\right],\label{expogammaresummed}\\
\nu&=&\f{1}{2}\left\{1+\f{(N+2)}{2(N+8)-\epsilon(N-3)}
\left[\epsilon+\epsilon^2\f{3(3N+14)}{(N+8)^2}\right]\right\}.\label{exponuresummed}
\eeqn
Their $\epsilon$-expansion are in agreement with $D=4-\epsilon$
results \cite{kleinert263,ks}. From these expressions we can
recover $\eta$ using the relation $\eta=2-\gamma/\nu$. The result
has now the correct $\epsilon$-expansion:
\be \eta=\f{N+2}{2(N+8)^2}\epsilon^2
\label{eta2loopsepsilon}.\ee

This calculation of $\eta$ via $\nu$ and $\gamma$ was made in
\cite{kleinert257,kleinert263} to compensate the lack of a linear
term in~(\ref{etagB}). Let us point out that, even if the
$\epsilon$-expansion is not recovered, it is nevertheless hidden
in a direct resummation of~(\ref{etagB}) to $\eta=0$. To see this,
we add a small dummy linear term $\zeta u$, to the defining
equation~(\ref{eta}), leading to the expansion
\be
\eta=\zeta\ubar_B+\left[8(N+2)-\zeta\f{4(N+8)}{\epsilon}\right]\ubar_B^2.
\ee
Using~(\ref{2loops}) and~(\ref{rhosquareover4}), this leads to
the strong-coupling value
\be \eta=-\f{\rho^2}{4}\f{\zeta^2}{8(N+2)- 4(N+8)\zeta/\epsilon},
\label{etazeta}
\ee
which is zero for $\zeta=0$. Consider however the
$\epsilon$-expansion of the right-hand-side performed at a finite
$\zeta$:
\be
\eta=\f{\rho^2}{4}\f{\zeta \epsilon}{4(N+8)}\left[
1+\f{2(N+2)}{N+8}\f{\epsilon}{\zeta}
\right].
\ee
If we now take the limit $\zeta\rightarrow0$, the right-hand-side
starts directly like $\epsilon^2$. Together with the lowest-order
value 1 of $\rho^2/4$, we obtain
correctly~(\ref{eta2loopsepsilon}).

For consistency, the different two-loop results for $\eta$, once
from~(\ref{expogammaresummed}) and~(\ref{exponuresummed}), and
once $\eta=0$ from~(\ref{etazeta}) should not be too far from
each other. This can indeed be verified by plotting the curves
$\eta=2-\gamma/\nu$ against a few values of $N$. The curves are
all close to the $\eta=0$ axis for all $N$, approaching it for
$N\rightarrow\infty$.

Also for higher-loop orders, $\eta$ could be obtained from the
strong-coupling limit $\gamma$ and $\nu$, or by taking a direct
strong-coupling limit.  Variational perturbation theory does not
know which of these approaches should be better. Ultimately, if we
know enough term in the series expansion, the extrapolation to
infinite order $L$ should certainly become insensible to which
function is resummed.

One may wonder if it is possible to set up a unique optimal
function of the critical exponents from which to derive the
strong-coupling limit. The answer to this question would improve
the theory considerably.

Collecting the different results of this section, we have  the
$D=3$ results
\beqn \omega&=&\f{1}{-1+2\sqrt{1+3(3N+14)/(N+8)^2}},
\label{resumeomega}\\
 \gamma&=& \f{2N^3+63N^2+540N+1492}{(N+8)^2(N+20)},
\label{resumegamma}\\
 \nu&=&\f{N^3+31N^2+262N+714}{(N+8)^2(N+19)},
\label{resumenu}\\
\eta_m&=&\f{2(N+2)}{2N+21}\left[ 1+\f{3(3N+14)}{(N+8)^2}\right],
\label{resumeetam}\\
\eta&=&\f{2(N+2)}{N+20}\f{(N+8)^2+3(3N+14)}{2(N+8)^3+5(N+8)^2+3(N+2)(3N+14)},
\label{resumeeta}\\
u^*&=& \f{1}{4(N+8)}+ \f{3}{4}\f{3N+14}{(N+8)^3},
\label{resumegstar}
 \eeqn
where we also included the value of the renormalized coupling
constant at the IR-fixed point. It is obtained from the one-loop
series in $u$ of the expansion ~(\ref{gseriesgB}):
\be u=\ubar_B-\f{4(N+8)}{\epsilon}\ubar_B^2. \label{uoneloop}
 \ee
We can restrict ourselves to one loop since it corresponds to a
power $\ubar_B^2$. The two-loop calculation was however needed to
get $\omega$ correctly, which itself enters~(\ref{uoneloop}).
With the help of~(\ref{2loops}), we obtain
\be u^*=\f{\rho^2}{4}\f{\epsilon}{4(N+8)}=\f{\epsilon}{4(N+8)}+
\f{3}{4}\f{3N+14}{(N+8)^3}\epsilon^2\ee

Since only two critical exponents are independent
\cite{zjbook,ks}, all other can be derived from
Eqs.~(\ref{resumeomega})--(\ref{resumeeta}). These two loop
expressions are only a lowest approximation to the exact results.
In the next section, we evaluate analytically the strong-coupling
limit of the exponents at the three-loop level.

\subsection{Critical exponents from three-loop expansions}

The three-loop calculations are algebraically more involved.
Moreover, as far as the critical exponents are concerned (we will
see later that this is not necessarily true for the amplitude
functions) the optimum of the function~(\ref{fresummed}) is not
given by the vanishing of the first derivative, but by a turning
point, i.e.,~by the vanishing of the second derivative. At the
three-loop order, this implies that the parameter $r$
in~(\ref{interm3loops}) is zero, leading to the three-loop
strong-coupling limit result~(\ref{3loops}). It is this feature
which renders the calculation analytically manageable, involving
only a  cubic equation for the determination of $\rho$ (without
$r=0$, we would have had to solve  an eight-order equation). In
order to obtain $\omega$ to three loop, we
generalize~(\ref{dlogubarB2loops}) to the same order, and find
\beqn -1&\equiv& \f{d}{d\log\ubar_B}\log
\f{u}{\ubar_B}=\f{-4(N+8)}{\epsilon}\ubar_B+ 16\left[
\f{(N+8)^2}{\epsilon^2}+\f{3(3N+14)}{\epsilon}\right]\ubar_B^2\nonumber\\
&&\mbox{}- 8 \left[
\f{8(N+8)^3}{\epsilon^3}+\f{60(N+8)(3N+14)}{\epsilon^2}+
\f{96\zeta(3)(5N+22)+33N^2+922N+2960}{\epsilon} \right]\ubar_B^3.
\eeqn
From this we extract  the coefficients $f_i~(i=0,\ldots,3)$
of~(\ref{interm3loops}). The argument of the square root $r$
turns then out to be negative, and the equation to be solved
is~(\ref{3loops}). This is true not only for $\epsilon=1$, but
also for  all $\epsilon\in[0,1]$. Since~(\ref{3loops}) is a cubic
equation for $\rho$, there are three solutions, one of which is
always negative, which we discard as unphysical, leaving us with
two solutions. Only one of them is connected smoothly to the
two-loop result. The purely algebraic form of the solution,
generalization of the square root coming from
solving~(\ref{rhosquareover4}), is somewhat too lengthily to be
written down here. As a check, we have derived its epsilon
expansion which reads
\be \rho_{\epsilon}=2+\f{3(3N+14)}{(N+8)^2}\epsilon-
\f{96\zeta(3)(5N+22)(N+8)+33N^3+214N^2+1264N+2512}{4(N+8)^4}\epsilon^2
\label{rhoeps3loops} \ee
and leads to the correct $\epsilon$-expansion for
$\omega=\epsilon/(\rho-1)$:
\be \omega_{\epsilon}= \epsilon-\f{3(3N+14)}{(N+8)^2}\epsilon^2
+\f{96\zeta(3)(5N+22)(N+8)+33N^3+538N^2+4288N+9568}{4(N+8)^4}\epsilon^3
\label{omegaeps3loops} \ee
which is the extension of~(\ref{omegaeps2}) to the order
$\epsilon^3$. This is to be compared with  Eq.~(17.15) of the
textbook \cite{ks}.

The trigonometric representation is however compact enough to be
written down here explicitly, at least for $\epsilon=1$.
Introducing an angle $\theta$ and two coefficient $a_0,b_0$
defined by
\beqn \small \theta&=&\arccos \Bigglb(\f{
           \left[13776+4738N+
               N^2(8N+405)+
               96(5N+22)
                \zeta(3)\right]^2}{2\left[106+
               N(N+28)\right]\left\{(N+8)\left[13776+4738N+N^2(8N+405)
+96(5N+22)\zeta(3)\right]\right\}^{3/2}}\nonumber\\
&&\hspace{0cm}\mbox{} \times\f{1}{
         \left[2209664+1040160N+162982N^2+9683N^3+184N^4+672(N+8)(5N+22)
                  \zeta(3)\right]^{3/2}}\nonumber\\
&&\hspace{0cm}\mbox{}\times\bigg\{
67181166592+64001040384N+25893312000N^2+5641828480N^3+
             713027988N^4+54733044N^5\nonumber\\
&&\hspace{0cm}\mbox{}+2760157N^6+
             88332N^7+1440N^8\nonumber\\
&&\hspace{0cm}\mbox{}-192(N+8)(5N+22)
              \left[4084864+1952480N+323706N^2
+20021N^3+514N^4\right]\zeta(3)\nonumber\\
&&\hspace{0cm}\mbox{}+746496\left[(N+8)(5N+22)
\zeta(3)\right]^2\bigg\}\Biggrb),\\
a_0&=&\f{1}{446336+213280N+35334N^2+
       2179N^3+56N^4-
       864(N+8)
        (5N+22)\zeta(3)}
\\
b_0&=&3
  \sqrt{(N+8)
         \left[13776+4738N+N^2(8N+405)+96(5N+22)\zeta(3)\right]
}\nonumber\\
&&\mbox{}\times
  \sqrt{
\left[2209664+1040160N+162982N^2+9683N^3+184N^4+672(N+8)(5N+22)\zeta(3)\right]},
 \eeqn
the relevant root of~(\ref{3loops}) can be written as
\be \rho= -\f{1}{6}+\f{256}{3}a_0\left[106+N(N+25)\right]^2
-a_0b_0\cos\left(\f{-2\pi+\theta}{3}\right). \label{rho3loops}
\ee

For the physically interesting cases $N=0,\ldots,4$, we obtain
the values for $D=3$ dimensions
\begin{center}
\begin{tabular}{|c|c|c|c|c|c|}
\hline
$N$ & 0 & 1 & 2 & 3 & 4 \\
\hline
$\rho$ & 2.41829 & 2.40384 & 2.38683 & 2.36910 & 2.35157 \\
\hline $\omega$ & 0.705073 & 0.712332 & 0.721069 & 0.730405 & 0.73988 \\
\hline $\omega$ (Ref.~\cite{zj98}) & 0.812 & 0.799 & 0.789 & 0.782 & 0.774 \\
\hline
\end{tabular}
\end{center}

Figure~1 illustrates the two- and three-loop critical exponent of
the approach to scaling $\omega=\epsilon/(\rho-1)$ as a function
of $N$ calculated from~(\ref{rhosquareover4})
and~(\ref{rho3loops}), respectively. For comparison, we also give
the three-loop unresummed result~(\ref{omegaeps3loops}), evaluated
at $\epsilon=1$ and the theoretical values given in Tables~1 and~3
of \cite{zj98}. The latter are based on a five-loop analysis
supplemented by a large loop order analysis.


\begin{figure}[thb]
\unitlength 1cm
\begin{center}
\begin{picture}(8,6)
\put(-5,0){\vbox{\begin{center}\psfig{file=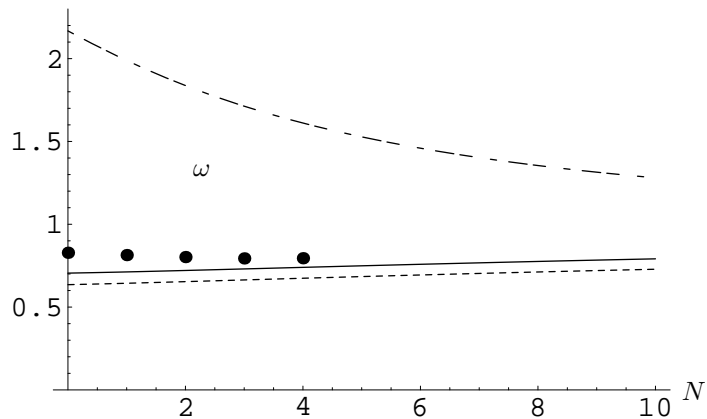,width=9cm}\end{center}}}
\put(8.5,.32){$N$}
\put(2,3.3){$ \omega $}
\end{picture}
\end{center}
\caption{Two-loop (short-dashed) and three-loop (solid) critical
exponent $\omega$ for different O($N$)-symmetries. For
comparison, the $\epsilon$-expansion (mixed-dashed)  and the
theoretical values of \protect\cite{zj98}  (dots) are also
given.} \label{figomega}
\end{figure}

Once $\omega$ is known to three loops, the other exponents and the
strong-coupling limit $u^*$ of the renormalized coupling constant
can be determined to the same order. To obtain $u^*$, the
two-loop expansion of $u$ in powers of $\ubar_B$ is enough since
it is of order ${\cal O}(\ubar_B^3)$. Recall that the three-loop
expansion of $u(\ubar_B)$ is needed only to calculate $\omega$.
From~(\ref{gseriesgB}) we identify $f_1,f_2,f_3$ and
use~(\ref{3loops}) (since the argument of the corresponding $r$
in~(\ref{interm3loops}) is negative) to obtain the critical value
\be u^*= \f{\epsilon(N+8)\rho(\rho+1)(2\rho-1)} {12\left[
2(N+8)^2+3\epsilon(3N+14) \right]}
-\f{2\epsilon(N+8)^3(2\rho-1)^3}{27 \left[2(N+8)^2
+3\epsilon(3N+14) \right]^2} \label{gstar3loopsresummed} \ee
with $\rho$ from~(\ref{rho3loops}). If we use instead the
$\epsilon$-expansion of $\rho$ given in~(\ref{rhoeps3loops}), we
obtain
\be u^*=\f{\epsilon}{4(N+8)}+ \f{3}{4}\f{3N+14}{(N+8)^3}\epsilon^2
+\f{4544+1760N+110N^2-33N^3-96(N+8)(5N+22)\zeta(3)}{32(N+8)^5}
\epsilon^3. \label{gstar3loopsexpanded} \ee

In the same way, we find the strong-coupling limit of the critical
exponents $\gamma$ and $\nu$, as defined in~(\ref{expoGamma})
and~(\ref{expoNu}) together with~(\ref{etam}) and~(\ref{eta}),
the latter two exponents being obtained from the
 the mass~(\ref{mseriesgB}) and wave function~(\ref{phiseriesgB})
  renormalization, respectively. The three-loop perturbative
  expansions are
 \beqn
\gamma&=&1+2(N+2)\ubar_B-4(N+2)\left[ \f{2(N+8)}{\epsilon}-(N-4)
\right]\ubar_B^2\nonumber\\
&&\mbox{}+4(N+2) \left[\frac{8(N+8)^2}
    {\epsilon^2}-\f{4(2N^2-N-106)}
    {\epsilon}+194+N(2N+17)\right]
\ubar_B^3,\\
\nu&=&\f{1}{2}+(N+2)\ubar_B-2(N+2) \left[\f{2(N+8)}{\epsilon}
-(N-3) \right]\ubar_B^2\nonumber\\
&&\mbox{}+4(N+2)\left[
\f{4(N+8)^2}{\epsilon^2}-\f{2(2N^2+N-90)}{\epsilon} +95+N(N+9)
\right]\ubar_B^3,
 \eeqn
from which it is immediate to identify the expansion coefficients
$f_0,\ldots,f_3$ which enter~(\ref{3loops}), to obtain
\beqn \gamma&=&
1-\f{\epsilon(N+2)\left[\epsilon(N-4)-2(N+8)\right]\rho(\rho+1)(2\rho-1)}
{3\left[8(N+8)^2-4\epsilon(2N^2-N-106)+\epsilon^2(2N^2+17N+194)\right]}\nonumber\\
&&\hspace{1cm}\mbox{}+\f{8\epsilon(N+2)\left[\epsilon(N-4)-2(N+8)\right]^3(2\rho-1)^3}
{27\left[8(N+8)^2-4\epsilon(2N^2-N-106)+\epsilon^2(2N^2+17N+194)\right]^2},
\label{resummedexpogamma3loops}\\
\nu&=&\f{1}{2}-
\f{\epsilon(N+2)\left[\epsilon(N-3)-2(N+8)\right]\rho(\rho+1)(2\rho-1)}
{12\left[4(N+8)^2-2\epsilon(2N^2+N-90)+\epsilon^2(N^2+9N+95)\right]}\nonumber\\
&&\hspace{1cm}\mbox{}+\f{\epsilon(N+2)\left[\epsilon(N-3)-2(N+8)\right]^3(2\rho-1)^3}
{27\left[4(N+8)^2-2\epsilon(2N^2+N-90)+\epsilon^2(N^2+9N+95)\right]^2},
\label{resummedexponu3loops} \eeqn
where $\rho$ for $\epsilon=1$ can be obtained
from~(\ref{rho3loops}). The associated $\epsilon$-expansions can
be obtained using~(\ref{rhoeps3loops}). They read
\beqn \gamma&=& 1+\f{N+2}{2(N+8)}\epsilon
+\f{(N+2)(N^2+22N+52)}{4(N+8)^3}\epsilon^2\nonumber\\
&&\mbox{}+
\f{(N+2)\left[3104+2496N+664N^2+44N^3+N^4-48(N+8)(5N+22)\zeta(3)\right]}
{8(N+8)^5}\epsilon^3,\label{expandedexpogamma3loops}\\
\nu&=& \f{1}{2}+\f{N+2}{4(N+8)}\epsilon
+\f{(N+2)(N+3)(N+20)}{8(N+8)^3}\epsilon^2\nonumber\\
&&\mbox{}+
\f{(N+2)\left[8640+5904N+1412N^2+89N^3+2N^4-96(N+8)(5N+22)\zeta(3)\right]}
{32(N+8)^5}\epsilon^3. \label{expandedexponu3loops}\eeqn

Figures~2 and 3 illustrate the two- and three-loop critical
exponents $\gamma$ and $\nu$, respectively, as a function of $N$.
They are given by Eqs.~(\ref{resumegamma})
and~(\ref{resummedexpogamma3loops}) for $\gamma$ and by
Eqs.~(\ref{resumenu}) and~(\ref{resummedexponu3loops}) for $\nu$.
For completeness, we also plot the
$\epsilon$-expansion~(\ref{expandedexpogamma3loops})
and~(\ref{expandedexponu3loops}) of the exponents, as well as the
theoretical values quoted in Tables~1 and~3 of ~\cite{zj98}.
Contrary to the case of the critical exponent $\omega$, we see
that the two- and three-loop critical exponents are very close
together. This is a virtue of working self-consistently with
$\omega$ obtained at the same loop order. In
\cite{kleinert257,kleinert263}, the extrapolated  $\omega$ to
infinite loop order was used instead. This  implies that each loop
order result for $\gamma$ and $\nu$ was not very close to its
asymptotic limit (contrary to what we get here). However, the
extrapolation formula~(\ref{largeL}) works precisely for this
case, and very precise extrapolated results for $\gamma$ and
$\nu$ could be obtained. In our present work, the critical
exponents are not very far from their asymptotic limit, already
at the two- and three-loop level. However, the extrapolation
formula~(\ref{largeL}) cannot be used. It is not yet clear to the
authors how it will be possible to extrapolate the five-loop
results obtained using the present formalism. This question is
left aside for a future work. We also note in passing that the
$\epsilon$-expansion result is not too far from the values
obtained in the strong-coupling limit.


\begin{figure}[thb]
\unitlength 1cm
\begin{center}
\begin{picture}(8,6)
\put(-5,0){\vbox{\begin{center}\psfig{file=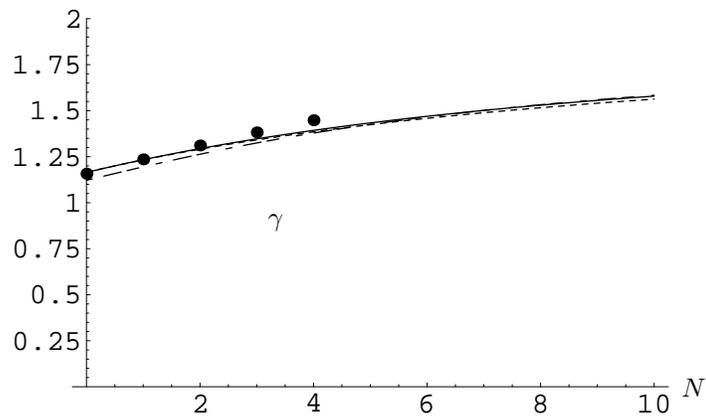,width=9cm}\end{center}}}
\put(8.5,.32){$N$}
\put(3,2.6){$  \gamma  $}
\end{picture}
\end{center}
\caption{Two-loop (short-dashed) and three-loop (solid) critical
exponent $\gamma$. For comparison, the $\epsilon$-expansion
(short- and long-dashed) and the theoretical values of
\protect\cite{zj98}  (dots) are also given.} \label{figgamma}
\end{figure}

\begin{figure}[thb]
\unitlength 1cm
\begin{center}
\begin{picture}(8,6)
\put(-5,0){\vbox{\begin{center}\psfig{file=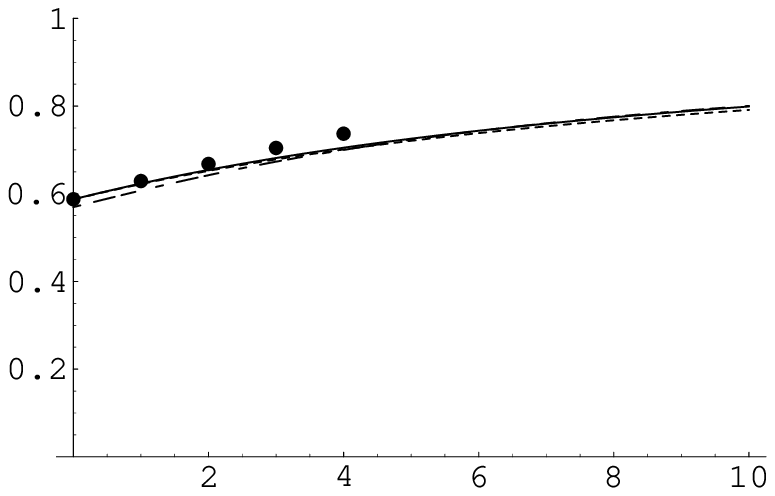,width=9cm}\end{center}}}
\put(8.5,.32){$N$}
\put(3,2.6){$  \nu $}
\end{picture}
\end{center}
\caption{Two-loop (short-dashed) and three-loop (solid) critical
exponent $\nu$. For comparison, the $\epsilon$-expansion (short-
and long-dashed) and the theoretical values of
\protect\cite{zj98}  (dots) are also given.} \label{fignu}
\end{figure}

The critical exponent  $\eta$ is obtained using $2-\gamma/\nu$,
with $\gamma$ and $\nu$ from~(\ref{resummedexpogamma3loops})
and~(\ref{resummedexponu3loops}), respectively. It has the
$\epsilon$-expansion
\be \eta= \f{(N+2)}{2(N+8)^2}\epsilon^2
-\f{(N+2)(N^2-56N-272)}{8(N+8)^4}\epsilon^3.
\label{expandedexpoeta3loops}
 \ee
Let us also calculate directly the strong coupling limit of
$\eta$ from its definition~(\ref{eta}):
\be \eta=8(N+2)\ubar_B^2-8(N+2)(N+8)\left( \f{8}{\epsilon}
+1\right)\ubar_B^3.\label{etadirectseries}\ee
At the two-loop level, the result was zero. At the three-loop
level, the calculation is different from that of $\gamma$ and
$\nu$ because there is no linear term in $\ubar_B$. This has
already been discussed after Eq.~(\ref{signofRoot}): although we
are working at the three-loop level, the optimum of the
variational perturbation theory is not governed by a turning point
but by an extremum for which the sign of the root $r$ is the
opposite to the usual case. The solution corresponds to $r=-1$ and
the optimum is $\uhat_B=-2\bar{f}_2/(3f_3)$, so that
\be \eta=\f{4}{27}\f{\bar{f}_2^3}{f_3^2}=
\f{32}{27}\f{(2\rho-1)^3(N+2)}{(N+8)^2(8+\epsilon)^2}\epsilon^2.
\label{eta3loopsdirect}
\ee
With~(\ref{rhoeps3loops}), this leads again to the correct
$\epsilon$-expansion~(\ref{expandedexpoeta3loops}). The
difference between  $\eta=2-\gamma/\nu$
and~(\ref{eta3loopsdirect}) at $\epsilon=1$ is illustrated on
Figure~4 which also  shows the direct evaluation of the
$\epsilon$-expansion series~(\ref{expandedexpoeta3loops}) as well
as the theoretical values quoted in Tables~2 and~3 of
~\cite{zj98}.


\begin{figure}[h]
\unitlength 1cm
\begin{center}
\begin{picture}(8,6)
\put(-5,0){\vbox{\begin{center}\psfig{file=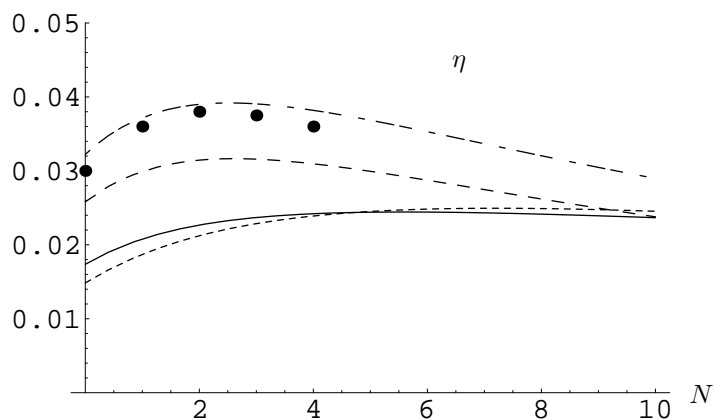,width=9cm}
\put(.15,.32){$N$}
\put(-3,4.8){$  \eta $}
\end{center}}}
\end{picture}
\end{center}
\caption{Two-loop (short-dashed) and three-loop (solid) critical
exponent $\eta$ from the definition $2-\gamma/\nu$. For
comparison, the $\epsilon$-expansion (short- and long-dashed),
$\eta$ from the strong-coupling limit of the direct
(medium-dashed) series~(\ref{eta3loopsdirect}) and the
theoretical values of \protect\cite{zj98} (dots) are also given.}
\label{figeta}
\end{figure}

It is amusing to see that the $\epsilon$-expansion is the best
approximation, followed by the strong-coupling limit of the direct
series~(\ref{eta3loopsdirect}). Comparing the different results,
we see that they differ by about 30\%. This is due to the
absolute smallness of $\eta$. The error is small compared to
unity.

To end up this section we also give the critical exponent
$\eta_m$. Up to three loops, the bare perturbation expansion
reads, from~(\ref{mseriesgB}) and~(\ref{etam}),
\be \eta_m= 4(N+2)\ubar_B -8(N+2) \left[\f{2(N+8)}{\epsilon}
+5\right]\ubar_B^2 + 16(N+2) \left[\f{4(N+8)^2}{\epsilon^2}
+\f{2(19N+122)}{\epsilon}+3(5N+37)\right]\ubar_B^3,\label{etamdirectseries}
\ee
from which we deduce the strong-coupling limit with $\rho$
from~(\ref{rho3loops})
\be \eta_m=\f{\epsilon(N+2)(2N+16+5\epsilon)\rho(\rho+1)(2\rho-1)}
{3\left[4(N+8)^2+\epsilon(38N+244)+3\epsilon^2(5N+37) \right]}
-\f{4\epsilon(N+2)(2N+16+5\epsilon)^3(2\rho-1)^3}
{27\left[4(N+8)^2+\epsilon(38N+244)+3\epsilon^2(5N+37) \right]^2}.
\label{etam3loopsdirect}\ee
Its $\epsilon$-expansion  is
\be \eta_m=\f{N+2}{N+8}\epsilon+
\f{(N+2)(13N+44)}{2(N+8)^3}\epsilon^2+
\f{(N+2)\left[5312+2672N+452N^2-3N^3-96(N+8)(5N+22)\zeta(3)\right]}
{8(N+8)^5}\epsilon^3. \label{etam3loopseps} \ee
The result~(\ref{etam3loopsdirect}) is analytically different but
numerically close to that obtained via the scaling
relation~(\ref{expoNu}), implying $\eta_m=2-\nu^{-1}$, as
illustrated in Figure~5. For completeness, the figure also shows
the $\epsilon$-expansion~(\ref{etam3loopseps}) and the theoretical
values quoted in Tables~2 and~3 of ~\cite{zj98}.


\begin{figure}[h]
\unitlength 1cm
\begin{center}
\begin{picture}(8,6)
\put(-5,0){\vbox{\begin{center}\psfig{file=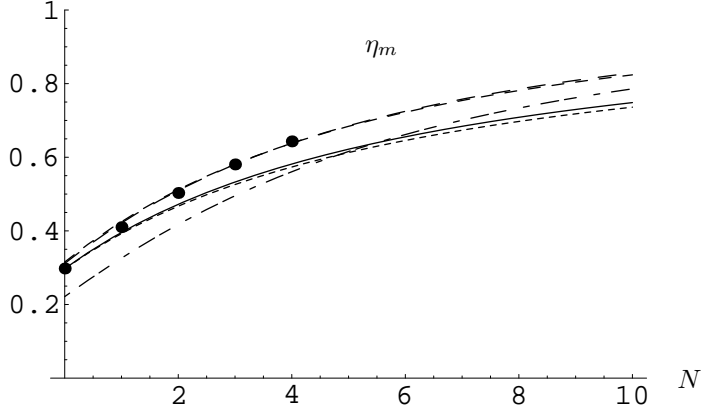,width=9cm}
\put(.15,.32){$N$}
\put(-4,4.8){$  \eta_m $}
\end{center}}}
\end{picture}
\end{center}
\caption{Two-loop (short-dashed) and three-loop (solid) critical
exponent $\eta_m$ from the definition $2-\nu^{-1}$. For
comparison, the $\epsilon$-expansion (short- and long-dashed),
$\eta_m$ from the strong-coupling limit of the direct
(medium-dashed) two-loop~(\ref{etam2loopsdirect}) and three-loop
(long-dashed) series~(\ref{etam3loopsdirect}) and the theoretical
values of \protect\cite{zj98} (dots) are also given.}
\label{figetam}
\end{figure}

We see a better agreement with the theoretical values quoted
from~\cite{zj98} when the exponent is evaluated in the
strong-coupling limit of the direct
series~(\ref{etam2loopsdirect}) and~(\ref{etam3loopsdirect}).
This was also the same for the exponent~$\eta$.

Collecting the different results of this section, we have the
analytical form of the  $D=3$-dimensions critical exponents in
the three-loop order

\beqn \omega&=&\f{1}{\rho-1},\\
 \gamma&=& 1+\f{(N+2)(N+20)\rho(\rho+1)(2\rho-1)}{3(2N^2+149N+1130)}
-\f{8(N+2)(N+20)^3(2\rho-1)^3}{27(2N^2+149N+1130)^2},\\
 \nu&=&\f{1}{2}+\f{(N+2)(N+19)\rho(\rho+1)(2\rho-1)}{12(N^2+71N+531)}
-\f{(N+2)(N+19)^3(2\rho-1)^3}{27(N^2+71N+531)^2},\label{resumenu3loops}\\
\eta_m&=&\f{(N+2)(2N+21)\rho(\rho+1)(2\rho-1)}{3(4N^2+117N+611)}
-\f{4(N+2)(2N+21)^3(2\rho-1)^3}{27(4N^2+117N+611)^2}, \\
\eta&=&\f{32}{2187}\f{(2\rho-1)^3(N+2)}{(N+8)^2},\\
u^*&=& \f{(N+8)\rho(\rho+1)(2\rho-1)} {12\left[ 2(N+8)^2+3(3N+14)
\right]} -\f{2(N+8)^3(2\rho-1)^3}{27 \left[2(N+8)^2 +3(3N+14)
\right]^2}, \label{resumeu3loops}
 \eeqn
where $\rho$  is given in Eq.~(\ref{rho3loops}). For $\eta$ and
$\eta_m$, we  took the strong-coupling limit of the direct
expansions:  Eq.~(\ref{eta3loopsdirect}) for $\eta$ and
Eq.~(\ref{etam3loopsdirect}) for $\eta_m$.  These results have to
be compared with the two-loop ones given
in~(\ref{resumeomega})--(\ref{resumeeta}).

For completeness, we give below the table of the critical
exponents to three loops:

\begin{center}
\begin{tabular}{|c|c|c|c|c|c|} \hline
$N$ & 0 & 1 & 2 & 3 & 4 \\
\hline
$\gamma$ & 1.16455 & 1.2338 & 1.29426 &1.34697 & 1.39307 \\
\hline $\gamma$ (Ref.~\cite{zj98}) & 1.1596 & 1.2396 & 1.3169 &1.3895 & 1.456 \\
\hline $\nu$ & 0.587376 & 0.623381 & 0.654552 & 0.681561 & 0.705071 \\
\hline $\nu$ (Ref.~\cite{zj98})& 0.5882 & 0.6304 & 0.6703 & 0.7073 & 0.741 \\
\hline $\eta_m$ & 0.311607 & 0.421796 & 0.509799 & 0.580684 & 0.638337 \\
\hline $\eta$ & 0.0258218 & 0.029917 & 0.031452& 0.0315846& 0.03096 \\
\hline
\end{tabular}
\end{center}

They cannot compete with the five-loop calculation
of~\cite{kleinert257,kleinert279,kleinert287,kleinert263,kleinert295,ks,zj98}.
 However, our results are analytical,
and already close to the asymptotic limit although we made no
assumption about the large order behaviour of the theory. We
consider this as promising. In a subsequent publication, we will
present a numerical calculation up to five loops, with
large-order behavior information included, of our self-consistent
formalism.

\section{Calculation of amplitude functions and ratios}
\label{section4}

From now on, we shall focus entirely upon the  $D=3$-dimensions
model. As we mentioned  in the introduction, it is only for the
critical exponents that the minimal subtraction scheme leads to
the same resummed values both for $D=3$ and $D=4-\epsilon$. For
this reason, it made sense to study  the $\epsilon$-expansions of
the critical exponents, which was also useful for comparing with
calculations in $4-\epsilon$ dimensions. The reason for this
equality is the mass independence  of the renormalization
constants in this MS scheme. The mass independence implies a
decomposition of the correlation functions into amplitude
functions and power parts, for which the latter can  be evaluated
in the symmetric phase. The amplitude functions, however, depend
on being in the ordered or disordered phase. Moreover, the
situation is complicated for $N>1$ by the presence of  Goldstone
singularities, most of which have to cancel at the end of the
calculations: only  the physical singularities, for example those
occurring in  the transverse susceptibilities,
should stay at the end of the calculations.

For this reason, apart from the  three-loop work \cite{dohm98},
no three- or higher-loop calculation has been done for
 $N>1$ below $T_c$, even numerically. The only relatively easy case is $N=1$
for which extensive numerical work has been done below $T_c$ up to
five-loop order  \cite{zj97,ber87,dohm92}.
 Above $T_c$, all $N$  can be treated in the same way
 \cite{nickel77,ber85,Li95}.  In the  latter reference
the critical exponents $\eta$ and $\eta_m$ have even been
obtained to seven loops, with resummation performed in
\cite{kleinert279,ks,zj97,antonenko95}.

We have explained in detail in the first part of this paper that
it is unnecessary to go to the renormalized theory since all
results can be obtained from the strong-coupling limit of the
bare theory. In the literature, the effective potential is given
in  terms of the renormalized quantities
\cite{dohm99,dohm98,dohm97}. To apply our theory, we shall
rewrite the expressions back in the bare form,
using~(\ref{gseriesgB}).

\subsection{Available expansions}

Let us list the most important available amplitude functions
derived from the minimally renormalized model at $D=3$ at
vanishing external magnetic field $h_B$. Up to two loops, they
can be found in Ref.~\cite{dohm97}:
\begin{itemize}
\item the square of the order parameter $M_B^2=\langle \phi_B^2\rangle$ below
  $T_c$:
\be
f_{\phi}=\f{1}{32\pi u}+\left[
\f{1}{27\pi}(160-82N)+\f{2}{\pi}(N-1)\ln3
\right]u,
\label{2loopsfphi}
\ee
\item the stiffness of phase fluctuations below $T_c$ (some authors call this
  helicity modulus \cite{fisher73})  $\Upsilon$:
\be
f_{\Upsilon}=\f{1}{8u}+\f{1}{3}+\left[
\f{1}{54}(2378-683N)+8(N-3)\ln3
\right]u,
\label{2loopsfupsilon}
\ee
\item the $q^2$ part of the transverse susceptibility $\chi_T$:
\be
f_{\chi_T}=1+\f{8}{3}u+\left[
\f{488}{3}-4N-128\ln3
\right]u^2,
\label{2loopschit}
\ee
\item the specific heat $C^{\pm}$ above and below $T_c$:
\beqn
F_+&=&-N-2N(N+2)u,
\label{2loopsF+}\\
F_-&=&\f{1}{2u}-4+8(10-N)u,
\label{2loopsF-}
\eeqn
\item  the isotropic susceptibility  above $T_c$ \cite{dohm90b}
\be f_{\chi_+}=1-\f{92}{27}(N+2)u^2, \label{fchiaboveTc}
\ee
\item the amplitude function of the susceptibility below $T_c$,
which we obtain taking the inverse of the two loop numerical
expansion given (up to five loops) in \cite{dohm92}: \be
 f_{\chi_-}=1+18u+164.44u^2.
 \label{fchibelowTc}
\ee
\end{itemize}
The latter quantity is restricted to  $N=1$.

From the series expansion of $f_{\chi_T}$ and $f_{\Upsilon}$, one sees that
the relation
\be
f_{\Upsilon}=4\pi
f_{\phi}f_{\chi_T}
\label{equalityfupsilon}
\ee
is satisfied to two loops. This is not a surprise: the bare
helicity modulus, defined as $\Upsilon=2\p\Gamma_B/\p q^2|_{q=0}$
where $\Gamma_B$ is the free energy, can be shown (at least to
two loops \cite{dohm97}) to be identical to $M_B^2(\p
\chi_T^{-1}/\p q^2)|_{q=0}$. This is a consequence of a Ward
identity for the broken O$(N)$-symmetry below $T_c$.

In Ref.~\cite{dohm99}, the perturbation expansions of the
amplitude functions for the order parameter and for the specific
heat have been carried to three loops. The additional terms are
(we use the notation $f_j=\sum_i f_j^{(i)}u^i$):
\beqn
f_{\phi}^{(3)}&=&-\f{1}{1080\pi}\bigg\{
2500N^2+65104N+29056+8640(5N+22)\zeta(3)+58320 c_1-15\pi^2(19N^2+643N+499)
\nonumber\\
&&\mbox{}-180(64N^2+640N+457)\Litwo\left(-\f{1}{3}\right)
-80(194N^2+1616N-1675)\ln 3+16(860N^2+8357N-7867)\ln 2\nonumber\\
&&\mbox{}+270(N-1)
\left[-8c_2+32\Litwo\left(-\f{1}{2}\right)
+42\Litwo\left(\f{1}{3}\right)
-64\Litwo(-2)+21(\ln 3)^2+16(\ln 2)^2
-96(\ln 2)(\ln 3)\right]
\bigg\},
\label{3loopsfphi}
\\
F_+^{(3)}&=&-4N(N+2)\left(
N-\f{7}{27}+4\ln\f{4}{3}
\right),\label{3loopsF+}\\
F_-^{(3)}&=&-\f{1}{27}(1080N^2+3464N+31120)-128(5N+22)\zeta(3)
-864 c_1+\f{2}{3}\pi^2(9N^2+N+17)\nonumber\\
&&\mbox{}+216\Litwo\left(-\f{1}{3}\right)
-32(4N+17)\ln 3+\f{32}{3}(31N+95)\ln 2\nonumber\\
&&\mbox{}+4(N-1)\left[-8c_2+16\Litwo\left(-\f{1}{2}\right)
+6\Litwo\left(\f{1}{3}\right)
-32\Litwo(-2)+3(\ln 3)^2+8(\ln 2)^2-48(\ln 2)(\ln 3)\right]
,
\label{3loopsF-}
\eeqn
where $\Litwo(x)=\sum_{n=0}^{\infty}x^n/n^2$ is the dilogarithmic
function \cite{gradshteyn},  and $c_1$ and $c_2$ are two
numerical constants  given by a single variable integration over
elementary functions \cite{rajantie96,dohm99}:
\beqn
c_1&=&\int_0^1\f{dx}{\sqrt{6-2x^2}}\left[
\ln\f{3}{4}+\ln\f{3+x}{2+x}+\f{x}{2+x}\left(
\ln\f{3+x}{3}+\f{x}{2-x}\ln\f{2+x}{4}
\right)
\right]\approx0.021737576333,
\label{constc1}\\
c_2&=&\f{\pi^2}{4\sqrt{2}}+\sqrt{2}\int_0^1\f{dx}{\sqrt{1+x^2}}\left[
\ln\f{x}{1+x}+\f{\ln(1+x)}{x}
\right]\approx0.973771427.
\label{constc2}
\eeqn

For completeness, we give in Appendix~\ref{appendixfreeenergy}
some hints on how to obtain these amplitude functions. For the
details, see Refs.~\cite{dohm98,dohm97}. Our own contribution
concerns the susceptibilities above and below $T_c$: Using the
three-loop integrals available in the literature
\cite{dohm99,rajantie96}, we have been able to calculate
analytically the thee-loop extension of the amplitude of the
isotropic susceptibility $f_{\chi_+}$:
\be
f_{\chi_+}^{(3)}=-\f{8}{27}
(N+2)(N+8)\left[
-21+12\pi^2+128\ln\f{3}{4}+144\Litwo\left(-\f{1}{3}\right)
\right],
\ee
as well as the three-loop amplitude function of the susceptibility below
$T_c$, for $N=1$:
\be
f_{\chi_-}=1+18u+\f{1480}{9}u^2+
\left[
1072-11664c_1+3\pi^2+10480\ln\f{4}{3}+36\Litwo\left(-\f{1}{3}\right)
\right]u^3.
\ee
Our analytical  two-loop coefficient $1480/9$ agrees with the
numerical coefficient given in (\ref{fchibelowTc}). We shall
comment on the three-loop one later. The details of the
calculation are given in Appendix~\ref{appendixfchiplus}
and~\ref{appendixfchiminus}.

\subsection{Amplitude ratios}
\label{sectionratio}

Besides the amplitude functions, we shall also evaluate three
important ratios: the amplitude ratio of the heat capacity, the
universal combination $R_C$, and the amplitude ratio of the
susceptibilities for $N=1$. For a review of amplitude ratios, see
\cite{aharony91}. The relevant equations for their determination
is given in  Appendix~\ref{appendixampratios}. One of the best
measured amplitude ratios was mentioned in the introduction: it is
the amplitude ratio of the specific heat of superfluid helium
above and below $T_c$, corresponding to $N=2$. It can however be
defined for all $N$ and, using our notation, can be written as \cite{dohm90a,dohm99}
\be \f{A^+}{A^-}=\left(\f{b^+}{b^-}\right)^{\alpha}\left( \f{4\nu
B^*+\alpha F_+^*}{4\nu B^*+\alpha F_-^*}\right),
\label{amplratioCv2loops} \ee where $\alpha$ and $\nu$  are
critical exponents and $B^*$ is the vacuum renormalization group
function associated with the additive renormalization constant of
the vacuum, evaluated at the critical point.  It is known to five
loops in the minimal subtraction scheme \cite{dohm99,kastening98}
and reads, up to three loops
\be uB=\f{N}{2}u+3N(N+2)u^3. \label{Bvacuumg} \ee
The ratio $b^+/b^-$ is equal to \cite{dohm98}:
 $b^+/b^-=2\nu P_+^*/[(3/2)-2\nu P_+^*]$, where $P_{+}$ is a polynomial in $u$,
 related to the  scale above $T_c$.
Its analytical derivation is given in
Appendix~\ref{appendixpplus}
 and reads, up to three-loops:
\beqn
P_+&=&1-2(N+2)u+4(N+2)u^2\nonumber\\
&&\mbox{}+\f{8}{27}(N+2)\left[-3(63N+572)+24(N+8)\pi^2
  +4(43N+182)\ln\f{3}{4}+288(N+8)\Litwo\left(-\f{1}{3}\right)
\right]u^3. \label{renormalizedPplus} \eeqn
The experimental test for the validity of the strong-coupling
expansion is to match~(\ref{amplratioCv2loops})
with~(\ref{lipaparam}) for $N=2$. We shall see in the next
subsection if this can be done.

The ratio $R_C$ is defined by the universal  combination of
amplitudes \cite{aharony91} $R_C=\Gamma^+A^+/A_M^2$ where
$\Gamma^+$ and $A_M$ are the leading amplitudes of the isotropic
susceptibility above $T_c$ and of the order parameter below
$T_c$, respectively.  This ratio has been written in
Ref.~\cite{dohm99} as
\be
R_C =\f{(2\nu P_+^*)^{2-2\beta}}{(3/2-2\nu P_+^*)^{-2\beta}}\f{4\nu B^*+\alpha
   F_+^*}{16\pi}\f{1}{f_{\phi}^*f_{\chi_+}^*}.
 \label{ratioRCdohm}
 \ee
All  the quantities have been defined previously, but for $\beta$
which may be taken from the hyperscaling relation $\beta=
\nu(D-2+\eta)/2=\nu(1+\eta)/2$ in $D=3$ dimensions. However, our
own calculation for $R_C$ gives a correction to
(\ref{ratioRCdohm}):
\be
R_C= \f{(2\nu P_+^*)^{2-\nu(D-2)}}{(3/2-2\nu P_+^*)^{-\nu(D-2)}}
\f{4\nu  B^*+\alpha F_+^*}{16\pi}\f{1}{f_{\phi}^*f_{\chi_+}^*}=
\f{(2\nu  P_+^*)^{2-2\beta}}{(3/2-2\nu P_+^*)^{-2\beta}}\left(
\f{b^+}{b^-}\right)^{\nu\eta}\f{4\nu B^*+\alpha
  F_+^*}{16\pi}\f{1}{f_{\phi}^*f_{\chi_+}^*}.
\label{ratioRC}
\ee
Since  this disagrees with~(\ref{ratioRCdohm}), we give our
derivation of this result in Appendix~\ref{appendixampratios}. We
have verified that the numerical values coming
from~(\ref{ratioRCdohm}) and~(\ref{ratioRC}) do agree within 1\%.
 This is
traced back to the small value of the exponent $\eta$. In the
following we shall however consider~(\ref{ratioRC}). We hope than
the analytical discrepancy  between~(\ref{ratioRCdohm})
and~(\ref{ratioRC}) will soon be resolved.

The third ratio to be investigated is the amplitude ratio of the
susceptibilities for $N=1$. Such a ratio can also be defined for
the longitudinal susceptibilities for $N>1$. This is a nontrivial
task
 requiring an appropriate description \cite{lawrie81}
 due to Goldstone singularities and this  will not be investigated here.
 Using the notation of \cite{dohm92,dohm90b},
the amplitude ratio can be written as
\be
\f{\Gamma^+}{\Gamma^-}=\f{f_{\chi_-}^*}{f_{\chi_+}^*}
\left(\f{\xi_+}{\xi_-}\right)^2=\f{f_{\chi_-}^*}{f_{\chi_+}^*}\left(
\f{b^+}{b^-}
\right)^{2\nu},
\label{ratiosusc}
\ee
where the ratio $b^+/b^-$ has been defined below
Eq.~(\ref{amplratioCv2loops}), and where the quantities are
restricted to $N=1$.

The question arises now to calculate the amplitude functions and ratios.
As for the case of
the critical exponents, we shall proceed also by order, starting with two
loops.

\subsection{Amplitude functions from two-loop expansions}
In order to apply strong-coupling theory to  the amplitude
functions
\comment{\footnote{More specifically, only $f_{\chi_T}$ will be resummed this
  way. The other amplitudes will be multiplied by $u$, the latter product
  being resummed.}}
 (\ref{2loopsfphi})--(\ref{2loopsF+}), we must reexpand them in powers of
the bare coupling $\ubar_B$ using~(\ref{gseriesgB}) up to two loops. The
strong-coupling  limit is then given by the general expression~(\ref{2loops}),
with $\rho^2/4$ given by Eq.~(\ref{rhosquareover4}) at $\epsilon=1$.

We start
considering  $f_{\phi}$.  To deal with a Taylor series, as assumed in the
general theory in Section~\ref{method}, we consider $uf_{\phi}$:
\be
uf_{\phi}=\f{1}{32\pi}+\left[
\f{1}{27\pi}(160-82N)+\f{2}{\pi}(N-1)\ln3
\right]\ubar_B^2.
\label{ufphi2loops}
\ee
This series is special because the linear term in $\ubar_B$, is absent:
the optimal value~(\ref{2loops}) is therefore given by $\uhat_B^*=0$, and the
two-loop value of $uf_{\phi}$ in the strong-coupling limit is
the same as the lowest-order value, which is  independent of $N$:
\be
u^*f_{\phi}^*=\f{1}{32\pi}.
\label{resummedufphi2loops}
\ee
It is worth pointing out here the effect of the special choice for
 $A_D$ in~(\ref{defAD}). We mentioned
there that this coefficient did not have any influence upon the
critical exponent. This is because the factor $A_D$ can be
absorbed in $u_B$ to give
 $\ubar_B$, implying the same strong-coupling limit. However, amplitude
 functions are  $A_D$-dependent. In particular, for $uf_{\phi}$, the chosen
 value  $A_3=1/(4\pi)$
 has made the linear term disappear.
 One sees that
 this choice correspond to an optimalization: the zero order, the one-loop
 and the two-loop  optimum values coincide. One expect then that the
 third-loop order  contributes only to a small deviation from it. This is
 indeed the
 case, as will be shown in the next section, and confirm previous expectations
 \cite{dohm90a,dohm98}.

The same situation holds for the amplitude function (\ref{fchiaboveTc}) of the
susceptibility
above $T_c$. The linear term in $u$ being absent,
the optimal value to two loops is independent of $N$ and is equal to
\be f_{\chi_+}^*=1. \label{fchi+*}
\ee

The strong-coupling limit of the amplitude function of the
stiffness of phase fluctuations   and of the $q^2$-dependent part
of the transverse susceptibility can also be  easily determined.
The bare expansion is obtained combining~(\ref{2loopsfupsilon}),
(\ref{2loopschit}) and~(\ref{gseriesgB}) to two loops:
\beqn
uf_{\Upsilon}&=&\f{1}{8}+\f{\ubar_B}{3}+\left[
\f{1}{54}(1802-755N)+8(N-3)\ln3
\right]\ubar_B^2,\label{2loopsfupsilongB}\\
f_{\chi_T}&=&1+\f{8}{3}\ubar_B-\f{4}{3}(
11N-58+96\ln3
)\ubar_B^2.
\label{2loopschitgB}
\eeqn
The corresponding optima are given by~(\ref{2loops})
with~(\ref{rhosquareover4})  from which we obtain
\beqn
u^*f_{\Upsilon}^*&=&\f{1}{8}+\f{6(N^2+25N+106)}{(N+8)^2\left[755N-1802-(432N-
1296)\ln  3
\right]},\label{resummed2loopsfupsilon}\\
f_{\chi_T}^*&=&1+\f{16(N^2+25N+106)}{3(N+8)^2\left[11N-58+96\ln3\right]}.
\label{resummed2loopschit}
\eeqn
The result~(\ref{resummed2loopsfupsilon}) has a pole for
$N=2(648\ln 3-901)/(432\ln 3-755)\approx1.349$, indicating that
the strong-coupling result is unreliable.  We expect the pole to
be an artifact of the limitation to two loops which disappears at
the three-loop level.
  Since $f_{\Upsilon}$ is not known to
three-loops,  we can only give plausible  arguments for this
expectation, suggested by the calculation of
 $u^*(F_-^*-F_+^*)$ up to
three loops in Eq.~(\ref{deltaFresummed3loops}), where  a similar
pole arises at  the two-loop level but disappears for three
loops  due to the interplay of the coefficients of the loop
expansion. The trouble  with~(\ref{resummed2loopsfupsilon})
derives from the fact that the term of order $\ubar_B^2$
in~(\ref{2loopsfupsilongB}) change sign  for the mentioned value
of $N\approx 1.349$, and at the two-loop level nothing can
compensate this. This is in contrast with critical exponents which
were observed to be alternating series in powers of $\ubar_B$. The
result~(\ref{resummed2loopschit}) for $f_{\chi_T}$ is smooth for
all positive  $N$. A more reliable result for $f_{\Upsilon}^*$
than the singular~(\ref{resummed2loopsfupsilon}) can therefore be
obtained by combining~(\ref{resummedufphi2loops})
with~(\ref{resummed2loopschit})  via
relation~(\ref{equalityfupsilon}), leading to
\be
f_{\Upsilon}^*=f_{\chi_T}^*/8.
\label{otherresummedfupsilon}
\ee
Note that for
  $N\geq 4$, far away from the pole, the two
  results~(\ref{resummed2loopsfupsilon}) and~(\ref{otherresummedfupsilon})
  agree within 2\%.

It is worth pointing out that an evaluation of the renormalized
expression~(\ref{2loopschit})  at
the critical point  $u^*$ given by~(\ref{resumegstar})  leads to a result
compatible  with~(\ref{resummed2loopschit}) within less that 1\%.
This is due to the fact that higher-order correction to the zero
order result $f_{\chi_T}=1$ are small for all $N$. This is in contrast to
$f_{\phi}$ and
$f_{\Upsilon}$, where  two-loop
corrections are important.

We now turn to the amplitude functions $F_{\pm}$ which enter the heat capacity
above and below $T_c$. At the two-loop level, they are given by
Eqs.~(\ref{2loopsF+}) and~(\ref{2loopsF-}), respectively.  With the relation
between the
renormalized and bare coupling constant~(\ref{gseriesgB}) to two loops, we
have the expansions
\beqn
uF_+&= &-N\ubar_B+2N(N+14)\ubar_B^2,\label{uFplusgB2loops}\\
uF_-&= &\f{1}{2}-4\ubar_B+8(N+26)\ubar_B^2.\label{uFminusgB2loops}
\eeqn
With the help of~(\ref{2loops})
and~(\ref{rhosquareover4}), we obtain
\beqn
u^*F_+^*&=& -\f{N}{2}\f{N^2+25N+106}{(N+8)^2(N+14)},
\label{resummeduFplus2loops}\\
u^*F_-^*&=& \f{1}{2}-2\f{N^2+25N+106}{(N+8)^2(N+26)}.
\label{resummeduFminus2loops}
\eeqn

In \cite{dohm98,dohm99}, $uF_+$ was not
 a good candidate for Borel resummation because its
 $u$-expansion~(\ref{2loopsF+}) lacking alternating signs of its coefficients.
 This
problem is absent in variational perturbation theory
since the  expansion~(\ref{uFplusgB2loops}) in term of the bare
coupling constant $\ubar_B$ does have alternating sign.  The latter
is then expected to lead to
  a reliable  result~(\ref{resummeduFplus2loops}). This
 will be confirmed by the three-loop result of the next section.

\begin{figure}[thb]
\unitlength 1cm
\begin{center}
\begin{picture}(8,6)
\put(-5,0){\vbox{\begin{center}\psfig{file=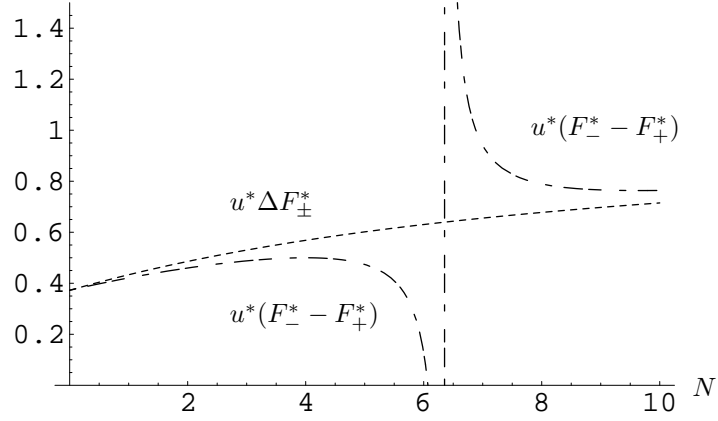,width=9cm}
\put(.15,.32){$N$}
\put(-6,2.8){$ u^*\Delta F_{\pm}^*  $}
\put(-2,3.8){$ u^*(F_-^*-F_+^*)  $}
\put(-6,1.3){$ u^*(F_-^*-F_+^*)  $}
\end{center}}}
\end{picture}
\end{center}
\caption{Comparison between the two-loop strong-coupling limit
of  $u^*(F_-^*-F_+^*)$
and $u^*\Delta
F_{\pm}^*$.}
\label{figdeltaF2loops}
\end{figure}

To apply the usual Borel resummation at the level of the
renormalized quantities, Refs.~\cite{dohm98,dohm99} wrote the
amplitude ratio of the heat capacity as
\be \f{A^+}{A^-}=\left(\f{b^+}{b^-}\right)^{\alpha}\left(
1-\alpha\f{F_-^*-F_+^*}{4\nu B^*+\alpha F_-^*}\right),
\label{cvsnddef}
\ee
 instead of~(\ref{amplratioCv2loops}),
 and resummed $u(F_--F_+)$ and $uF_-$, avoiding the direct resummation of
 $uF_+$.
For comparison, we give below the optimal value of the difference
 $u(F_--F_+)$. It is  is determined from the expansions~(\ref{2loopsF+})
 and~(\ref{2loopsF-}). Using~(\ref{gseriesgB}),
 it yields
\be
u(F_--F_+) =\f{1}{2}+(N-4)\ubar_B-2(N^2+10N-104)\ubar_B^2.
\label{deltaFgB}
\ee
Its strong-coupling limit is, from~(\ref{2loops})
and~(\ref{rhosquareover4}):
\be
u^*(F_-^*-F_+^*)&=&\f{1}{2}+\f{(N-4)^2(N^2+25N+106)}{2(N+8)^2(N^2+10N-104)}.
\label{deltaFresummed}
\ee
The latter expression diverge for a positive value  of
$N=-5+\sqrt{129}\approx6.358$. Then, the difference $u(F_+-F_-)$
is not the good quantity for the strong-coupling limit at the
two-loop level. We should rather evaluate $uF_+$ and $uF_-$
separately in the amplitude ratio~(\ref{amplratioCv2loops}),
instead of using the equivalent expression~(\ref{cvsnddef}).  We
shall see in the next section that the pole of $u^*(F_+^*-F_-^*)$
is an artifact of the two-loop calculation.
 A similar
conclusion was also obtained for the strong-coupling limit of
$f_{\Upsilon}$, see Eqs.~(\ref{resummed2loopsfupsilon})
and~(\ref{otherresummedfupsilon}). For $N\ll 4$ and for
$N\gg-5+\sqrt{129}$, the two-loop expansion~(\ref{deltaFgB}) is
alternating, and we expect that the strong-coupling
result~(\ref{deltaFresummed}) is reliable. As an indication for
this, we compare~(\ref{deltaFresummed}) with the difference of
the optimized $u^*F_{\pm}^*$ values given in
Eqs.~(\ref{resummeduFplus2loops})
and~(\ref{resummeduFminus2loops})
\be
u^*\Delta F_{\pm}^*=\f{1}{2}-\f{N^2+25N+106}{(N+8)^2}\left(
\f{2}{N+26}-\f{N}{2(N+14)}
\right).
\label{deltaFresummedequality}
\ee
In Figure~\ref{figdeltaF2loops}, we compare the two
curves~(\ref{deltaFresummed}) and~(\ref{deltaFresummedequality}).


 As far as the amplitude ratio of the heat capacity~(\ref{amplratioCv2loops}),
 or~(\ref{cvsnddef}), is concerned,
 we still need to determine the strong-coupling limit of the renormalization
 group function $B(u)$  of the vacuum~(\ref{Bvacuumg}) and of the polynomial
 $P_+$ defined in~(\ref{renormalizedPplus}). Because there is no
 contribution of the two-loop order to~(\ref{Bvacuumg}), its
 strong-coupling limit is
 \be
u^*B^*=u^*\f{N}{2}. \label{uBu2loops}
 \ee
Since the optimal two-loop result is identical to the one-loop
result, it is clear that we may expect the large order limit
$L\rightarrow\infty$ to differ only
 little from $N/2$. This has been confirmed in the
five-loop resummation performed in \cite{dohm98}, and will be
also seen in our three-loop calculation in the next section.

The polynomial $P_+$ given in~(\ref{renormalizedPplus}) is
evaluated in the strong-coupling limit using the same lines. The
starting point is the expansion in powers of the bare coupling
constant given in Eq.~(\ref{renPplusgB}) of
Appendix~\ref{appendixpplus}. Its two-loop part combined with
Eqs.~(\ref{2loops}) and~(\ref{rhosquareover4}) leads to
\be
P_+^*=1-\f{(N+2)(N^2+25N+106)}{(N+8)^2(2N+17)}.
\label{resummedPplus2loops}
\ee

The last amplitude we shall calculate using strong-coupling
theory is $f_{\chi_-}$. From~(\ref{2loops}),
(\ref{rhosquareover4}) at $N=1$ and from the two-loop part
of~(\ref{barefchi-}), we find, with Eqs.~(\ref{2loops})
and~(\ref{rhosquareover4}),
\be f_{\chi_-}^*=1+9\f{\rho^2}{4}\f{18^2}{4532}=211/103.
\label{fchi-*} \ee

Combining with the unit value~(\ref{fchi+*}) of $f_{\chi_+}^*$, we
have a ratio $f_{\chi_-^*}/f_{\chi_+^*}$ identical
to~(\ref{fchi-*}). However, this ratio might as well be
determined as the strong-coupling limit of its perturbative
expansion, instead of evaluating independently the
strong-coupling limit of the numerator and the denominator. The
relevant equation is given in Appendix~\ref{appendixfchiminus}:
Using the two-loop expansion of~(\ref{expandedratiosusc}), we
have, with Eqs.~(\ref{2loops}) and~(\ref{rhosquareover4}),
\be
\left(\f{f_{\chi_-}}{f_{\chi_+}}\right)^*=1+\f{\rho^2}{4}\f{3\times18^2}{1420}=
\f{751}{355}. \label{fchi-fchi+2loops} \ee

\subsection{Amplitude functions  from three-loop expansions}
\label{3loopsampl}

Some of the amplitude functions have been obtained to three
loops. We now turn to their strong-coupling limit. This is done
by applying  Eqs.~(\ref{interm3loops}), (\ref{3loops})
and~(\ref{rho3loops}) to the different amplitude expansions.

\begin{figure}[thb]
\unitlength 1cm
\begin{center}
\begin{picture}(8,6)
\put(-5,0){\vbox{\begin{center}\psfig{file=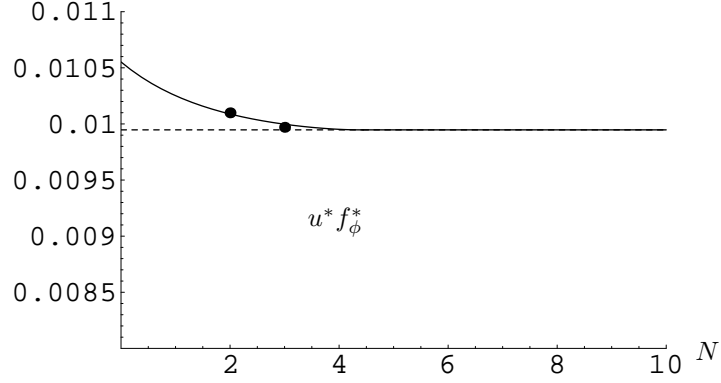,width=9cm}
\put(.15,.52){$N$}
\put(-5,2.3){$ u^*f_\phi^* $}
\end{center}}}
\end{picture}
\end{center}
\caption{Comparison between the two-loop (short-dashed) and
three-loop (solid) amplitude function of the order parameter. The
resummed values \protect\cite{dohm99}
   obtained  using a Borel resummation  are
   indicated by
   the dots
   for values of $N$ available.}
\label{figfphi}
\end{figure}

We start with the amplitude function of the square of the order
parameter. Combining the two-loop expansion~(\ref{2loopsfphi})
with the three-loop term $f_{\phi}^{(3)}$~(\ref{3loopsfphi}), and
using also the relation between the  bare and renormalized
coupling constant~(\ref{gseriesgB}), we have the three-loop
expansion
\be
uf_{\phi}=\f{1}{32\pi}+\left[
\f{1}{27\pi}(160-82N)+\f{2}{\pi}(N-1)\ln3
\right]\ubar_B^2+\left\{f_{\phi}^{(3)}-8(N+8)
\left[
\f{1}{27\pi}(160-82N)+\f{2}{\pi}(N-1)\ln3
\right]
\right\}\ubar_B^3.
\label{ufphi3loopsgB}
\ee
From this, we read off the expansion coefficients
$f_0,f_1,f_2,f_3$ entering Eqs.~(\ref{interm3loops}),
(\ref{3loops}) and~(\ref{rho3loops}). Since the linear term $f_1$
vanishes, we have to follow the development below
Eq.~(\ref{signofRoot}), adapting it to the present case. This
development was done assuming a series with alternating sign
since the expansions of the critical exponents had this property.
Here, this is no longer true. Consider once more the derivation of
the strong-coupling limit following from the  optimal value of
$f=f_0+\bar{f}_2\uhat_B^2+f_3\uhat_B^3$:
$\uhat_B^*(2\bar{f}_2+3f_3\uhat_B^*)=0$. Two solutions are
possible: $\uhat_B^*=0$ and $\uhat_B^*=-2\bar{f}_2/(3f_3)$.  The
latter was relevant for the critical exponent $\eta$. This does
not mean that the other solution has to be rejected. In fact,
looking at the nature of the extremum (minimum or maximum), we see
directly that the first solution corresponds to
\be
\left.\f{\p^2f}{\p\uhat_B^2}\right|_{\uhat_B=\uhat_B^*=0}=2\bar{f}_2,
\label{plus2f2}
\ee
while the other  leads to
\be
\left.\f{\p^2f}{\p\uhat_B^2}
\right|_{\uhat_B=\uhat_B^*=-2\bar{f}_2/(3f_3)}=-2\bar{f}_2.
\label{minus2f2}
\ee
If one solution is a maximum, the other one is a minimum. Looking
for the sign of $\bar{f}_2$ in Eq.~(\ref{ufphi3loopsgB}), we see
that it is positive for $N<N_{\phi}$, with $N_{\phi}=
(80-27\ln3)/(41-27\ln3)\approx4.43992$, corresponding to a
maximum, and negative for $N$ greater, corresponding to a
minimum. Variational perturbation theory at loop order $L>1$ says
nothing about the nature of the extremum. It might be a minimum
or a maximum.
\comment{variational
perturbation theory is a principle of minimum sensitivity.}
In quantum mechanics, this has been explained  in the book
\cite{kleinertpi}. In quantum field theory, the exponent $\eta$
illustrates this: we had chosen the maximum (recall
Eq.~(\ref{eta3loopsdirect})). In this way, the
$\epsilon$-expansion was obtained. Taking the solution
$\uhat_B^*=0$, corresponding to the minimum, we would have
obtained the three-loop result $\eta=0$. The lack of reproducing
the $\epsilon$-expansion gives
 a hint that the maximum solution has to be chosen. In the case
of $f_{\phi}$ we can also argue that the maximum solution has to
be chosen, although there is here no $\epsilon$-expansion
available, by definition of the model. However, at the point were
$f_3=0$, we have to recover an optimization problem of a quadratic
equation in $\uhat_B$, see Eq.~(\ref{ufphi3loopsgB}). We know for
this function that, because no linear term is present, the
strong-coupling limit is $u^*f_{\phi}=1/(32\pi)$. This implies
that $\uhat_B^*=0$ at this point, i.e., that the maximum solution
has to be chosen. By continuity, this remains true in a
neighborhood. The nature of the solution can only be changed when
both solutions are equal, i.e., for $N$ smaller than its value
$N_{\phi}$ making $f_2$  vanish. Below  $N_{\phi}$, we can
imagine that we have an interchange of solutions, and that the
minimum has to be chosen. In this case, we would have
$u^*f_{\phi}^*=1/(32\pi)$ for all $N$. If we decide to keep  the
maximum for all $N$, which we could prove to be true only for
$N\geq N_{\phi}$,  this would imply that
$\uhat_B^*=-2\bar{f}_2/(3f_3)$ has to be chosen below $N_{\phi}$
and  $\uhat_B^*=0$ above. Below $N_{\phi}$ we have
$f^*=f_0+4\bar{f}_2^3/(27f_3^2)$, as was the case for  the
critical  exponent $\eta$, while above  $N_{\phi}$, the solution
is $f^*=f_0$. The strong-coupling limit of $f_{\phi}$ to three
loops is then
\be
u^*f_{\phi}^*=\f{1}{32\pi}+\f{4}{27}(2\rho-1)^3
\f{\left[
(160-82N)/(27\pi)+2(N-1)(\ln3)/\pi
\right]^3}
{\left\{f_{\phi}^{(3)}-8(N+8)
\left[
(160-82N)/(27\pi)+2(N-1)(\ln3)/\pi
\right]\right\}^2}\Theta\left(\f{80-27\ln3}{41-27\ln3}-N\right)
\label{resummedufphi}
\ee
with $\rho$ given by Eq.~(\ref{rho3loops}), and where $\Theta(x)$
is the step function of Heaviside, being equal to 1 for $x>0$ and
being vanishing for $x<0$. As mentioned, we cannot be assured
that for $N<N_{\phi}$ the maximum has still to be chosen. The
possibility that the three-loop result is identical to the
two-loop result remains. Would the above analysis not be
performed, i.e., choosing the minimum solution everywhere, we
would have obtained~(\ref{resummedufphi}) without the step
function, meaning the presence of a pole at the vanishing of the
coefficient of the cubic term in~(\ref{ufphi3loopsgB}), i.e., for
$N\approx4.92915$. We have checked that the solution is sharply
peaked near this value so that it would  appear that the optimal
value is valid almost everywhere. In fact, since the pole gives a
very  peaked contribution, a calculation at fixed integer
  value of $N$ would have
missed it completely, making one to believe that the resummation
was correct. But  this would not be true, the true solution
being~(\ref{resummedufphi})  everywhere.
 We give in Figure~\ref{figfphi} the comparison between our two- and
three-loop results. Our values  for $N<N_{\phi}$ lie above the
two-loop result $1/(32\pi)$ obtained
in~(\ref{resummedufphi2loops}).


This is also the case for the resummed values given in
\cite{dohm99} for $N=2,3$ as can be seen in the following table:
\begin{center}
\begin{tabular}{|c|c|c|c|c|c|c|} \hline
$N$ & 0 & 1 & 2 & 3 & 4 & $N>N_{\phi}=(80-27\ln3)/(41-27\ln3)$ \\
\hline
$u^*f_{\phi}^*$ (2 loops) & $1/(32\pi)$ & $1/(32\pi)$ &
$1/(32\pi)$
&$1/(32\pi)$ & $1/(32\pi)$ & $1/(32\pi)=0.00994718$ \\
\hline $
u^*f_{\phi}^*$ (3 loops) & 0.0105523 & 0.0102518 &
0.0100884 &
 0.00999735 & 0.00995195 & $1/(32\pi)$ \\
\hline
$u^*f_{\phi}^*$ (Ref.~\cite{dohm99}) &  &  & 0.010099 & 0.00997 & &  \\
\hline
\end{tabular}.
\end{center}
The agreement between our two- and three-loop order, and between
our work and \cite{dohm99}, is excellent. It is due to the fact
that the term of order zero contains almost all information on
this amplitude.

The three-loop amplitude functions $uF_+$ and $uF_-$ are given by
Eqs.~(\ref{2loopsF+}), (\ref{3loopsF+}), (\ref{2loopsF-}) and
(\ref{3loopsF-}). As in the  case of the previous amplitudes, the
present expansions may not be alternating. This may make the
argument of the parameter $r$ in Eq.~(\ref{interm3loops})
positive, so that~(\ref{interm3loops}) has to be used to obtain
the strong-coupling limit rather than~(\ref{3loops}). However,
(\ref{3loops}) remains correct for all $N$ for $uF_+$, while the
alternating property is lost for $uF_-$ for $N\gtrsim 40$. Since
the physical cases corresponds to  $N=0,1,2,3,4$, we  can ignore
the alternative~(\ref{interm3loops}) and~(\ref{3loops}) is used
throughout. Using the relation~(\ref{gseriesgB}) between the bare
and renormalized coupling constant, the three-loop bare extension
of Eqs.~(\ref{uFplusgB2loops}) and~(\ref{uFminusgB2loops}) are
\beqn
uF_+&=&-N\ubar_B+2N(N+14)\ubar_B^2+\left[
F_+^{(3)}-24N(7N+46)
\right]\ubar_B^3,\label{uFplususgB3loops}\\
uF_-&=&\f{1}{2}-4\ubar_B+8(N+26)\ubar_B^2
+\left[
F_-^{(3)}-480(3N+22)\right]\ubar_B^3.
\label{uFminusgB3loops}
\eeqn
This allows to identify the appropriate $f_0,f_1,f_2,f_3$
functions to enter Eq.~(\ref{3loops}). In the strong-coupling
limit, we obtain
\beqn
u^*F_+^*&=&\f{2N^2(N+14)\rho(\rho+1)(2\rho-1)}
{6\left[F_+^{(3)}-24N(7N+46)\right]}
+\f{2\left[2N(N+14)\right]^3(2\rho-1)^3}{27\left[F_+^{(3)}-24N(7N+46)\right]^2}
\label{resummeduFplus3loops},\\
u^*F_-^*&=&\f{1}{2}+\f{32(N+26)\rho(\rho+1)(2\rho-1)}{6\left[
F_-^{(3)}-480(3N+22)\right]}+
\f{2\left[8(N+26)\right]^3(2\rho-1)^3}{27\left[F_-^{(3)}-480(3N+22)\right]^2}
\label{resummeduFminus3loops},
\eeqn
with $\rho$ from Eq.~(\ref{rho3loops}).

\begin{figure}[thb]
\unitlength 1cm
\begin{center}
\begin{picture}(8,6)
\put(-5,0){\vbox{\begin{center}\psfig{file=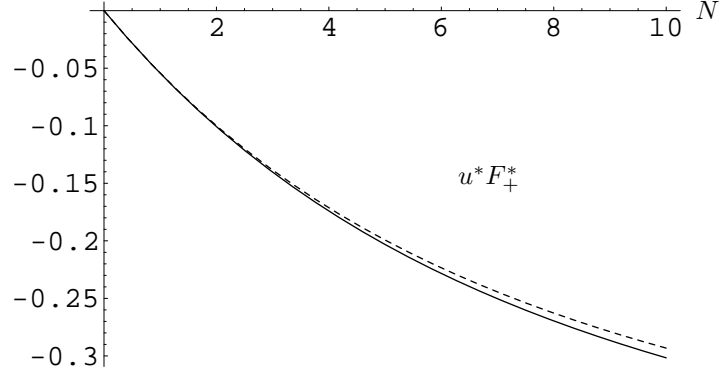,width=9cm}
\put(.15,4.99){$N$}
\put(-3,2.8){$ u^*F_+^*$}
\end{center}}}
\end{picture}
\end{center}
\caption{Comparison between the strong-coupling limit of the
two-loop (shot-dashed) and three-loop amplitude
    function~(solid) $u^*F_+^*$.}
\label{figuFplus2and3loops}
\end{figure}

\begin{figure}[thb]
\unitlength 1cm
\begin{center}
\begin{picture}(8,6)
\put(-5,0){\vbox{\begin{center}\psfig{file=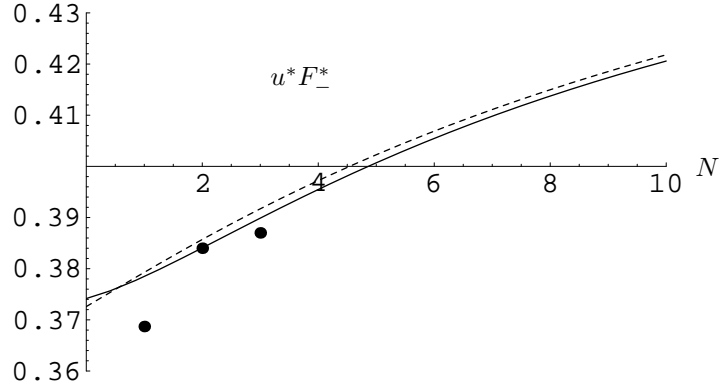,width=9cm}
\put(.15,2.99){$N$}
\put(-5.5,4.2){$ u^*F_-^*$}
\end{center}}}
\end{picture}
\end{center}
\caption{Comparison between the strong-coupling limit of the
two-loop (shot-dashed) and three-loop amplitude
    function~(solid) $u^*F_-^*$. The
resummed values \protect\cite{dohm99}
   obtained  using a Borel resummation  are
   indicated by
   the dots
   for values of $N$ available.}
\label{figuFminus2and3loops}
\end{figure}

Figures~\ref{figuFplus2and3loops} and~\ref{figuFminus2and3loops}
show the comparison between the
two-loop~(\ref{resummeduFplus2loops})
and~(\ref{resummeduFminus2loops}) results of the previous section
and the corresponding three-loop~(\ref{resummeduFplus3loops})
and~(\ref{resummeduFminus3loops}) results, as well as a
comparison with values given in \cite{dohm98}, when available.


To be more precise concerning the comparison with~\cite{dohm98},
we give the appropriate values of $u^*F_-^*$ in the next table:
\begin{center}
\begin{tabular}{|c|c|c|c|c|c|} \hline
$N$ & 0 & 1 & 2 & 3 & 4 \\
\hline $u^*F_-^*$ (2 loops) & 0.372596 & 0.379287 &
 0.385714 & 0.391707 & 0.397222 \\
\hline $u^*F_-^*$ (3 loops) & 0.374166 & 0.378474 &
 0.384065 & 0.389883 & 0.395484 \\
\hline
$u^*F_-^*$ (Ref.~\cite{dohm98}) &  & 0.3687 & 0.384 & 0.387 &  \\
\hline
\end{tabular}.
\end{center}

From this table, we see that that the strong-coupling limit
results for $u^*F_-^+$ at the three-loop level differ only a
little from their two-loop counterpart. This was also see on
Figure~\ref{figuFminus2and3loops} and on
Figure~\ref{figuFplus2and3loops} for $u^*F_+^*$. For $N=1$, we
can also infer from the table that the results coming from
variational perturbation theory and from a Borel resummation
\cite{dohm98} are not in excellent agreement, not even within the
error-bars of the latter: $u^*F_-^*(N=1)=0.3687\pm0.0040$. The
agreement is however recovered for the values $N=2,3$.

For $u^*F_+^*$, there is no available comparison between our work
and others. The authors of \cite{dohm98} could not performed a
reliable Borel resummation, presumably because of the lack of an
alternating series. A comparison is however possible for the
difference $u^*(F_-^*-F_+^*)$. We have seen in the previous
section that the two-loop evaluation of this difference in the
strong-coupling limit did not work well in our case because the
second-order term in the bare expansion changes sign for some
value of $N$. Let us see how the situation changes at the
three-loop level, which has the expansion (see
Eqs.~(\ref{uFplususgB3loops}) and~(\ref{uFminusgB3loops}))
\be
u(F_--F_+)=\f{1}{2}+(N-4)\ubar_B-2(N^2+10N-104)\ubar_B^2+\left[
F_-^{(3)}-F_+^{(3)}+24(7N^2-14N-440) \right]\ubar_B^3.
\label{bareudeltaf} \ee
The coefficient of the second-order term  vanishes for
$N=-5+\sqrt{129}$. This is not anymore a problem since there is a
three-loop order term preventing a $1/f_2$-behavior, see the
comparison of~(\ref{2loops}) and~(\ref{3loops}). The coefficient
of the three-loop order can itself vanish.
For~(\ref{bareudeltaf}), this happens for
$N\equiv\bar{N}\approx10.5324$. Since~(\ref{interm3loops})
and~(\ref{3loops}) imply a behavior like $1/f_3$, it is
legitimate to wonder about poles. The answer is simple: if the
coefficient of the three-loop term vanishes, then the problem is
formally equivalent to evaluating the strong-coupling limit of a
two-loop series. The coefficient of the linear and quadratic
terms are however different from the two-loop result since the
linear term has a factor $\rho(\rho+1)/2$ instead of $\rho$ and
the quadratic one a coefficient $(2\rho-1)$ instead of a factor
1. We conclude that when the three-loop term vanishes, the
strong-coupling limit should be well-behaved, giving a smooth
curve around $\bar{N}$. This discussion shows that the function
$r$ in~(\ref{interm3loops}) is not always zero here because $r$
contains the coefficient $f_2$ of the two-loop term, and its zero
govern the behavior of the solution~(\ref{interm3loops}).
 We note here
the important following point: the positive square root $+r$ was
chosen in~(\ref{interm3loops}) in order to match with a vanishing
$f_3$. We explained, and this was used when evaluating $\eta$,
that the negative root might play a role as well. For $\eta$ to
three-loop order, we had a negative $r$. Let us see what happens
for $f_2=0$. The expansion to be optimized is
$f=f_0+\bar{f}_1\uhat_B+f_3\uhat_B^3$, such that we have to solve
$\bar{f}_1+3f_3(\uhat_B^*)^2=0$. For the critical exponents, the
signs of $\bar{f}_1$ and $f_3$ are the same because the series are
alternating. For this reason, this equation has no real solution,
and we must solve the turning-point equation $6f_3\uhat_B^*=0$,
which is $\uhat_B^*=0$, leading to the optimized result $f^*=f_0$.
For the amplitude functions, we have already seen that the
alternating property is not necessarily true, so that the solution
$\uhat_B^2=-\bar{f}_1/(3f_3)$ is real. At the point where $f_2$
vanishes, we can see that the optimal value is
\be
f^*=f_0\pm\f{2}{3}\bar{f}_1\sqrt{-\f{2\bar{f}_1}{3f_3}}
\ee
the positive or negative sign being chosen to get a continuity of the
 solution around $f_2$.

 For the difference function $u(F_--F_+)$,
it is possible to follow exactly the strong-coupling limit as a
function of $N$. Depending on $N$, there are four
 different solutions: below $\bar{N}_1\approx2.48527$ and above $\bar{N}_3\approx16.6066$,
 the argument of the square root of $r$ is negative, and one uses
 Eq.~(\ref{3loops}). For $N\in[\bar{N}_2,\bar{N}_3[$, one uses
 Eq.~(\ref{interm3loops}) with the positive root ($r=|r|$), where
 $\bar{N}_3=-5+\sqrt{129}\approx6.35782$ is the value of $N$ for which $f_2$ changes sign.
 The zero of $f_3$ lies within the same region.
 Finally, the last region is within the range
 $N\in]\bar{N}_1,\bar{N}_2]$, for which we use   Eq.~(\ref{interm3loops}),
 but with the negative root $r=-|r|$.
 More precisely,
\beqn
u^*(F_-^*-F_+^*)&=&\f{1}{2}+\f{2(N-4)(N^2+10N-104)\rho(\rho+1)(2\rho-1)}
{6\left[F_-^{(3)}-F_+^{(3)}+24(7N^2-14N-440)
\right]}\left(1-\f{2}{3}r\right)\nonumber\\
&&\mbox{}-\f{2[2(N^2+10N-104)]^3(2\rho-1)^3}
{27\left[F_-^{(3)}-F_+^{(3)}+24(7N^2-14N-440)\right]^2}(1-r),
\label{deltaFresummed3loops}
\ee
with $r=0$ for  $N\lesssim \bar{N}_1$ and $N\gtrsim \bar{N}_3$,
and $r$ the  negative or positive square root of
\be
r^2=1-3\f{(N-4)\left[F_-^{(3)}-F_+^{(3)}+24(7N^2-14N-440)
\right]\rho(\rho+1)}{2[2(N^2+10N-104)]^2(2\rho-1)^2}
\label{paramRdeltaFresummed3loops}
\ee
for $N\in]\bar{N}_1,\bar{N}_2]$ or $N\in[\bar{N}_2,\bar{N}_3[$,
respectively.

Thus, the pole in $N$ of the two-loop approximation to
$u^*(F_-^*-F_+^*)$ was only an artifact of the low order. At the
three-loop level, the singularity is avoided by the interplay
between the different possible solutions of~(\ref{interm3loops})
arising from the different branches of $r$:  $r=0,\pm|r|$, with
$|r|$ to be identify with
 the function $r$ defined below Eq.~(\ref{interm3loops}).
 This possibility was not exploited in  previous works
 \cite{kleinert257,kleinert263} because of the
 alternating signs for the critical exponents. (See however $\eta$ which required
 $r=-1$ for  the strong-coupling limit of (\ref{etadirectseries}).)

\begin{figure}[thb]
\unitlength 1cm
\begin{center}
\begin{picture}(8,6)
\put(-5,0){\vbox{\begin{center}
      \psfig{file=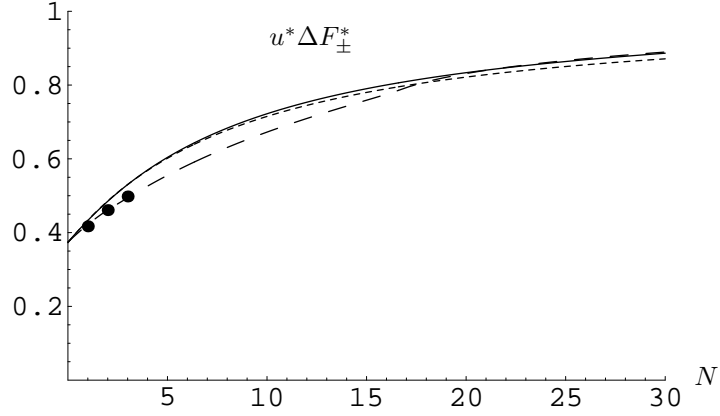,width=9cm}
\put(.15,.4){$N$}
\put(-5.5,4.92){$ u^*\Delta F_{\pm}^*$}
\end{center}}}
\end{picture}
\end{center}
\caption[]{Comparison between the strong-coupling limit of the
two-loop (shot-dashed) and three-loop amplitude
    function~(solid) $u^*\Delta F_{\pm}^*=u^*F_-^*-u^*F_+^*$. The
    three-loop (long-dashed) evaluation of $u^*(F_-^*-F_+^*)$ is
    in better agreement with
the resummed values (dots) obtained in \protect\cite{dohm99}
   using a five-loop ($N=1$) or three-loop ($N\ne0$) Borel resummation.
    }
 \label{figdeltaF3loops}
\end{figure}

In Figure~\ref{figdeltaF3loops},
 we show  the strong-coupling limit of $u^*(F_-^*-F_+^*)$. For comparison,
we also give the direct difference $u^*\Delta F_{\pm}^*$ between
$u^*F_-^*$ and $u^*F_+^*$, as obtained from
Eqs.~(\ref{resummeduFminus3loops})
and~(\ref{resummeduFplus3loops}), as well as its two-loop
counterpart~(\ref{deltaFresummedequality}). The range for $N$ has
been increased to 30 in order to investigate the regions
delimited by $\bar{N}_1$, $\bar{N}_2$ and $\bar{N}_3$.


For the direct difference, the changes brought about by the
three-loop is very small, as before in
Figs.~\ref{figuFplus2and3loops} and~\ref{figuFminus2and3loops}.
The difference between  $u^*(F_-^*-F_+^*)$ and  $u^*\Delta
F_{\pm}^*$ is however somewhat larger. To facilitate the
comparison, the following table should be of help:
\begin{center}
\begin{tabular}{|c|c|c|c|c|c|} \hline
$N$ & 0 & 1 & 2 & 3 & 4 \\
\hline $u^*\Delta F_{\pm}^*$ (2 loops) & 0.372596 & 0.433608 &
 0.485714 & 0.530258 & 0.568519 \\
\hline $u^*\Delta F_{\pm}^*$ (3 loops) & 0.374166 & 0.432926 &
 0.484899 & 0.530224 & 0.569615 \\
\hline $u^*(F_-^*-F_+^*)$ (3 loops) & 0.374166 & 0.421864 &
 0.461436 & 0.489995 & 1/2 \\
\hline
$u^*(F_-^*-F_+^*)$ (Ref.~\cite{dohm98}) &  &  0.4179 & 0.461 & 0.498 &  \\
\hline
\end{tabular}.
\end{center}

The simple value
  $1/2$ for the case $N=4$ comes from Eqs.~(\ref{deltaFresummed3loops})
  and~(\ref{paramRdeltaFresummed3loops}): for $N=4$, $r$ is vanishing, meaning
  the third term of~(\ref{deltaFresummed3loops}) does not contribute. Since
  the second term is also proportional to $N-4$, only the zero loop order
  survives for the Higgs case.
Our results for $u^*(F_-^*-F_+^*)$ are in good agreement with the
Borel results of Ref.~\cite{dohm98}. This is probably not a
coincidence since  we now resum the same function as they did. We
note however that, for the Ising model ($N=1$), we are not within
the error bars of~\cite{dohm98}. We have already noted this for
the strong-coupling limit of $u^*F_-^*$.

\comment{CAREFUL : MAYBE NOT TRUE ANYMORE BECAUSE OF
SELFCONTENTNESS. For the comparison of the two- and three-loop
value of $u^*\Delta F_{\pm}^*$, we note that except for $N=0$
which is different from the two loop result by about 1\%, the
values for $N=1,2,3$ are within 0.2\%. Since our two and three
loop calculation based on the difference of the resummed
$u^*F_-^*$ and $u^*F_+^*$ functions are almost identical, it may
be tempting to believe this is a sign that we are very close to
the extrapolated $L\rightarrow\infty$ limit. However, this may be
misleading because, as it was the case also for the critical
exponents, odd- and even-loop order for $u^*F_-^*$ and $u^*F_+^*$
come from different equations: the even orders are solution of an
extremum condition, see Eq.~(\ref{interm3loops}), while the odd
orders come from the turning point condition given by
Eq.~(\ref{3loops}). For the critical exponents, it meant that odd
and even approximations were on different converging lines, and
this fact was used in \cite{kleinert257,kleinert263,ks} to
extrapolate the results to the infinite-loop limit. The resummed
values based on $u^*(F_-^*-F_+^*)$ may be more precise. However,
it is not possible to connect with the two-loop order which
showed a pole, except for low $N\lesssim2$. We remark however
that~(\ref{deltaFresummed3loops}) has then to be chosen with
$r=0$, i.e., we face again a turning point condition and the two-
and three-loops values should be on different converging lines.
Clearly, a higher-order calculation is needed to answer definitely
the question.}

Before cloturing the investigation of $u(F_--F_+)$, we recall the
case of $f_{\Upsilon}$, whose direct two-loop strong-coupling
limit gave
 Eq.~(\ref{resummed2loopsfupsilon}), exhibiting a pole. We know
 the strong-coupling limit should
not have been far from $f_{\chi_T}^*/8$, see the discussion
leading to Eq.~(\ref{otherresummedfupsilon}). We have shown in
this section how a pole in $u^*(F_-^*-F_+^*)$ at the two-loop
level might disappear at the three-loop one. This is probably the
case for $f_{\Upsilon}$. It would be very useful to get its
three-loop order.

We can now turn to the strong-coupling limit of the
renormalization group constant of the vacuum $B(u)$. Its
three-loop value has been given in Eq.~(\ref{Bvacuumg}).

\comment{We can resum $B(u^*)$, or $u^*B(u^*)$. The former stops
at order $u^2*$, although this is the three loop term. It might
not be clear how to resum in this case. The trivial way would be
just to stop the reexpansion at the same order, leading then to
\be
B=\f{N}{2}+3N(N+2)\ubar_B^2.
\ee
Applying what we know already for a vanishing linear term, this
leads to a resummed value $B^*=N/2$, as for the zero to two loop
order. However, one may argue that the series of $B$ starts in
fact with a power $1/u$ although with a vanishing coefficient.
Then, as for the case of $F_-$ and $F_+$, we have to multiply by
$u$: in this way, we only have to resum series with positive
powers of $u$, which we know how to handle thanks to the algorithm
explained in detail in this paper. }

\begin{figure}[thb]
\unitlength 1cm
\begin{center}
\begin{picture}(8,6)
\put(-5,0){\vbox{\begin{center}\psfig{file=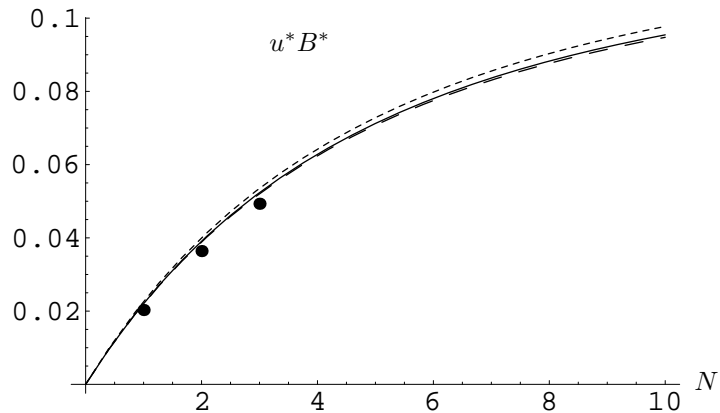,width=9cm}
\put(.15,.4){$N$}
\put(-5.5,4.92){$  u^*B^* $}
\end{center}}}
\end{picture}
\end{center}
\caption{Comparison between the strong-coupling limit of the
two-loop (short-dashed)  and three-loop
      (solid) renormalization group constant of
      the vacuum $u^*B^*$. For completeness, we also
      give the approximate three-loop (long-dashed)
      result $u^*_{(3)}N/2$.  A five-loop calculation
      \protect\cite{dohm98} using Borel resummation
      is included (dots) for values of $N$ available.}
 \label{uBufig2and3loops}
\end{figure}

Upon inserting the relation between the renormalized and
the bare coupling constant~(\ref{gseriesgB}), we obtain
\be
uB(u)=\f{N}{2}\ubar_B-2N(N+8)\ubar_B^2+N(8N^2+167N+686)\ubar_B^3.
\ee
The series is alternating and behaves as for the critical
exponents. No subtleties arises here as in the case of $f_{\phi}$
and $u^*(F_-^*-F_+^*)$. In particular, the argument of the square
root of $r$ in Eq.~(\ref{interm3loops}) is negative for all $N$
and we have to work with~(\ref{3loops}). Using~(\ref{3loops}),
the strong-coupling limit is
\be
u^*B^*=\f{N(N+8)\rho(\rho+1)(2\rho-1)}{6(8N^2+167N+686)}-
\f{16N(N+8)^3(2\rho-1)^3}{27(8N^2+167N+686)^2}.
\label{resummeduBu}
\ee
This result is plotted in Figure~\ref{uBufig2and3loops} together
with the two-loop result $u^*N/2$, see Eq.~(\ref{uBu2loops}). We
also indicate the approximate result $u^*B^*=u^*N/2$ with $u^*$
from the three-loop expansion~(\ref{resumeu3loops}). There is no
visible difference between the latter and~(\ref{resummeduBu}).

\comment{Our two loop results for $u^*$ are
   for $N=0,1,2,3,4$: $53/1024\approx0.0517578,11/243\approx
   0.0452675,0.4,95/2662\approx0.0356875,37/1152\approx0.0321181$. The three
   loop counter part
   is:$0.0502627,0.0439949,0.0388885,0.0346961,0.0312193$. The   five
   loop  values, resummed using a Borel technique, are given in the text.}


To facilitate the comparison between the different
approximations, we recapitulate the numerical results in the next
table:
\begin{center}
\begin{tabular}{|c|c|c|c|c|c|} \hline
$N$ & 0 & 1 & 2 & 3 & 4 \\
\hline $u^*B^*$ (2 loops) & 0 & 0.0226337 &
0.04 & 0.0535312 & 0.0642361 \\
\hline $u^*B^*$ (3 loops) & 0 & 0.0221074 &
 0.0391089 & 0.0523643 & 0.0628447 \\
 \hline $u^*B^*=u^*_{(3)}N/2$ & 0 & 0.0219975 &
 0.0388885 & 0.0520441 & 0.0624386 \\
\hline
$u^*B^*$ (Ref.~\cite{dohm98}) & 0 &  0.020297 & 0.0363919 & 0.049312 &  \\
\hline
\end{tabular}.
\end{center}
For the comparison with \cite{dohm98}, we have multiplied their
five-loop results  for $B^*$ with their five-loop $u^*$. These
five-loop results are within 7\% from our three-loop
strong-coupling calculation. This confirms that  $B^*\approx N/2$
to all orders. We shall however see in the next section that this
7\% difference leads to a non-negligible difference in the
universal combination $R_C$.

\comment{In particular, we shall show in the last section that,
although our two and three loop result are close, the asymptotic
limit is not yet reached. One again, this is traced back to the
fact that odd and even orders are on different converging lines
and should not be compared.}

The next quantity we shall resum to three loops is the polynomial
$P_+$. The three-loop bare expansion of the renormalized $P_+$
was given in Eq.~(\ref{renPplusgB}) and resummed to two loops
in~(\ref{resummedPplus2loops}). The series in the bare coupling
constant is alternating, and behaves as for the critical
exponents. The strong-coupling limit of $P_+^*$ to three loops is
then given by
\beqn P_+^*&=&1+\f{9}{2}\f{(N+2)(2N+17)\rho(\rho+1)(2\rho-1)}
{\left[-3(36N^2+837N+3920)+24(N+8)\pi^2+4(43N+182)\ln(3/4)+288(N+8)
\Litwo(-1/3) \right]}\nonumber\\
&&\mbox{}+54
\f{(N+2)(2N+17)^3(2\rho-1)^3}{\left[-3(36N^2+837N+3920)
+24(N+8)\pi^2+4(43N+182)\ln(3/4)+288(N+8)\Litwo\left(-1/3\right)
\right]^2}, \label{resummedPplus3loops} \eeqn
with $\rho$ from Eq.~(\ref{rho3loops}).

\begin{figure}[thb]
\unitlength 1cm
\begin{center}
\begin{picture}(8,6)
\put(-5,0){\vbox{\begin{center}\psfig{file=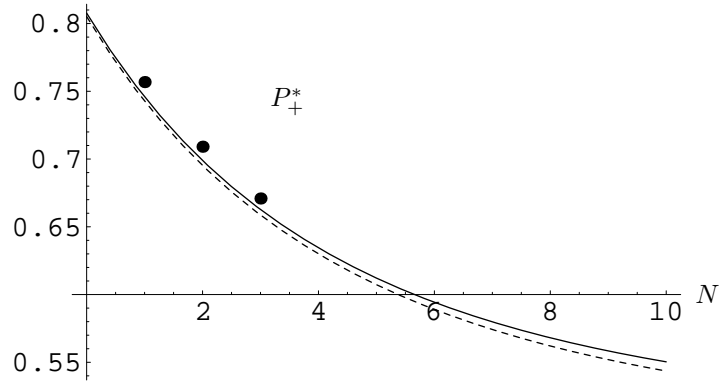,width=9cm}
\put(.15,1.3){$N$}
\put(-5.5,3.92){$  P_+^*$}
\end{center}}}
\end{picture}
\end{center}
\caption{Comparison between the strong-coupling limit of the
two-loop (short-dashed)
  and three-loop (solid) polynomial $P_+^*$.
  The values \protect\cite{dohm99}
   obtained  using a five-loop Borel resummation  are
   indicated by
   the dots
   for values of $N$ available.}
 \label{figPplus2and3loops}
\end{figure}

In Figure~\ref{figPplus2and3loops}, we
compare~(\ref{resummedPplus3loops})  with the two-loop result from
Eq.~(\ref{resummedPplus2loops}). Almost no difference is found
between our two- and three-loop expansions.


For a better comparison, we quote the numerical values for
$N=0,1,2,3,4$ in the next table:
\begin{center}
\begin{tabular}{|c|c|c|c|c|c|} \hline
$N$ & 0 & 1 & 2 & 3 & 4 \\
\hline $P_+^*$ (2 loops) & 0.805147 & 0.74269 &
 0.695238 & 0.658642 & 0.63 \\
\hline $P_+^*$ (3 loops) & 0.807683 & 0.745874 &
 0.698901 & 0.662717 & 0.63447 \\
\hline
$P_+^*$ (Ref.~\cite{dohm98}) &  &  0.7568 & 0.7091 & 0.6709 &  \\
\hline
\end{tabular}.
\end{center}
The two- and three-loop results agree within 1\%. The results
agree fairly well with the five-loop Borel resummation performed
in~\cite{dohm98}. We shall however see later that amplitude
ratios depends crucially on the exact value of $P_+^*$. For this
reason, our three-loop calculation is probably not precise enough.
We shall present in the last section a numerical five-loop
strong-coupling evaluation of $P_+^*$ to more firmly settle this
statement.

\begin{figure}[thb]
\unitlength 1cm
\begin{center}
\begin{picture}(8,6)
\put(-5,0){\vbox{\begin{center}\psfig{file=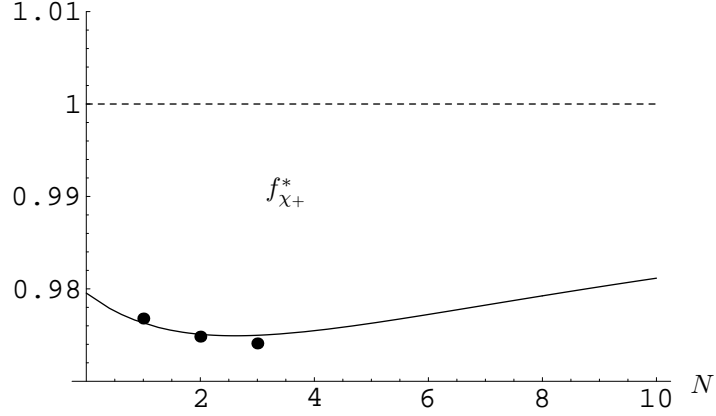,width=9cm}
\put(.15,.3){$N$}
\put(-5.5,2.92){$  f_{\chi_+}^*$}
\end{center}}}
\end{picture}
\end{center}
\caption{Comparison between the two-loop strong-coupling limit
(short-dashed) of the amplitude
$f_{\chi_+}^*$ of the susceptibility above $T_c$
and the second possible solution~(\ref{fchisecondsolution3loops})
at the three-loop level (solid).
   The values \protect\cite{dohm90b}
   obtained  using a five-loop Borel resummation  (dots) are
 given for values of $N$ available.}
 \label{fchifig2and3loops}
\end{figure}

To conclude this section, we discuss the amplitude of the
susceptibilities above and below $T_c$
 to three loops. We already know from the previous section that the
two-loop amplitude above $T_c$ is identical to the order zero:
$f_{\chi_+}^*=1$, see Eq.~(\ref{fchi+*}). As for the case of
$uf_{\phi}$, we then expect a very small deviation from the
zero-order value as well as a very smooth $N$-dependence. The
series to evaluate in the strong-coupling limit is given in
Eq.~(\ref{fchi3loops}). It is alternating and behaves like the
series of the critical exponents. Moreover, with a vanishing
linear term, but with a negative coefficient of the quadratic
term, the solution of the optimalization problem is at variance
with the case of the exponent $\eta$ or the amplitude $f_{\phi}$
if, as for these quantities, we admit that the solution is a
maximum. From Eq.~(\ref{plus2f2}), we determine that the optimal
value is $\uhat_B^*=0$, so that
\be f_{\chi_+}^*=1 \label{fchifirstsolution3loops} \ee
remains true at the three-loop level: The amplitude of the
susceptibility above $T_c$  at the three-loop level does not
depend on $N$! This is in contrast to \cite{dohm90b}, where a
$N$-dependent fit, using Borel resummation, has been performed.
Because  our
 resummed value up to three loops is
$f_{\chi_+}=1$, it is tempting to conjecture that this is true
for all orders. However, contrary to the case of $\eta$ and
$f_{\phi}$, we have here no argument to tell that the maximum has
to be chosen instead of the minimum. Only when going to higher
orders, then having more expansion coefficients, can we decide
which solution is the right one. For this reason, we also mention
below this other solution, which differs from unity for at most
2.5\%:
\be
f_{\chi_+}^*=1-\f{48668(N+2)^3(2\rho-1)^3}{[27(N+2)(N+8)]^2\left[
-113+12\pi^2+128\ln(3/4)+144\Litwo(-1/3) \right]^2}.
\label{fchisecondsolution3loops} \ee
The comparison between the two curves is given in
Figure~\ref{fchifig2and3loops}, as well as a comparison with the
fit
\be f_{\chi_+}=1-92(N+2)u^2(1+b_{\chi_+}u)/27 \ee
taken  from Table~1 of \cite{dohm90b}, with $b_{\chi}=9.68\
(N=1), 11.3\ (N=2), 12.9\ (N=3)$, combined with the five-loop
$u^*$ of Ref.~\cite{dohm98}.


More precisely, we have the table:
\begin{center}
\begin{tabular}{|c|c|c|c|c|c|} \hline
$N$ & 0 & 1 & 2 & 3 & 4 \\
\hline $f_{\chi_+}^*$ (2 loops)& 1 & 1 &
 1 & 1 & 1 \\
\hline $f_{\chi_+}^*$ (3 loops)
from~(\ref{fchifirstsolution3loops}) & 1 & 1 &
 1 & 1 & 1 \\
\hline $f_{\chi_+}^*$ (3 loops)
from~(\ref{fchisecondsolution3loops}) & 0.979543 & 0.976298 &
 0.975082 & 0.974977 & 0.975472 \\
 \hline
$f_{\chi_+}^*$ (Refs.~\cite{dohm90b,dohm98}) &  &  0.976791 & 0.9748331 & 0.9740978 &  \\
\hline
\end{tabular}.
\end{center}

The fact that our three-loop
calculation~(\ref{fchisecondsolution3loops}) agrees very well
with Refs.~\cite{dohm98,dohm90b} might be an indication
that~(\ref{fchisecondsolution3loops}) should be preferable
to~(\ref{fchifirstsolution3loops}). However, from a variational
perturbation theory point of view, nothing can be said. Only the
determination of the next order might resolve the ambiguity.

Finally, we determine the strong-coupling limit of the amplitude
of the susceptibility below $T_c$ for $N=1$. We have checked that
the parameter $r$ in~(\ref{interm3loops}) is zero, i.e., we have
to work with the turning-point equation~(\ref{3loops}). Applying
it to~(\ref{renfchi-}), we have
\beqn
f_{\chi_-}^*&=&1+\f{4352\rho(\rho+1)(2\rho-1)}{3\left[
19904-11664c_1+3\pi^2+10480\ln(4/3)+36\Litwo(-1/3)
\right]}\nonumber\\
&&\mbox{}-\f{164852924416(2\rho-1)^2}{19683\left[
19904-11664c_1+3\pi^2+10480\ln(4/3)+36\Litwo(-1/3)
\right]^2},
\label{fchi-resummed3loops}
\ee
with $\rho$ from Eq.~(\ref{rho3loops}). Numerically, this is
evaluated as $f_{\chi_-}^*\approx2.09387$, to be compared with
the two-loop~(\ref{fchi-*}) result $211/103\approx2.048544$. They
agree within 3\%.

For the ratio~(\ref{ratiosusc}), the calculation of
$f_{\chi_-}/f_{\chi_+}$ is needed. Its strong-coupling limit can
be determined using the individual strong-coupling limit of the
numerator and the denominator. In that case, the ambiguity on
$f_{\chi_-}^*$ at the three-loop level is relevant. According to
the choice $f_{\chi_+}^*=1$ from~(\ref{fchifirstsolution3loops})
or $f_{\chi_+}^*\approx0.976298$
from~(\ref{fchisecondsolution3loops}) with $N=1$, we have
\beqn  \f{f_{\chi_-}^*}{f_{\chi_+}^*}&=& 2.09387,\label{209}\\
\f{f_{\chi_-}^*}{f_{\chi_+}^*}&=&2.1447,\label{2144}
 \eeqn
respectively.

The strong-coupling limit of the ratio $f_{\chi_-}/f_{\chi_+}$
can also be computed from its perturbative expansion. It has been
derived in  Appendix~\ref{appendixfchiminus}, see
Eq.~(\ref{expandedratiosusc}). The strong-coupling limit reads
\beqn
\left(\f{f_{\chi_-}}{f_{\chi_+}}\right)^*&=&1+\f{1420\rho(\rho+1)(2\rho-1)}{3\left[
19184-11664c_1+99\pi^2+9456\ln(4/3)+1188\Litwo(-1/3)
\right]}\nonumber\\
&&\mbox{}-\f{5726576000(2\rho-1)^2}{729\left[
19184-11664c_1+99\pi^2+9456\ln(4/3)+1188\Litwo(-1/3)
\right]^2}.
\label{ratiofchi3loops}
\ee
Its numerical value is 2.11227, to be compared with the
two-loop~(\ref{fchi-fchi+2loops}) result $751/355\approx2.11549$.
The three-loop level is in very good agreement with the two-loop
result, within less than 0.2\%. However, this by no means signify
that the asymptotic limit has been reached, and the
ratios~(\ref{209}) or~(\ref{2144}) might be closer to the true
ratio than~(\ref{ratiofchi3loops}). This is due to the fact the
even and odd orders are on different converging lines because odd
(even) terms come from an extremum (turning-point) condition or
vice-versa. To see the speed of convergence, it would be
necessary to compare the fourth-loop order with the two-loop
order, and the fifth-loop order with the third-loop order.

\subsection{Amplitude ratios from two- and three-loop expansions}

We have now everything in hand in order to compute the ratio of
the heat capacity $A^+/A^-$, the universal combination $R_C$ and
the ratio of the susceptibilities $\Gamma_+/\Gamma_-$. This
section is restricted to a full too- and three-loop calculation.
In order to improve the ratios, we shall break our rule of being
selfconsistent  in the next section and use there the maximum
information available.

We start with  the heat capacity $A^+/A^-$. Since we have a
preference for~(\ref{amplratioCv2loops}) over~(\ref{cvsnddef}),
we shall work with the separate strong-coupling limit evaluation
of $u^*F_-^*$ and $u^*F_+^*$. We have checked that the effect on
the ratio $A_+/A_-$ is negligible. The exponent  $\alpha$
entering it is calculated from the two- or three-loop result for
$\nu$ given in ~(\ref{resumenu}) and~(\ref{resumenu3loops}),
respectively, using the hyperscaling relation $\alpha=2-D\nu$.

Combining the different results derived previously, we have
\begin{center}
\begin{tabular}{|c|c|c|c|c|c|} \hline
$N$ & 0 & 1 & 2 & 3 & 4 \\
\hline $A_+/A_-$ (2 loops)& 0 & 0.489106 & 0.843065
& 1.12691 & 1.37015 \\
\hline $A_+/A_-$ (3 loops) & 0 & 0.491088 &
 0.862439 & 1.16719 & 1.43243 \\
 \hline
\end{tabular}.
\end{center}
 Regarding the fact that the critical exponent
 $\alpha$ is
 far away from its asymptotic limit (it is still positive for $N=2$, while
 the shuttle experiment \cite{lipa96} shows clearly a negative
 value), the results of this table are promising: For $N=2$, we obtain
 $A_+/A_-\approx0.862439$ at the three-loop level, while the shuttle experiment
 \cite{lipa96} gives $A_+/A_-\approx 1.0442$, see Eq.~(\ref{lipaparam}).
 We shall see in
 the next section that working with asymptotic critical exponents
 leads to a better agreement with experiments.

 \comment{Note on resumming alpha: For the expansion of $\exp(\alpha)$, we
could get a negative value
 $\alpha\approx-0.0108256$, already very closed to the shuttle experiment. The
 three loop result give a slightly more negative value, still for $N=2$:
 $\alpha\approx-0.0116664$. This is once again a nice example that different
 series lead to different resummed results. The good point here is that we can
 get a negative $\alpha(N=2)$ already at the two loop level. However, it is
 necessary to perform a resummation up to  five loops to see if summing
 $\exp(\alpha)$ really brings something, i.e., if the $\alpha$ obtained here
 really converges faster than the one from the direct resummation of
 $\nu$. This form the subject of a separate publication
 \cite{kleinertvdbalpha}.}

The next ratio we examine is~(\ref{ratioRC}), the universal
combination $R_C$. The results are best displayed in a table:
\begin{center}
\begin{tabular}{|c|c|c|c|c|c|} \hline
$N$ & 0 & 1 & 2 & 3 & 4 \\
\hline $R_C$ (2 loops)& 0 & 0.062474 & 0.124819 & 0.184355 & 0.239967 \\
\hline $R_C$ (3 loops, $f_{\chi_+}^*=1$) & 0 & 0.05944 &
0.121628 & 0.182413 & 0.239691 \\
 \hline
$R_C$ (3 loops, $f_{\chi_+}^*$
from~(\ref{fchisecondsolution3loops})) & 0 & 0.060883 &
0.124736 & 0.187094 & 0.245718 \\
 \hline
\end{tabular}.
\end{center}
We see an overall agreement between the two- and three-loop
results. We have also checked that the ratio $R_C$ calculated
with the formula~(\ref{ratioRCdohm}) used in \cite{dohm99} is
within less than 1\%. Moreover, our results are in agreement with
the values $R_C(N=2)=0.123$ and $R_C(N=3)=0.189$ given in
Ref.~\cite{dohm99}.
Since we expect that using the true critical exponent leads to a
better ratio $A_+/A_-$, it is important to see how $R_C$ evolves.
Will the agreement with \cite{dohm99} be lost? This issue is
investigated in the next section.

To end this section, we  study the ratio of the susceptibilities
$\Gamma_+/\Gamma_-$ for the Ising model, see
Eq.~(\ref{ratiosusc}).

The two-loop result for the amplitude ratio~(\ref{ratiosusc}) is,
 with $\nu=56/90$ from~(\ref{resumenu}), with $P_+^*=127/171$
from~(\ref{resummedPplus2loops}) and with $f_{\chi_+}^*=1$:
\be \f{\Gamma_+}{\Gamma_-} =\f{211}{103}\left(\f{4\nu
P_+^*}{3-4\nu P_+^*} \right)^{2\nu}=\f{211}{103}\left(
\f{14224}{8861} \right)^{56/45}\approx3.69171. \ee
This is still far from the value $\approx4.7$ quoted in the
literature \cite{zj98,aharony91}. A small improvement is obtained
using the direct strong-coupling evaluation of
$f_{\chi_-}/f_{\chi_+}$ of Eq.~(\ref{fchi-fchi+2loops}):
\be \f{\Gamma_+}{\Gamma_-} =\f{751}{355}\left(\f{4\nu
P_+^*}{3-4\nu P_+^*} \right)^{2\nu}=\f{751}{355}\left(
\f{14224}{8861} \right)^{56/45}\approx3.81236.
\label{2loopsratiosc}\ee
This value is still far from the expected ratio $4.7$.
However, the ratio depends sensibly on the value of the critical
exponent $\nu$. For example, using $\nu=0.63$, we
increase~(\ref{2loopsratiosc}) to $\Gamma_+/\Gamma_-=4.002$. The
sensibility  is also seen when calculating the three-loop value of
the ratio:
\be \f{\Gamma_+}{\Gamma_-}\approx3.88785, \ee
where the ratio $(f_{\chi_-}/f_{\chi_+})^*$ has been obtained
from~(\ref{ratiofchi3loops}).

 \comment{To two loops, neither to three loops, we do not expect $\alpha$
 and $\nu$ to
 be good extremely precise: We did  not try
 to extrapolate the results to the infinite order limit
 $L\rightarrow\infty$. We shall see in the next section the effect
 on the amplitude ratios
 of using accurate values for the critical exponents.}

\subsection{Amplitude ratios using maximum
  information}
\label{maxinfo}

Up to now, we have followed the strategy to make a fully
consistent two- and three-loop calculation. The comparison between
the two- and three-loop amplitude functions has made us believed
that the resummed values are close to the extrapolated limit
$L\rightarrow\infty$, although one has to take care that odd and
even approximations are on different converging lines. For the
critical exponents, it is primordial going to the asymptotic
limit. For example, we have $\alpha(N=2)$ still positive at the
three-loop level, while the shuttle experiment, see second
reference of \cite{lipa96} and Eq.~(\ref{lipaparam}), shows a
value of $\alpha(N=2)=-0.01056$.

In this section, we shall relax our constrain of working only
with two- and three-loop quantities and will take the maximum
available information, i.e., our three-loop result for the
amplitudes and extrapolated, or experimental, value for the
critical exponents. We shall also see the effect of using $uB$ to
five loops.

Except for $\alpha(N=2)$ that we took from the shuttle experiment
\cite{lipa96}, the exponents are taken from the $D=3$ tables of
\cite{zj98}, i.e., we are working with, for $N=0,1,2,3,4$:
\be
\begin{array}{ccr@.lr@.lr@.lr@.lr@.l}
\nu&= &0&5882 & 0&6304 & 0&6703 & 0&7073 & 0&741\\
\alpha&=&0&235 & \phantom{-}0&109 & -0&01056 & -0&122 & -0&223
\end{array}.
\label{fiveloopcritexp}
\ee
Combining the three-loop strong-coupling limit of the amplitudes
performed in  section~\ref{3loopsampl} with these exponents, we
obtain, for $A_+/A_-$:

\begin{center}
\begin{tabular}{|c|c|c|c|c|c|}
\hline
$N$ & 0 & 1 & 2 & 3 & 4 \\
\hline
$A_+/A_-$ & 0 & 0.543406 & 1.04516 & 1.54386 & 2.0444 \\
$A_+/A_-$ (Ref.~\cite{dohm98})& 0 & 0.540  & 1.056 & 1.51 &
\\
\hline
\end{tabular}.
\end{center}

We have checked that the increase from the three-loop value (for
$N=2$, this ratio was $0.862439$) is mainly due to using  the
correct $\alpha$. For example, with the correct $\alpha$ but still
using the three-loop $\nu$ of~(\ref{resumenu3loops}), we would
have obtained, for $N=2$, a ratio $1.04711$. It also does not
depend too sensitively  on using the five-loop strong-coupling
limit of $u^*B^*$ and $P_+^*$, neither on using $u^*(F_-^*-F_+^*)$
instead of the separate calculation  of $u^*F_-^*$ and
$u^*F_+^*$. For example, playing with all these quantities, the
ratio, for $N=2$, could be changed from $A_+/A_-=1.04516$ to, at
most, $A_+/A_-=1.049$, depending which quantities are taken to
five loops.  A complete numerical study of this ratio, using
variational perturbation theory up to five loops, will be
presented elsewhere \cite{kleinertvdbratio5loops}.

For $N=2$, our result  $1.04516$ coincide remarkably well with
the shuttle experiment, see second reference of \cite{lipa96}.
For $N=1$, we have $0.543406$, which agrees reasonably well with
the values quoted in Table~5 of \cite{zj98}, values which are
both experimental and theoretical. In Ref.~\cite{dohm98}, the
authors obtained $1.056$ for $N=2$. Their Table~4 make a
comparison between their result and other works and experiments,
for $N=1,2,3$. We see that the agreement is good. In our table, we
have listed only the values calculated in the work~\cite{dohm98}
since the model is the same.

For the universal combination $R_C$, we obtain, using the
five-loop critical exponents~(\ref{fiveloopcritexp}) and the
three-loop amplitudes of section~\ref{3loopsampl}
\be R_C=0,0.0616257,0.130341,0.201404,0.270882
\ee
for $N=0,1,2,3,4$.

Here also, we have checked that the main effect is due to
choosing the correct $\alpha$. Working with $\nu$ at the
three-loop level only modifies slightly the result. While working
with the true exponents for the ratio $A_+/A_-$  had considerably
improved it, making it coincide with the experimental values, we
see for $R_C$ that the values of the previous section, with a
wrong $\alpha$ were in better agreement with the quoted values in
\cite{dohm99}: $R_C=0.123,0.189$ for $N=2,3$, respectively. We
have checked that our result for $N=2$ is not changed if we take
the values of $\alpha$ and $\nu$ taken in \cite{dohm99}. Also,
the result does not depend sensibly on $u^*F_+^*$, although our
value differs from their. We have traced back the difference
between our result and \cite{dohm98} to  $uB$ at the critical
point: limiting ourselves to $N=2$, we have
$u^*B^*\approx0.0391089$ while \cite{dohm98} gives a value
$u^*B^*\approx0.0363919$. This difference is all is needed to
explain the difference between our result and the result of
Ref.~\cite{dohm99}, apart from a very small difference coming
also from our use of Eq.~(\ref{ratioRC}) instead
of~(\ref{ratioRCdohm}). Since  $u^*B^*$ has been obtained in
\cite{dohm98} using a five-loop Borel resummation, it is tempting
to believe it is more accurate. For this reason, we have also
determined numerically the five-loop strong-coupling limit of
$uB$. We shall show a detailed numerical resummation in
\cite{kleinertvdbratio5loops}, showing here only the main steps.
Starting from the five-loop expansion \cite{dohm98,kastening98}
\beqn uB(u)&=&
\f{N}{2}u + \f{N(N+2)}{48}u^3+
  \f{N(N+2)(N+8)
  \left[-25+12\zeta(3)\right]}{648}u^4+
  N(N+2)\nonumber\\
  &&\hspace{-1cm}\mbox{}\times
      \f{(-319N^2+13968N+64864+16(3N^2-382N-1700)\zeta(3)
        96(4N^2+39N+146)\zeta(4)
    -1024\left(5N+22)\zeta(5)\right)
}{41472}u^5,\nonumber\\ \eeqn
and using the algorithm given by Eq.~(\ref{fresummed}), the
corresponding strong-coupling limit is:
\begin{center}
\begin{tabular}{|c|c|c|c|c|c|}
\hline
$N$ & 0 & 1 & 2 & 3 & 4 \\
\hline
$u^*B^*$ (5 loops) & 0 & 0.0209552 & 0.0372717 & 0.0502225 & 0.0605918 \\
\hline
$u^*B^*$ (Ref.~\cite{dohm98}) & 0 &  0.020297 & 0.0363919 & 0.049312 &  \\
\hline
\end{tabular}
\end{center}
Our five-loop result is now much nearer to the Borel resummed
values of \cite{dohm98} than our three loop order of
Section~\ref{3loopsampl}. For this reason, we believe our
five-loop result is near the infinite-loop limit extrapolation.
More details will be given in \cite{kleinertvdbratio5loops},
which also contains the effect of variations of $P_+^*$, which is
the second source, after $u^*B^*$, of error for $R_C$.

Finally, our best values for the ratio $R_C$ are collected in the
next table:
\begin{center}
\begin{tabular}{|c|c|c|c|c|c|}
\hline
$N$ & 0 & 1 & 2 & 3 & 4 \\
\hline
$R_C$ & 0 & 0.05803 & 0.12428 & 0.19402 & 0.26285 \\
\hline
$R_C$ (Ref.~\cite{dohm98}) & 0 & & 0.123 & 0.189 &  \\
\hline
\end{tabular}
\end{center}

To our knowledge no experimental value of this ratio is known for
$N=2$. The case $N=3$ is presented in Table 7.6 of
Ref.~\cite{aharony91}. For $N=1$, the value of the ratio has only
slightly changed compared to the results based on three-loop
$\alpha$ and $\nu$. This is due to the fact that, for $N=1$,
$\alpha$ is positive and its effect on $R_C$ is less sensitive.
In the work \cite{zj98}, the theoretical and experimental values
of $R_C$ are also given for $N=1$. The theoretical values seem to
prefer a value around $0.057$ while the experimental values are
around $0.050$. From Table~7.1 of \cite{aharony91}, we however
see that values close to $0.06$ might as well be obtained.

Better experiments or other theoretical studies are needed in
order to see if our predictions are correct or have to be ruled
out.

Finally, we conclude this section with  the ratio of the
susceptibilities for the Ising model. Using the critical exponent
$\nu$ to five loops~(\ref{fiveloopcritexp}), we obtain
\be \Gamma_+/\Gamma_-= 4.06419,
\ee
where we took the ratio $(f_{\chi_-}/f_{\chi_+})^*$
from~(\ref{ratiofchi3loops}). We might have slightly increased
$\Gamma_+/\Gamma_-$ using the value  $2.1447$ of
Eq.~(\ref{2144}). However, we would still be far from the value
$4.77$  of \cite{zj98}. \comment{As a last argument against it,
we mention that with the value of $f_{\chi_+}^*$ which would give
$2.1447$, i.e., $f_{\chi_+}^*$ given by
Eq.~(\ref{fchisecondsolution3loops}), we would have obtained
higher values for the ratio $R_C$.} The only possible quantity we
may still vary in the ratio~(\ref{ratiosusc}) is $P_+^*$. Our
three-loop value is 0.745874, while the five-loop result given in
\cite{dohm98} using Borel resummation is 0.7568. Using this value
in our formula for the ratio, we find
\be \Gamma_+/\Gamma_-= 4.27154. \label{extrapolatedsus} \ee
The ratio of the susceptibilities depends sensitively on $P_+^*$.
We postpone to \cite{kleinertvdbratio5loops} the application of
variational perturbation theory up to five loops for the
resummation of $P_+^*$ and $\Gamma_+/\Gamma_-$. We do not however
expect a resummed $P_+^*$ different from \cite{dohm98}. For this
reason, the ratio~(\ref{extrapolatedsus}) is probably  the best
we can obtain. A ratio of $4.77$  obtained in \cite{zj98} and
references therein seems to be ruled out from our analysis.

\section{Conclusion}
\label{section5}

In this paper, we have shown that variational strong-coupling
theory \cite{kleinert257,kleinert263} can be applied not only to
critical exponents, but also to
 various amplitude ratios. We have focused on two- and three-loop
results were analytical results for the amplitude functions are
known \cite{dohm98,dohm99,dohm97} for all $N$  both above and
below $T_c$. Our results are analytical expressions, except in the
last section were we used more information to find $A_+/A_-, R_C$
and $\Gamma_+/\Gamma_-$. The results are quite sensitive to the
precise value of the critical exponents. In addition, a five-loop
evaluation of the renormalization constant $B^*$ was necessary.
The ratio $R_C$ was so sensitive to it that a three-loop
calculation was not sufficient. The same remark holds for $P_+^*$,
which affects mainly $\Gamma_+/\Gamma_-$. A numerical study of
the known five-loop amplitudes will be done in
\cite{kleinertvdbratio5loops}, which will contain refined results
compared to Section~\ref{maxinfo}

One interesting observation of our work is that we can evaluate
series which
 have caused problem in previous Borel resummations when the expansion coefficients
  in term of the renormalized coupling constant are not alternating to
 low orders.  For
 these functions, strong-coupling theory turned out to work well.

Having obtained analytical expressions in  $N$, we have shown that
the coefficient of the series in the bare coupling constant may
vanish and change sign. At the two-loop level, this lead to
 diverging results near certain value of $N$.
We have seen that the problem disappears at the three-loop level,
because of the interplay of the different coefficients of the
series. We could show precisely how it works because all our
results were analytical and not restricted to integer values of
$N$.

When using variational perturbation theory, nothing is known on
the nature of the optimal variational parameter, which can be a
minimum,  a maximum, or a turning point. The analysis performed
here should help to identify the correct (numerical) solution at
higher-loop order. See for example the amplitude $f_{\chi_+}^*$,
for which it is not yet clear which of the
solutions~(\ref{fchifirstsolution3loops})
or~(\ref{fchisecondsolution3loops}) has to be chosen.

\begin{acknowledgments}
We thank  B. Kastening, F. Nogueira and A. Pelster for valuable
discussions
during the  completion of this work and M. Bachmann for help speeding up the
calculations.
\end{acknowledgments}

\appendix

\section{Exponent $\omega$ from strong-coupling theory}
\label{appendixmethod}

\comment{Now comes the question to identify properly the function
$f$ to be resummed. More appropriately, we should talk about
functions to be evaluated in the infinite coupling limit since,
doing so, they are automatically resummed. Contrary
to~\cite{kleinert257,kleinert263,ks} which use the resummed
series in the form~(\ref{fstar}) to obtain the equations which
follow below, see in particular~(\ref{omegaovereps})
and~(\ref{intermomegaovereps}), we shall present an alternative
derivation, working with the unresummed series. We shall see that
care has to be taken when trying to resum, due to the fact that
the functions should be blockwise resummed.}

The  power $p/q$ of the leading power
behavior $\ubar_B^{p/q}$ of a function $W_L$ whose perturbative expansion has
been given in~(\ref{functionWL})
can be obtained taking the
logarithmic derivative, giving~(\ref{poverq}). A subtlety arises for
functions going to a constant in the strong-coupling limit. For
such functions,
$p$ vanishes and the corresponding $f^*$ in Eq.~(\ref{fresummed})
vanishes. Care has to be taken: the limit $f^*\rightarrow0$ is
different from imposing $f^*=0$. In the former case, we can
identify $q$ (or $\omega$) by matching the series to achieve
$f^*=0$. Working directly with a series which has $f^*=0$ implies
a leading behavior  $p'/q=-\omega/\epsilon$. The
algorithm~(\ref{fresummed}) serves then to identify the
coefficient $c_0$ of the right-hand-side of Eq.~(\ref{fstar}). As
an example how to use the series, let us derive the relation
\cite{kleinert257,kleinert263,ks}
\be -\f{\omega}{\epsilon}-1 =\f{d\log W_L'}{d\log \ubar_B}.
\label{omegaovereps} \ee
The left-hand-side is  of the type (\ref{fstar}), and the
algorithm~(\ref{fresummed}) can be applied.
Formula~(\ref{omegaovereps}) follows directly from~(\ref{fstar}).
Alternative derivation starts from~(\ref{poverq}): If $p/q$ is
vanishing, this means that its series has a leading exponent
$p'/q=-\omega/\epsilon$, which we derive in the following manner.
Start from formula~(\ref{poverq}) with its exponent $p'/q$ which
we know from the general behavior~(\ref{fstar}) with $f^*\equiv
p/q=0$, i.e.,
\be \f{p'}{q}=-\f{\omega}{\epsilon}=\f{d\log (p/q)}{d\log\ubar_B}
\ee
where $p/q$ is not yet taken at its asymptotic zero value, but is
given as the right-hand-side of~(\ref{poverq}). It then follows
\be  -\f{\omega}{\epsilon}=1+\ubar_B\left(
\f{W_L''}{W_L'}-\f{W_L'}{W_L} \right)=1+\ubar_B
\f{W_L''}{W_L'}-\f{p}{q}. \label{intermomegaovereps}\ee
Taking the limit $\ubar_B\rightarrow\infty$, the term $p/q$
vanishes by hypothesis, and we end up once more with
formula~(\ref{omegaovereps}).

Although the algorithm (\ref{fresummed}) cannot be applied
directly for the right-hand-side of (\ref{poverq}) if $p/q$ is
vanishing exactly but only in the limit $p/q\rightarrow0$,
\comment{(if $p/q=0$, we said that the algorithm allows to
extract the constant $c_0$ on the right-hand-side of
Eq.~(\ref{fresummed}))} we can nevertheless use a trick to
circumvent this problem: If the series for $W_L$  has a vanishing
leading power $p/q$, then $W_L/\ubar_B$ has a power $p'/q=-1$.
This allows to deduce
\be \f{p'}{q}\equiv-1=\f{d\log(W_L/\ubar_B)}{d\log\ubar_B}
=\ubar_B\f{W_L'}{W_L}-1=\f{p}{q}-1. \label{intermpoverq} \ee
This shows that the right-hand-side of (\ref{poverq}) can  be
used to reach the limit 0. Then, $\omega$ can be extracted either
from~(\ref{poverq}) or from~(\ref{omegaovereps}). It is also
clear from  the expression (\ref{intermomegaovereps})  that the
right-hand-side has to be resummed blockwise: we have to use the
intermediate result $p/q=0$ before tempting to resum. Using a full
resummation of the right-hand-side of the latter equation would
lead to badly resummed results (although the underlying
$\epsilon$ expansion would be the same): It is necessary to use
$p/q=0$ in Eq.~(\ref{intermomegaovereps}), and not its analytical
form which would have been  mixed up with the power series of
$W_L{''}/W_L'$.

\section{Free energy to three loops} \label{appendixfreeenergy}

From the model Hamiltonian~(\ref{glfunctional}), the analytical
calculation of the Gibbs free energy $\Gamma_B({m'}_B^2,u_B,M_B
)$ near the coexistence  curve below $T_c$ and for $M_B^2\equiv
\langle\phi_B^2\rangle=0$ above $T_c$ has been obtained at the
two-loop order in \cite{dohm97} and at the three-loop order
in~\cite{dohm98}, thus extending the $N=1$ calculation of
Rajantie~\cite{rajantie96}. We write directly the three-loop
result: \be
\Gamma_B=\f{1}{2}{m'}_B^2M_B^2+u_BM_B^2+\sum_{b=1}^3\sum_{l=0}^{b-1}
\sum_{k=0}^1
(-1)^k2^{-l-k}F_{blk}(\bar{\omega},N)(24u_B)^{3-l}(M_B^2)^l \left[
\f{r_{0L}}{(24u_B)^2}\right]^{\f{4-b-2l}{2}} \ln\left[
\f{r_{0L}}{(24u_B)^2} \right]^k, \ee where
$r_{0L}={m'}_B^2+12u_BM_B^2$ is the longitudinal bare mass, the
transverse one $r_{0T}={m'}_B^2+4u_BM_B^2$ being included in the
parameter $\bar{\omega}=r_{0T}/r_{0L}$. The nonanalyticity in the
coupling constant is seen in the last term.

The functions $F_{blk}$ can be found in \cite{dohm98,dohm97}.
Since we need them later on in this Appendix, we shall write the
nonzero components:
\beqn
F_{100}
&=& -\f{1}{12\pi}\left[ 1+(N-1)\bar{w}^{3/2} \right],
\label{F100}
\\
F_{200}
 &=& \f{1}{384\pi^2} \left[3+2(N-1)\bar{w}^{1/2}+(N^2-1)\bar{w}
            \right],
\label{F200}\\
F_{210}
 &=& \f{1}{288\pi^2} (N-1)\ln\frac{1+2\bar{w}^{1/2}}{3},
\label{F210}\\
F_{211}
 &=& -\f{1}{288\pi^2} (N+2).
\label{F211}\\
F_{300}
 &=& \f{1}{18432\pi^3} \Bigglb(15+24\ln\frac{3}{4}
        -(N-1)\bigg\{\bar{w}^{-1/2} +2N -6+8\ln\f{2+2\bar{w}^{1/2}}{3}
\nonumber\\
&& \mbox{}
        +\bar{w}^{1/2} \left[N^2-6N-9
        +4(N+1) \ln\frac{16\bar{w}}{9}
        +8\ln \f{2+2\bar{w}^{1/2}}{3} \right]
        +\bar{w}(N-1) \bigg\} \Biggrb),
\label{F300}\\
F_{301}
 &=& \f{1}{2304\pi^3} \left\{3 +(N-1)\left[ 1 +(N+2)\bar{w}^{1/2}
            \right] \right\},
\label{F301}\\
F_{310}
 &=& \f{1}{27648\pi^3} \Bigglb(9\pi^2-18
            +108\Litwo\left(-\f{1}{3}\right)
 -(N-1)\bigg\{4\bar{w}^{-1/2}   +4N+2 -(N+2)\pi^2 \nonumber\\
&& \mbox{} -12\Litwo\left(\f{1}{3}\right)
            -32\ln2 -6(\ln3)^2
            + \bar{w}^{1/2} \left[10N+32
            -16(2N+3)\ln2 +48\ln3 -8(N+1)\ln\bar{w} \right]
             \nonumber\\
&& \mbox{}  + \frac{1}{3}\bar{w} \Big(84N-100-128\ln2 \Big)
            \bigg\} \Biggrb),\label{F310}
\\
F_{320}
 &=& \f{1}{165888\pi^3} \Bigglb[ 432\ln\frac{4}{3}
       -324\Litwo\left(-\f{1}{3}\right)-432c_1-27\pi^2 \nonumber\\
&&\mbox{}  - (N-1) \Bigglb( 16\bar{w}^{-1/2}+\f{3N+14}{3} \pi^2
       +18(\ln3)^2 +36\Litwo\left(\frac{1}{3}\right) \nonumber\\
&&\mbox{}  +16\left[ c_2 +4\Litwo{-2} -2\Litwo\left(-\f{1}{2}\right)
       +\left(6\ln3-\ln2-\f{13}{3}\right)\ln2\right] \nonumber\\
&&\mbox{} -\f{128}{3}
      +16\bar{w}^{1/2} \left[7-N+(N+1)\ln(16\bar{w}) +2\ln2-6\ln3
      \right] \nonumber\\
&&\mbox{}
      +4\bar{w} \left\{4c_2-12N -\frac{224}{5}+\pi^2
      +6\left(6\ln3 -\ln2 -\frac{16}{15}\right)\ln2 +
 12\left[2\Litwo{-2}-\Litwo\left(-\f{1}{2}\right) \right]\right\}
      \Biggrb) \Biggrb].
\label{F320}
\eeqn
where, in the  coefficients $F_{310}$ and $F_{320}$ below $T_c$, terms of
order $O(\bar{w}^{3/2},\bar{w}^{3/2}\ln\bar{w})$
 have been neglected (the coefficients are calculated
in the vicinity of the coexistence curve where an expansion
with respect to $\bar{w}$ is justified). The constants $c_1$ and $c_2$ have
been defined in the main text, see Eqs.~(\ref{constc1}) and~(\ref{constc2}).

Inverting the equation
of state
\be
h_B=\f{\p}{\p M_B}\Gamma_B
\label{eos}
\ee
gives, in the limit $h_B\rightarrow0$, and to this order, the square of the
magnetization:
\beqn
M_B^2 &=& \f{1}{8u_B}(-2{m'}_B^2)+\f{3}{4\pi}(-2{m'}_B^2)^{1/2}
+\f{u_B}{8\pi^2} \left[10-N +4(N-1)\ln 3 -2(N+2)
          \ln \f{-2{m'}_B^2}{(24u_B)^2} \right]
          \nonumber\\
      &&  \mbox{}
          +\f{u_B^2(-2{m'}_B^2)^{-1/2}}{1920\pi^3} \,\Bigg\{
          -2736N -5904 -6480c_1 +240(N-1)c_2-(75N^2-5N+875)\pi^2\nonumber\\
      &&  \mbox{}
      -1260 \left
            [ (N-1)\Litwo\left(\frac{1}{3}\right)
          +9\Litwo\left(-\f{1}{3}\right) \right]
 +960(N-1) \left[2\Litwo\left(-2\right)
            -\Litwo\left(-\frac{1}{2}\right)  \right]
          \nonumber\\
      &&  \mbox{}
          -630(N-1) (\ln 3)^2     -48\ln2 \Big[
          10(N-1)\ln2 -60(N-1)\ln3 +111N-561 \Big]\nonumber\\
 &&\mbox{}+240(12N-57)\ln3
          -1440(N+2) \ln \f{-2{m'}_B^2}{(24u_B)^2}\Bigg\}.
\label{magnetization}
\eeqn

The logarithmic terms in $u_B$ are nonanalyticities which can be
removed using the length  $\xi_-$  instead of ${m'}_B^2<0$, see
\cite{dohm90a}. Up to the three-loop order, the relation between
$\xi_-$ and ${m'}_B^2<0$ is
\beqn
-2{m'}_B^2&=&
\xi_-^{-2}\bigg\{1+\f{N+2}{\pi}u_B\xi_--\f{N+2}{\pi^2}(u_B\xi_-)^2
\left[\f{1385}{108}+4\ln(24u_B\xi_-)\right]\nonumber\\
&&\mbox{}+\f{N+2}{108\pi^3}(u_B\xi_-)^3\left[
3(438N+4349)+576(N+8)\Litwo\left(-\f{1}{3}\right)+48(N+8)\pi^2+8(43N+182)
\ln\f{3}{4} \right]\bigg\}. \label{mBasximinus} \eeqn

Using~(\ref{mBasximinus}) in~(\ref{magnetization}) one obtains an
analytic function of $u_B$, from which one extract the amplitude
$f_{\phi}$ of Eqs.~(\ref{2loopsfphi}) and~(\ref{3loopsfphi}),
after proper normalization with the help of $\Zphi$. This has
been done in~\cite{dohm99,dohm97} and will not be repeated here.
The equations we have quoted here are mentioned because we shall
need them below for  obtaining the amplitude functions
$f_{\chi_+}$ and $f_{\chi_-}$ which, to our knowledge, have not
been determined analytically within this model.

\section{Three-loop amplitude function of the isotropic
susceptibility
  above
  $T_c$}
\label{appendixfchiplus}

By definition, the amplitude of the susceptibility above $T_c$ is
obtained from the susceptibility at zero momentum
$f_{\chi_+}^B=\xi_+^2\chi_{+,B}^{-1}$, where the inverse
susceptibility is given by the two-point function $\Gamma_B^{(2)}$
at zero momentum.  The  correlation length above the critical
temperature $\xi_+$ is defined as in
Refs.~\cite{dohm98,dohm99,dohm97}:
\be
\xi_+^2=\left.\chi_{+,B}(q)\p\chi_{+,B}^{-1}(q)/\p q^2
 \right|_{q^2=0}.
 \label{barechiplus}
\ee
Combining with the definition of $f_{\chi_+}^B$, we have
\be f_{\chi_+}^B=\left.\f{\p \chi_{+,B}^{-1}}{\p
q^2}\right|_{q^2=0}=\left.\f{\p \Gamma_B^{(2)}}{\p
q^2}\right|_{q^2=0}. \label{fchiB} \ee
The derivative of $\Gamma_B^{(2)}$ with respect to $q^2$ is
needed. This is in contrast to Refs.~\cite{dohm98,dohm99,dohm97}
where only the combination~(\ref{barechiplus}) was needed. For
this reason, the intermediate result leading to Eq.~(\ref{fchiB})
was not published . Being needed to determine the ratio $R_C$
in~(\ref{ratioRC}) and the ratio of the
susceptibilities~(\ref{ratiosusc}), we derive it in the following.
The two-point function can be written as
$\Gamma_B^{(2)}=r_0+q^2-\Sigma_B(q,r_0,\ubar_B)$, where the
self-energy has the expansion
$\Sigma_B(q,r_0,\ubar_B)=\sum_{m=1}^{\infty}(-\ubar_B)^m\Sigma_B^{(m)}(q,r_0)$.
The two-loop results have first been given in Appendix~A of
Ref.~\cite{dohm97}, with the result, up to order $q^2$:
\be \Gamma^{(2)}=q^2+r_0-4(N+2)A_D u_B r_0^{1/2}+8(N+2)^2A_D^2
u_B^2-32\f{(N+2)}{(4\pi^3)}\left[
\f{2\pi}{D-3}-\f{2\pi}{27}\f{q^2}{r_0}\right]u_B^2.
 \ee
The pole at $D=3$ can be eliminated by subtraction, leading to
the masses $m_B^2$ and ${m'}_B^2$. This is however of no concern
here since we are interested in
 taking the derivative with
respect to $q^2$:
\be \left.\f{\p \Gamma_B^{(2)}}{\p
  q^2}\right|_{q^2=0}=1+\f{N+2}{27\pi^2}\f{u_B^2}{r_0}.
\label{fchi2loops} \ee
For the three-loop expansion, one must calculate the diagrams in
 Appendix~B of \cite{dohm99}. Again,
we concentrate on the derivative of the susceptibility at zero
momentum, focusing on the diagrammatic Eq.~(B5) of \cite{dohm99}.
The corresponding vacuum diagrams have been given by Rajantie in
\cite{rajantie96}, see in particular its Eqs.~(15) and~(25) and,
taking the appropriate derivative with respect to the mass, we
obtain the contribution of the three-loop diagrams:
\be \left.\f{\p \Sigma_B^{(3)}}{\p q^2}\right|_{q^2=0}=\left\{
\f{(N+2)^2}{27\pi^3}-\f{(N+2)(N+8)}{54\pi^3}\left[
-8+3\pi^2+32\ln\f{3}{4}+36\Litwo\left(-\f{1}{3}\right) \right]
\right\}r_0^{-3/2}. \ee
This has to be combined with (\ref{fchi2loops}) to yield the
expansion
\be \left.\f{\p \Gamma_B^{(2)}}{\p
  q^2}\right|_{q^2=0}=1+\f{N+2}{27\pi^2}\f{u_B^2}{r_0}+
\left\{ \f{(N+2)^2}{27\pi^3}-\f{(N+2)(N+8)}{54\pi^3}\left[
-8+3\pi^2+32\ln\f{3}{4}+36\Litwo\left(-\f{1}{3}\right) \right]
\right\}\f{u_B^3}{r_0^{3/2}}. \label{eqgammaB2} \ee

Since there is no linear term $u_B$, the amplitude  of the
susceptibility above $T_c$ to three loops requires only  the
one-loop order of the  correlation-length $\xi_+$, i.e., the
one-loop order of the susceptibility. To three loops, the
following expression
\beqn
{m'}_B^2&=&\xi_+^{-2}\bigg\{1+\f{N+2}{\pi}u_B\xi_++\f{N+2}{\pi^2}(u_B\xi_+)^2
\left[\f{1}{27}+2\ln(24u_B\xi_+)\right]\nonumber\\
&&\mbox{}+\f{N+2}{\pi^3}(u_B\xi_+)^3\left[
3(3N+22)-144(N+8)\Litwo\left(-\f{1}{3}\right)-12(N+8)\pi^2-2(43N+182)
\ln\f{3}{4} \right]\bigg\}. \label{mBasxi} \eeqn
is found in the literature, see \cite{dohm99}. This is the
analogue of~(\ref{mBasximinus}) above $T_c$. At the one-loop
level, there is no distinction between $r_0$ and ${m´}_B^2$, and
we identify $r_0=\xi_+^{-2}[1+(N+2)u_B\xi_+/\pi]$. Together with
Eq.~(\ref{eqgammaB2}), we arrive at
\be f_{\chi_+}^B\equiv\left.\f{\p \Gamma_B^{(2)}}{\p
  q^2}\right|_{q^2=0}=1+\f{N+2}{27\pi^2}(u_B\xi_+)^2
-\f{(N+2)(N+8)}{54\pi^3}\left[
-8+3\pi^2+32\ln\f{3}{4}+36\Litwo\left(-\f{1}{3}\right)
\right](u_B\xi_+)^3. \label{barefchi3loops}
\ee
This has to be compared with the numerical coefficients
$a_m^{(2)}$ of Table~2 in \cite{dohm90b}.

Having calculated the bare amplitude function $f_{\chi_+}^B$, we
can now turn to the normalized one $f_{\chi_+}$. Since the latter
is related to a two-point function, the normalization factor is
equal to the wave function renormalization constant $\Zphi$:
$f_{\chi_+}=\Zphi f_{\chi_+}^B$, with $\Zphi$ being supplied
by~(\ref{phiseriesgB}). Using the relation~(\ref{ubardef})
between $u_B$ and $\ubar_B=u_B\xi_+/(4\pi)$, we arrive at the
normalized amplitude function of the susceptibility above $T_c$,
expressed in terms of the reduced bare coupling constant
$\ubar_B$:
\be f_{\chi_+}=1-\f{92}{27}(N+2)\ubar_B^2-\f{8}{27}
(N+2)(N+8)\left[
-113+12\pi^2+128\ln\f{3}{4}+144\Litwo\left(-\f{1}{3}\right)
\right]\ubar_B^3. \label{fchi3loops} \ee
The corresponding expansion  in terms of the renormalized coupling
constant gives
\be f_{\chi_+}=1-\f{92}{27}(N+2)u^2-\f{8}{27} (N+2)(N+8)\left[
-21+12\pi^2+128\ln\f{3}{4}+144\Litwo\left(-\f{1}{3}\right)
\right]u^3, \label{renfchiplus}
\ee
where we used~(\ref{gseriesgB}). Contrary to~(\ref{fchi3loops})
which is well-behaved regarding strong-coupling theory,
Eq.~(\ref{renfchiplus}), which coincide with the numerical
coefficients $c_m^{(2)}$ of Table~4 of \cite{dohm90b}, is
problematic when considering the Borel resummation scheme: All its
coefficients are negative. For this reason, we have not been able
to reproduce the Borel resummation made in Ref.~\cite{dohm90b}.
We shall however make, in the main text,  a comparison between
the strong-coupling limit of~(\ref{fchi3loops}) and the
resummation performed in \cite{dohm90b}.

\section{Three-loop amplitude function of the
$N=1$-susceptibility below
  $T_c$}
\label{appendixfchiminus}

In Ref.~\cite{dohm92}, the amplitude function of the
susceptibility below $T_c$ for $N=1$ has been calculated
numerically to five loops. We have quoted in~(\ref{fchibelowTc})
the corresponding two-loop part. This amplitude function enters
the ratio of the susceptibilities~(\ref{ratiosusc}). Since
$f_{\chi_+}$ has been obtained to three loops in the previous
section, it is also interesting to obtain $f_{\chi_-}$
analytically: The ratio~(\ref{ratiosusc}) will thus be analytical.

 In Ref.~\cite{dohm99},  the free energy
$\Gamma_B$ has been given analytically up to three loops. We
shall use this knowledge to determine $f_{\chi_-}$. We have
recalled the relevant equations in the first part of this
Appendix, which have to be evaluated for $N=1$ and
$\bar{\omega}=0$. The derivative of the free energy with respect
to the magnetization leads to the equation of state~(\ref{eos})
which can be inverted to obtain the magnetization~\cite{dohm99}.
We have recalled its expression in~(\ref{magnetization}). The
equation of state can itself be derived with respect to the
magnetization, defining the inverse susceptibility below $T_c$:
$\chi_{-,B}^{-1}=\p h_B/\p M_B$. Only at this stage is the
external field $h_B$ taken to be vanishing. The  length $\xi_-$
\cite{dohm99}, which we recalled in~(\ref{mBasximinus}), is then
used to remove the nonanalyticity coming from logarithms of the
coupling constant. Doing so, and using the magnetization given by
the equation of state, we have been able to obtain the inverse
bare susceptibility below $T_c$:
$\chi_{-,B}^{-1}=\xi_-^{-2}f_{\chi_-}^B$, with
\be f_{\chi_-}^B=
1+\f{9}{2\pi}(u_B\xi_-)-\f{1061}{36\pi^2}(u_B\xi_-)^2 +\left[
19472-11664c_1+3\pi^2+10480\ln\f{4}{3}+36\Litwo\left(-\f{1}{3}\right)
\right]\f{(u_B\xi_-)^3}{64\pi^3}. \ee
Numerically, this expansion
reads
\be 1+1.43239(u_B\xi_-)-2.98616(u_B\xi_-)^2+11.2134(u_B\xi_-)^3.
\ee This result  agrees perfectly with the numerical expansion
given in the last column of Table~2 in \cite{dohm92}. Using the
relation~(\ref{gseriesgB}) between $u_B$ and $\ubar_B$, we obtain
\be f_{\chi_-}^B= 1+18\ubar_B-\f{4244}{9}\ubar_B^2+ \left[
19472-11664c_1+3\pi^2+10480\ln\f{4}{3}+36\Litwo\left(-\f{1}{3}\right)
\right]\ubar_B^3. \label{barefchi-}
\ee
The
renormalized version of~(\ref{barefchi-}) is found by
multiplying it with $\Zphi$ from~(\ref{phiseriesgB}):
\be
f_{\chi_-}=1+18\ubar_B-\f{4352}{9}\ubar_B^2+ \left[
19904-11664c_1+3\pi^2+10480\ln\f{4}{3}+36\Litwo\left(-\f{1}{3}\right)
\right]\ubar_B^3.
\label{renfchi-}
\ee
This is the amplitude to
be evaluated in the strong-coupling limit and entering the
 ratio~(\ref{ratiosusc}).

The bare amplitudes~(\ref{barefchi3loops}) and~(\ref{barefchi-})
might as well be chosen to enter the amplitude
ratio~(\ref{ratiosusc}) since the renormaliazation constant
$\Zphi$ drops out, being the same above and below $T_c$. We have
however chosen to work with the renormalized
quantities~(\ref{fchi3loops}) and~(\ref{renfchi-}).

For completeness, we also state the expansion of $f_{\chi_-}$
 in terms of the renormalized coupling constant.
Using~(\ref{gseriesgB}) and~(\ref{renfchi-}), we obtain
\be
f_{\chi_-}=1+18u+\f{1480}{9}u^2+ \left[
1072-11664c_1+3\pi^2+10480\ln\f{4}{3}+36\Litwo\left(-\f{1}{3}\right)
\right]u^3.
\ee
Taking the inverse of this equation, we recover the coefficients
of the second column of Table~3 of Ref.~\cite{dohm92}.

For an application to the evaluation of the amplitude ratio of the
susceptibilities, we also give the perturbative expansion of the
ratio $f_{\chi_-}/f_{\chi_+}$ at $N=1$.
Combining~(\ref{fchi3loops}) with~(\ref{renfchi-}), we obtain
\be
\f{f_{\chi_-}}{f_{\chi_+}}=
1+18\ubar_B-\f{1420}{3}\ubar_B^2+
\left[
19184-11664c_1+99\pi^2+9456\ln\f{4}{3}+1188\Litwo\left(-\f{1}{3}\right)
\right]\ubar_B^3.
\label{expandedratiosusc}
\ee

\section{Amplitude ratios} \label{appendixampratios}

To get the different amplitude ratios of
Section~\ref{sectionratio}, we make use of the relations
\beqn
\chi_+&=&\Zphi\f{\xi_+^2}{f_{\chi_+}}\exp\left[-
\int_{u(l_+)}^u\f{\gamma_{\phi}}{\beta_u}du'
\right],
\label{chi+}\\
\chi_-&=&\Zphi\f{\xi_-^2}{f_{\chi_-}}\exp\left[-
\int_{u(l_-)}^u\f{\gamma_{\phi}}{\beta_u}du'
\right],\label{chi-}\\
A^{\pm}&=&\f{(b^{\pm})^2}{(\xi_{\pm}^0)^D}\f{A_D}{4}(4\nu B^*+\alpha
F_{\pm}^*),
\label{apm}
\\
\langle\phi_B\rangle^2&=&\Zphi \f{f_{\phi}}{\xi_-^{D-2}}\exp\left[-
\int_{u(l_-)}^u\f{\gamma_{\phi}}{\beta_u}du'
\right],
\label{phib2}
\eeqn
which were derived in \cite{dohm89} for (\ref{chi+}),
\cite{dohm92} for (\ref{chi-}) and~(\ref{phib2}), and
\cite{dohm90a} for (\ref{apm}). All the quantities have been
defined in the main text, except for
\be
l_{\pm}=\exp\left(\int_u^{u(l_{\pm})}\f{du'}{\beta_u}\right),
\ee
 with $u(1)=l_{\pm}$ and the flow parameter chosen as $l_{\pm}\mu\xi_{\pm}=1$,
 and with $\xi_{\pm}=\xi_{\pm}^0|t|^{-\nu}$.

The amplitude ratio of the heat
capacity~(\ref{amplratioCv2loops}) follows trivially
from~(\ref{apm}), while the amplitude ratio for the
susceptibilities~(\ref{ratiosusc}) is a direct consequence
of~(\ref{chi+}) and~(\ref{chi-}). The only missing information is
the ratio $\xi_{+}^0/\xi_-^0$, given explicitly in \cite{dohm90a}
as
\be
\f{\xi_+^0}{\xi_-^0}=\left(\f{b^+}{b^-}\right)^{\nu}.
\label{ratioxi}
\ee

Because our derivation (\ref{ratioRC}) of the universal
combination $R_C$ does not coincide with~(\ref{ratioRCdohm})
derived by the authors of \cite{dohm99}, we reproduce below our
calculation. We need the amplitude $A_M$, related
to~(\ref{phib2}) by \cite{aharony91} $\langle\phi_B\rangle\equiv
M_B\approx A_M|t|^{\beta}$. We deduce
\be
A_M^2=\left.\Zphi \f{f_{\phi}}{(\xi_-^0)^{(D-2)}}|t|^{\nu(D-2)-2\beta}\exp\left[
-\int_{u(l_-)}^u\f{\gamma_{\phi}}{\beta_u}
\right]\right|_{l_-\rightarrow0},
\ee
where we have specified that the right-hand-side is evaluated at the critical
point.

In the same way, the amplitude of the susceptibility is obtained
from~(\ref{chi+}) using \cite{aharony91}
$\chi_+\approx\Gamma^+|t|^{-\gamma}$. We deduce
\be
\Gamma^+=\left.\Zphi\f{(\xi_+^0)^2}{f_{\chi_+}}|t|^{\gamma-2\nu}\exp\left[
-\int_{u(l_+)}^u\f{\gamma_{\phi}}{\beta_u}
\right]\right|_{l_+\rightarrow0}.
\ee
Taking the ratio $\Gamma^+/A_M^2$, we have directly
\be
\f{\Gamma^+}{A_M^2}=\f{(\xi_+^0)^2}{f_{\chi_+}^*f_{\phi}^*}
(\xi_-^0)^{(D-2)}.
\ee
The dependence in $|t|$ has disappeared, as it should, due to the
identity $\gamma-2\nu+2\beta-\nu(D-2)=0$.

Combining with the definition of $A^+$ in~(\ref{apm}), we get
\be
R_C\equiv\f{\Gamma^+A^+}{A_M^2}=\f{(b^+)^2}{f_{\chi_+}^*f_{\phi}^*}
\left(\f{\xi_-^0}{\xi_+^0}\right)^{(D-2)}\f{A_D}{4}(4\nu B^*+\alpha
F_{\pm}^*)=\f{(b^+)^{2-\nu(D-2)}}{(b^-)^{-\nu(D-2)}}\f{A_D}{4}(4\nu B^*+\alpha
F_{\pm}^*)\f{1}{f_{\chi_+}^*f_{\phi}^*}.
\ee
where we used~(\ref{ratioxi}) to obtain the last equality. Using
$b^+=2\nu P_+$ and $b^-=3/2-2\nu P_+$ \cite{dohm90a}, as well as
$A_3=1/(4\pi)$ from~(\ref{defAD}), we arrive to the amplitude
ratio $R_C$ given in~(\ref{ratioRC}).

\section{Determination of the polynomial $P_+$ to three loops}
\label{appendixpplus}

In this section, we want to derive the analytical expression for
the polynomial $P_+$ up to three loops. It has been given
numerically, and resummed, for $N=1,2,3$ up to five loops
in~\cite{dohm90b}, so that our analytical result will have to
match this reference. Above $T_c$, the relation between ${m'}_B^2$
and the correlation length has been given in~(\ref{mBasxi}) at the
three-loop level. A polynomial $P^B_+$ in powers of $u_B$ is
defined through the relation
\be P^B_+=\p {m'}_B^2/\p \xi_+^{-2}, \label{bareP0} \ee
leading to
\beqn
P^B_+&=&1+\f{N+2}{2\pi}(u_B\xi_+)-\f{N+2}{\pi^2}(u_B\xi_+)^2\nonumber\\
&&\mbox{}+\f{N+2}{108\pi^3}\left[
-3(3N+22)+12(N+8)\pi^2+2(43N+182)\ln\f{3}{4}+144(N+8)\Litwo
\left(-\f{1}{3}\right) \right](u_B\xi_+)^3. \eeqn
The numeric coefficient $b_m$ of Table~2 of \cite{dohm90b}
coincide perfectly, up to three loops, with our analytical
expression, which has the advantage of being valid for all $N$.
Its renormalized counter part is defined by
\be P_+=Z_r^{-1}P^B_+, \label{renP} \ee
 where the renormalization constant $Z_r^{-1}$ has
been given to three loops in Eq.~(\ref{mseriesgB}).
The corresponding  power series in $\ubar_B$ follows readily:
\beqn
P_+&=&1-2(N+2)\ubar_B+4(N+2)(2N+17)\ubar_B^2\nonumber\\
&&\mbox{}+\f{8}{27}(N+2)\left[-3(36N^2+837N+3920)+24(N+8)\pi^2
+4(43N+182)\ln\f{3}{4}+288(N+8)\Litwo\left(-\f{1}{3}\right)
\right]\ubar_B^3, \label{renPplusgB} \eeqn
where we have used the relation between $u_B$ and $\ubar_B$ given
in Eq.~(\ref{ubardef}), the scale $\mu$ being identified with the
inverse of the correlation length: $\ubar_B=u_B\xi_+A_3$.
Eq.~(\ref{renPplusgB}) is the polynomial whose strong-coupling
expansion has to be calculated. The corresponding power series in
the renormalized coupling constant $u$ follows from
Eq.~(\ref{gseriesgB}):
\beqn
P_+&=&1-2(N+2)u+4(N+2)u^2\nonumber\\
&&\mbox{}+\f{8}{27}(N+2)\left[-3(63N+572)+24(N+8)\pi^2
+4(43N+182)\ln\f{3}{4}+288(N+8)\Litwo\left(-\f{1}{3}\right)
\right]u^3. \eeqn
The reader can verify that the analytical result coincide, for $N
=1,2,3$, with the numerical values in Table~4 of
Ref.~\cite{dohm90b}. It differs only in the fifth decimal place
of the cubic term $c_{P3}$ of this table.


\newpage




\begin{thebibliography}{99}
\bibitem{conventions}
Note that the calculation we do in this paper are based on the
normalization of the coupling constant of the Aachen group
\cite{dohm99,dohm98,dohm97} which differ from Kleinert´s
\cite{kleinert263,ks}. The difference in conventions disappears
at the level of the exponents, and the value of $u^*$ is only
different by a factor 12, as mentioned after Eq.~(\ref{EqZphi}).
\end{thebibliography}
\end{document}